\documentclass[11pt,a4paper]{article}
\usepackage[pagella]{preprint-style}
\usepackage[a4paper,margin=25mm]{geometry}
\usepackage{graphicx,booktabs,array}
\usepackage[export]{adjustbox}
\usepackage{xurl,hyperref}
\hypersetup{hypertexnames=false}
\DeclareUnicodeCharacter{2212}{\ensuremath{-}}
\newcommand{\NoteLeft}{\leftskip=0pt\rightskip=0pt\parfillskip=0pt plus 1fil\relax}
\newcommand{\ParenEnum}{\renewcommand{\labelenumi}{(\arabic{enumi})}}
\newsavebox{\MainTextTableBox}
\setcitestyle{authoryear,round,aysep={},citesep={;},notesep={, }}
\makeatletter
\renewcommand{\NAT@spacechar}{\space}
\makeatother
\makeatletter
\newcommand{\AppendixSectionEntry}[2]{%
  \par\addvspace{.65em}%
  \begingroup\bfseries
  \@dottedtocline{1}{0pt}{6.5em}{\numberline{Appendix~#1}\hyperref[sec:#1]{#2}}{\pageref{sec:#1}}%
  \endgroup}
\newcommand{\AppendixSubsectionEntry}[2]{%
  \@dottedtocline{2}{1.5em}{2.75em}{\numberline{#1}\hyperref[sec:#1]{#2}}{\pageref{sec:#1}}}
\makeatother

\DeclareUnicodeCharacter{03C3}{\ensuremath{\sigma}}
\newcommand{\RepeatedNumber}{}

\theoremstyle{plain}
\newtheorem{fixedassumption}{Assumption}

\theoremstyle{plain}
\newtheorem{fixedtheorem}{Theorem}

\theoremstyle{plain}
\newtheorem{fixedproposition}{Proposition}

\theoremstyle{definition}
\newtheorem{fixedexample}{Example}

\theoremstyle{definition}
\newtheorem{fixedremark}{Remark}

\theoremstyle{plain}
\newtheorem{fixedlemma}{Lemma}

\theoremstyle{plain}

\theoremstyle{definition}

\begin{document}
\title{What Should an Experiment Measure? Combining a Short-Term Experiment with Confounded Observational Data to Identify Long-Term Treatment Effects\tnoteref{tn}}
\tnotetext[tn]{Date: \today. I am grateful to Ryo Okui for his continued guidance and support. All errors are my own.}
\author{Kentaro Kawato\fnref{fa}}
\address{Graduate School of Economics, The University of Tokyo}
\fntext[fa]{\raggedright Graduate School of Economics, The University of Tokyo, 7-3-1 Hongo, Bunkyo-ku, Tokyo 113-0033, Japan.\\ Email address: \href{mailto:kawato-kentaro380@g.ecc.u-tokyo.ac.jp}{kawato-kentaro380@g.ecc.u-tokyo.ac.jp}}
\makeatletter\pps@maketitle\makeatother
\begin{abstract}
Observational data often record long-term outcomes but do not identify long-term treatment effects when selection depends on unrecorded characteristics, while randomized experiments remove this selection but require years of follow-up. We ask what a short-term experiment should measure so that, combined with observational data, it identifies the average long-term treatment effect in the observational population. In our main setting, the observational data record a coarse post-treatment measurement of the unrecorded characteristics but not a detailed measurement that would remove confounding. In this setting, the experiment should (1) collect the same coarse measurement, (2) measure the characteristics in enough detail that conditioning on this detailed measurement and recorded covariates removes confounding and leaves the coarse measurement uninformative about selection and the treatment-specific long-term outcome mean, and (3) distinguish no more levels of the characteristics than the coarse measurement can separate. We characterize the treatment effect through linear inverse problems based on exclusion restrictions and completeness, and thereby show that combining such an experiment with the observational data identifies the effect. We propose an estimator that is based on the efficient influence function and removes the first-order effect of nuisance estimation errors. We give conditions under which this estimator is asymptotically normal and attains the semiparametric efficiency bound. When the exclusion restriction for the long-term outcome mean may not be plausible, we also characterize sharp bounds on the effect by an optimal transport problem. In simulations and a Job Corps illustration, the proposed Tikhonov-penalized series estimator has smaller absolute bias and coverage closer to the nominal level than competing estimators.
\end{abstract}

\section{Introduction}\label{sec:1}

Many policy decisions turn on the long-term effects of a treatment, such as the effect of job training on earnings several years later or the effect of smaller classes on high school graduation. For many policies, observational data that record the long-term outcome already exist, but these data do not identify the long-term effect because selection into treatment depends on characteristics that the data do not record. Administrative records, for instance, link past job-training spells to later earnings and past class sizes to high school graduation. Workers may enroll in training when their earnings prospects are poor, and schools may place students who need academic support in small classes. Treated and untreated units in these data then differ in characteristics that also affect the long-term outcome, so comparisons between them remain biased after adjustment for the recorded covariates.

This paper asks what an experiment should measure so that, combined with existing observational data, it identifies the long-term effect. The main constraint on such an experiment is time. A randomized experiment removes selection into treatment, but following its participants until the long-term outcome is realized takes years. What a short-term experiment contributes therefore depends on what it measures, and the researcher who designs the experiment chooses these measurements depending on the observational data. We study the combination of the two data sources as a problem of experimental design that overcomes the limitations of the observational data, and we give conditions on the measurements under which a short-term experiment and the observational data together identify the average treatment effect on the long-term outcome. Figure 1 summarizes this design problem. Recent work also combines short-term experiments with observational data to estimate long-term effects \citep{ref4,ref5}, and we discuss its relation to our results at the end of this section.

\begin{figure}[htbp]
\centering
\includegraphics[max width=\textwidth]{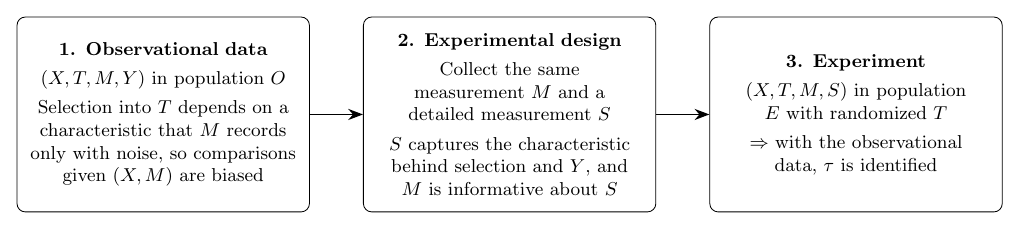}
\setcounter{figure}{0}
\caption{The observational data, the experimental design, and the experiment}
\label{figure:1}
\par\smallskip{\small\NoteLeft Note: $X$ denotes pretreatment covariates, $T$ the treatment, $M$ the coarse measurement shared by the two data sources, $S$ the detailed measurement of the characteristic, and $Y$ the long-term outcome. $O$ and $E$ denote the observational and experimental populations, and $\tau$ is the average treatment effect on $Y$ in the observational population.\par}
\end{figure}

What the experiment should measure depends on what the observational data record after treatment, and we consider observational data that record, in addition to the long-term outcome, a coarse post-treatment measurement of each unit, such as a routine standardized test score or the share of weeks worked in administrative employment records. Let $G\in\{E,O\}$ index the experimental and observational populations, $X$ denote pretreatment covariates, $T\in\{0,1\}$ the treatment, and $Y$ the long-term outcome. We write $M$ for the coarse post-treatment measurement and $S$ for a detailed post-treatment measurement of each unit that only the experiment collects. The experimental data record $(X,T,M,S)$ and the observational data record $(X,T,M,Y)$, so the detailed measurement and the long-term outcome are recorded in different samples.

We show that the experiment should collect the same coarse measurement $M$ that the observational data record and, in addition, the detailed measurement $S$. Our main identification result gives conditions under which these data identify the average treatment effect in the observational population, $\tau=\mathbb E[Y(1)-Y(0)\mid G=O]$, where $Y(t)$, $S(t)$, and $M(t)$ denote the potential values under treatment $t$. Besides consistency and overlap, the conditions require that the treatment be randomized in the experiment given $X$ and that, given $X$, the joint distribution of $(S(t),M(t))$ be the same in the experimental and observational populations. Three further conditions carry the identification.

The first condition is latent unconfoundedness given the detailed measurement,
\begin{equation}\tag{1.1}\label{eq:1.1}
T\perp Y(t)\mid S(t),X,G=O
\end{equation}
for each treatment $t$, which states that units in the observational population with the same covariates and the same value of $S(t)$ are comparable across treatment arms. Condition (1.1) suggests one interpretation of $S$. If the detailed measurement captures a latent characteristic of each unit that drives both selection into treatment and the long-term outcome, units with the same value of this characteristic are comparable, and we therefore call $S$ the characteristic. In the job-training example, $S$ is a detailed measurement of the earnings capacity of a worker after the program, and condition (1.1) requires that, once this capacity and the covariates are fixed, the decision to enroll be unrelated to the potential earnings. In the education example, $S$ is a detailed assessment of the ability of a student, and condition (1.1) requires that, once ability and the covariates are fixed, placement in a small class be unrelated to the potential graduation outcome.

The other two conditions link the characteristic to the coarse measurement. The second condition consists of two exclusion restrictions in the observational population. The first exclusion restriction states that, among units with the same covariates and the same characteristic, the probability of receiving a treatment does not vary with the coarse measurement under that treatment. Because the coarse measurement is taken after treatment, selection cannot respond to it directly, and the restriction requires that, once the characteristic is fixed, it carry no information about selection. In the education example, two students of the same ability are then equally likely to be placed in a small class whatever their routine test scores. The second exclusion restriction states that, among units with the same treatment, covariates, and characteristic, the mean of the long-term outcome does not vary with the coarse measurement. In the education example, the routine score then has no bearing on graduation once ability is fixed. Both restrictions hold when the coarse measurement is a noisy record of the characteristic and the noise affects neither selection nor the long-term outcome. The third condition is completeness, which requires the conditional distribution of $S(t)$ given $M(t)$ in the observational population to vary enough with $M(t)$. When $S(t)$ and $M(t)$ take finitely many values, completeness holds if and only if the matrix of the conditional probabilities of the values of $S(t)$ given the values of $M(t)$ has full column rank, so $M(t)$ must take at least as many values as $S(t)$.

A measurement error model gives one special case in which these conditions hold. Let $U_1$, $U_2$, and $U_3$ be mutually independent given $X=x$ and $G=O$, where $U_1$ represents the latent characteristic of a unit, $U_2$ is measurement error, and $U_3$ is a disturbance of the long-term outcome. For each $t\in\{0,1\}$, suppose that
\[
\begin{aligned}
&S(t)=f(t,U_1),\\
&M(t)=g(t,U_1,U_2),\\
&Y(t)=h(t,U_1,U_3),
\end{aligned}
\]
where $f(t,\cdot)$ is one-to-one with a measurable inverse, and that selection into treatment in the observational population depends only on the characteristic in the sense that
\[
P(T=1\mid U_1,U_2,U_3,X=x,G=O)=P(T=1\mid U_1,X=x,G=O).
\]
The characteristic $U_1$ then drives both selection and the long-term outcome, the experiment measures it without error through $S$, and the observational data record it with noise through $M$. Because $M(t)$ contains the noise $U_2$, conditioning on $M(t)$ leaves variation in $U_1$, and latent unconfoundedness given $M(t)$ fails in general. Conditioning on $S(t)$ fixes $U_1$, so latent unconfoundedness (1.1) and both exclusion restrictions hold, and completeness holds under conditions on the distribution of the measurement error $U_2$.

Identification rests on two moment equations that link the two populations through the coarse measurement. For each $t$ and $x$, define the selection probability and the long-term outcome mean in the observational population as functions of the characteristic by
\[
\begin{aligned}
&\pi_t(s,x)=P(T=t\mid S(t)=s,X=x,G=O),\\
&\mu_t(s,x)=\mathbb E[Y\mid T=t,S=s,X=x,G=O].
\end{aligned}
\]
Both functions describe the observational population, but they are functions of the characteristic, which the observational data do not record, and by the exclusion restrictions they do not depend on the coarse measurement. The observational data alone therefore cannot recover them, and the experiment, which records the characteristic together with the coarse measurement, supplies the missing link. Under our conditions, for each $t$, $F_{X\mid G=O}$-almost every $x$, and every bounded measurable function $h$, these functions satisfy
\[
\begin{aligned}
\mathbb E[\mathbf 1\{T=t\}h(M)\mid X=x,G=O]
&=\mathbb E[\pi_t(S,x)h(M)\mid T=t,X=x,G=E],\\
\mathbb E[\mathbf 1\{T=t\}Yh(M)\mid X=x,G=O]
&=\mathbb E[\pi_t(S,x)\mu_t(S,x)h(M)\mid T=t,X=x,G=E].
\end{aligned}
\]
The left sides are moments of the observational data, and the right sides involve the joint distribution of $(S,M)$ in the experimental data for arm $t$. Each choice of $h$ weights the units by a function of the coarse measurement. Because the coarse measurement is excluded from both unknown functions, changing $h$ changes the weight that each value of the characteristic receives without changing the functions, so different choices of $h$ give different equations for the same unknown function. Completeness requires that these equations leave only one solution. Recovering a function from such equations is a linear inverse problem, and nonparametric instrumental variable models are one example \citep{ref28,ref21}. The ratio of the two solutions gives $\mu_t$, and averaging $\mu_1$ and $\mu_0$ over the distribution of the characteristic in the corresponding experimental arm and over the covariates of the observational population gives $\tau$.

These conditions give concrete guidance to a researcher who plans the experiment. Suppose that the observational data are school records that contain a routine standardized test score and high school graduation, and that an experiment randomly assigns an educational intervention. The experiment should then be designed as follows.

\begin{enumerate}\ParenEnum
\item Administer the same routine test and score it in the same way as the school records do, so that the coarse measurement $M$ is shared by the two data sources.
\item Measure the ability of the students in more detail than the routine test, for example through a longer assessment that the researchers administer at the same time, so that, once this detailed measurement $S$ is fixed, the routine score carries no further information about class placement or graduation.
\item Do not distinguish more levels of ability than the routine score can separate. If the researcher records $S$ in a given number of categories, the routine score must take at least as many values in the observational data, so a finer detailed measurement helps only as long as the shared measurement can still distinguish its categories.

\end{enumerate}
For estimation and inference, we assume that the covariates take finitely many values. Because the outcome mean is the ratio of two unknown functions, estimation errors in both functions shift a plug-in estimator of $\tau$ to first order. Our estimator removes this first-order effect by adding correction terms built from two weight functions. The nuisance functions solve linear inverse problems and can be estimated by series methods \citep{ref1,ref6,ref8}, and the simulation and the application use series estimates with a Tikhonov penalty. Tikhonov regularization of linear inverse problems in econometrics is studied by \citet{ref7} and \citet{ref13}, and penalizing the coefficients of a series estimate is related to the penalized sieve minimum distance estimators of \citet{ref10}. When the weight functions are unique and a tangent space condition holds, the resulting influence function is the efficient influence function of $\tau$ in the sense of \citet[Chapter 25]{ref32}, relative to submodels along which the two unknown functions are differentiable in mean square.

Under two kinds of conditions on the estimates of the nuisance functions, the estimator is asymptotically normal with a consistent variance estimator, and it attains the efficiency bound when, in addition, the weight functions are unique and the tangent space condition holds. The first kind are rate conditions on products of estimation errors, which take the form of \citet{ref11} and hold, for example, when each error vanishes faster than the inverse of the fourth root of the sample size, so nonparametric estimates that converge slowly can be used. The second kind control the error from averaging the plugged-in estimates over the same sample that produced them. They hold when, with probability approaching one, the plugged-in functions lie in fixed Donsker classes and their squared errors in fixed Glivenko–Cantelli classes \citep[Chapter 19]{ref32}, and these classes can be too small to contain nonparametric or machine learning estimates. A cross-fitted version, which estimates the nuisance functions on folds that exclude the observations it averages over, reaches the same conclusions under the rate conditions alone, as in the double/debiased machine learning of \citet{ref11}.

Several approaches relax the assumptions of our main setting and partially identify the effect, and we consider one that drops the exclusion restriction for the outcome mean. Without it, the selection probability remains identified, and the mean of each potential outcome is the mean of the observed outcome weighted by the inverse of the selection probability. Within each treatment arm, value of the covariates, and value of the coarse measurement, the data identify the distribution of the outcome and that of the inverse selection probability but not how the two are paired. When nothing else restricts the outcome mean, the largest and smallest values of the mean of their product are therefore the values of optimal transport problems between two distributions on the real line, whose solutions pair the quantiles of the two distributions in the same and in the opposite order. Weighting these values by the observational probabilities of the treatment arm and the coarse measurement given the covariates, summing over the coarse measurement, and averaging over the covariates gives the ends for the mean of each potential outcome, and the bounds on the effect combine the upper end for one arm with the lower end for the other. Restrictions on the outcome mean, such as monotonicity in the characteristic and in the coarse measurement, can be added, and the bounds then solve linear programs that are easy to compute. 

In a simulation with a continuous characteristic in one design and a characteristic that takes three values in the other, the proposed estimator with a Tikhonov penalty has the smallest absolute bias of the estimators that we compare at every sample size, and the coverage of its confidence intervals rises toward the nominal level as the samples grow, while the comparison estimators remain biased. In both designs, selection into treatment in the observational population depends on the characteristic, and the coarse measurement records the characteristic with error. The first comparison estimator is the experimental selection correction of \citet{ref4} computed with the coarse measurement in place of the characteristic, and the designs are cases in which this replacement is insufficient, because units with the same coarse measurement still differ in the characteristic that drives selection. The second comparison estimator is the regression of the outcome on the treatment and the covariates in the observational sample, which is biased by the selection. When the bounds without the exclusion restriction are computed in the same designs, the confidence intervals for the effect cover it in almost every replication in the continuous design, and in the discrete design their coverage rises with the sample size and exceeds the nominal level in the two larger samples.

We illustrate the method with the public data of the National Job Corps Study, in which the offer to enroll in a job-training program was randomly assigned. Because these data come from an experiment, we construct both data sources from them. From the persons in the study, we draw an experimental sample that records categories of year-3 earnings as the characteristic and an observational sample that records weekly earnings in year 4 as the long-term outcome, both with levels of the share of weeks worked in year 3 as the coarse measurement. In the observational sample, persons with lower year-3 earnings under the offer are more likely to receive the offer, a choice motivated by the negative selection into training programs, and this builds selection on the characteristic into the observational data by construction. Compared with the average effect computed from the experiment, the proposed estimator with a Tikhonov penalty has a smaller absolute bias than the comparison estimators, and its confidence intervals have coverage close to the nominal level. In these data, the mean of year-4 earnings differs across the levels of the share of weeks worked among persons with the same earnings category, so the exclusion restriction for the outcome mean fails, and we also compute the bounds that drop it. Computed from all persons in the study, the optimal transport bounds contain both negative and positive values, whereas the bounds that also require the outcome mean to be nondecreasing in the earnings category, a restriction that holds in these data, are positive and thus determine the sign of the effect. In samples of the size that we draw, the confidence intervals for these bounds exclude zero in about half of the replications.

\subsection{Related literature}\label{sec:1.1}

This paper relates to work that combines experimental and observational data to estimate long-term treatment effects. \citet{ref4} consider a short-term outcome observed in both samples, with the long-term outcome observed only in the observational sample and the treatment randomized in the experiment, so that the experimental data record $(X,T,S)$ and the observational data record $(X,T,S,Y)$. Their key identifying assumption is latent unconfoundedness given the short-term outcome \citep[Assumption 4 of][]{ref4}, which takes the form of condition (1.1). Under this assumption and conditions that make the two populations comparable, they identify the average treatment effect in the observational population \citep[Theorem 1 of][]{ref4} and estimate it by the experimental selection correction (ESC). The surrogate index of \citet{ref5} relies instead on surrogacy, under which the treatment and the long-term outcome are independent given the short-term outcomes and the covariates, and its observational data need not record the treatment.

In our setting, the detailed measurement is observed only in the experiment. The observational data record only the coarse measurement $M$, while the experiment records both $M$ and the detailed measurement $S$, so applying their approach with $M$ in place of $S$ would require latent unconfoundedness given $M(t)$. In the measurement error model above, this condition fails because $M$ records the characteristic with noise, whereas condition (1.1) given $S(t)$ holds, and in our simulation and application ESC computed with $M$ is biased. Instead of conditioning on a shared short-term outcome, we recover the outcome mean for each treatment as a function of the characteristic and average it over the distribution of the characteristic in the corresponding experimental arm.

Table 1 compares these data structures. In the settings of \citet{ref4} and \citet{ref5}, the short-term outcome $S$, on which their identifying assumptions condition, is recorded in both the experimental and the observational data. In our setting, the variable $S$ that carries the information on selection is recorded only in the experimental data, and the coarse measurement $M$ is the only post-treatment variable that both data sources record.

\begin{table}[tbp]
\centering
\def\MainTextTableBody{
\begin{tabular}{lccccc}
\toprule
Setting & $X$ & $T$ & $M$ & $S$ & $Y$ \\
\midrule
Latent unconfoundedness \citep{ref4} & $E, O$ & $E, O$ & — & $E, O$ & $O$ \\
Surrogacy \citep{ref5} & $E, O$ & $E$ & — & $E, O$ & $O$ \\
Shared post-treatment measurement $M$ (this paper) & $E, O$ & $E, O$ & $E, O$ & $E$ & $O$ \\
\bottomrule
\end{tabular}
}
\setlength{\tabcolsep}{4pt}\small
\sbox{\MainTextTableBox}{\MainTextTableBody}
\def\MainTextTableFont{small}
\ifdim\wd\MainTextTableBox>\textwidth
  \footnotesize\sbox{\MainTextTableBox}{\MainTextTableBody}
  \def\MainTextTableFont{footnotesize}
\fi
\typeout{TABLE-WIDTH 1 font=\MainTextTableFont; width=\the\wd\MainTextTableBox; limit=\the\textwidth}
\ifdim\wd\MainTextTableBox>\textwidth
  \PackageError{main-text}{Table 1 remains too wide}{Stop and report; no scaling allowed}
\fi
\usebox{\MainTextTableBox}
\setcounter{table}{0}
\caption{Variables recorded in the experimental ($E$) and observational ($O$) data in prior work and in this paper}
\label{table:1}
\par\smallskip{\small\NoteLeft Note: In each cell, $E, O$ indicates that the variable is recorded in both data sources, $E$ that it is recorded only in the experimental data, $O$ that it is recorded only in the observational data, and a dash that the variable is not used in that setting. In the first two rows, $S$ is the short-term outcome of the respective paper.\par}
\end{table}

Supplementing the information on the characteristic that the observational data lack with measurements from another sample connects this paper to the measurement error literature. \citet{ref29} reviews identification and estimation with validation data and with auxiliary variables such as repeated measurements, multiple indicators, and instruments. In our setting, the sample that provides the joint distribution of the detailed and the coarse measurements is a randomized experiment, and we use this joint distribution separately in each treatment arm. Combined with selection into treatment in the observational population and latent unconfoundedness, it identifies the long-term causal mean.

Recovering unknown functions of the characteristic from the variation of an excluded measurement is a linear inverse problem, as in nonparametric instrumental variable models. \citet{ref28} characterize identification of the structural function in these models by completeness of a conditional distribution, \citet{ref7} survey linear inverse problems in structural econometrics, and \citet{ref21} reviews nonparametric instrumental variables estimation for applied research. In our moment equations, the coarse measurement plays the role of the instrument and the characteristic that of the endogenous variable, and the operator that maps a function of the characteristic to its conditional mean given the measurement is identified from the experiment. \citet{ref14} combines completeness with independence between an instrument and selection given the outcome to handle endogenous selection, and our exclusion restriction for selection restricts the coarse measurement in the same way given the characteristic.

Proximal causal inference also identifies causal effects through solutions of integral equations in observed proxies of an unobserved variable \citep{ref24,ref31}. Our weight functions solve integral equations of the same type, with the coarse measurement as the proxy of the characteristic. In our setting, the characteristic is observed in the experimental sample, so the conditional distribution that links the characteristic and the proxy is identified from a second sample. \citet{ref12} derive the efficient influence function of the average treatment effect in proximal causal inference, and our derivation of the efficient influence function follows a similar approach.

In estimation, we plug estimated nuisance functions into a moment whose first-order sensitivity to their estimation errors is zero, as in the double/debiased machine learning of \citet{ref11}, who combine Neyman orthogonal moments with cross-fitting. Because we combine averages over two samples and average the ratio of two unknown functions, the conditions on the estimators take the form of rates for products of the error in the ratio and the errors in the two weight functions. Our main estimator averages over the same sample that estimates the nuisance functions and imposes conditions on this averaging, and our cross-fitted estimator replaces these conditions by sample splitting.

Our bounds without the exclusion restriction for the outcome mean relate to work on partial identification with combined data. \citet{ref18} derive sharp bounds on counterfactual distributions and treatment effects when the outcomes and the covariates are observed in separate data sets, applying the monotone rearrangement inequality, and \citet{ref19} review the use of optimal transport in econometrics, including the characterization of identified sets. In our bounds, the two variables whose pairing is unknown are the outcome and the inverse of the selection probability within each treatment arm, value of the covariates, and value of the coarse measurement, and the experimental observations, reweighted by the selection probability that the two samples identify, give the distribution of the latter. Our confidence intervals cover the effect rather than the interval of bounds, as in \citet{ref22}.

\subsection{Outline}\label{sec:1.2}

Section 2 presents identification with a shared post-treatment measurement, Section 3 presents estimation and inference, and Section 4 bounds the effect without the exclusion restriction for the outcome mean. Section 5 reports a simulation study, and Section 6 illustrates the method with the Job Corps data. Section 7 concludes. The appendix contains the proofs for Sections 2, 3, and 4 and ends with a guide to two supplementary materials.

\section{Identification with a shared post-treatment measurement}\label{sec:2}

This section identifies the long-term treatment effect when the experimental and observational data share a post-treatment measurement. Section 2.1 sets up the model and the target, Section 2.2 states the assumptions, and Section 2.3 gives the identification result.

\subsection{Setting}\label{sec:2.1}

This subsection sets up the two populations, the target, and the functions of the characteristic that the identification result recovers. Two populations, experimental and observational, are indexed by $G\in\{E,O\}$. The experimental data record $(X,T,S,M)$, where $X$ denotes pretreatment covariates, $T\in\{0,1\}$ a binary treatment, $S$ the detailed measurement, and $M$ the coarse but shared measurement. The observational data record $(X,T,M,Y)$, where $Y$ is the long-term outcome. As in the introduction, we call $S$ the characteristic. We write $S(t),M(t),Y(t)$ for the potential values under treatment $t$. The target is the average treatment effect in the observational population,
\begin{equation}\tag{2.1}\label{eq:2.1}
\tau=\mathbb E[Y(1)-Y(0)\mid O]
\end{equation}
where, here and below, $\mid E$ and $\mid O$ denote conditioning on the population and $\mid x,G$ further conditions on $X=x$. The variables take values in Euclidean spaces, and $Y$ is real valued. We write $F$ for conditional distributions and $L^2(F)$ for the space of equivalence classes of functions that are square integrable with respect to $F$.

We describe selection into treatment and the long-term outcome in the observational population as functions of the characteristic, which the observational data lack, through
\begin{equation}\tag{2.2}\label{eq:2.2}
\begin{aligned}
&\pi_t(s,x)=P(T=t\mid S(t)=s,x,O),\\
&\mu_t(s,x)=\mathbb E[Y\mid T=t,S=s,x,O],\\
&\nu_t(s,x)=\pi_t(s,x)\mu_t(s,x)
\end{aligned}
\end{equation}
These functions are features of the distribution of the observational population that involve the missing characteristic. Measuring $S$ and $M$ together in the experiment turns moments of $M$ in the observational sample into equations for them.

\subsection{Identification assumptions}\label{sec:2.2}

This subsection states the assumptions in three groups. The first group makes the joint distribution of the characteristic and the measurement in the experiment informative about the observational population, the second links the recovered outcome mean to the potential outcomes, and the third allows the two unknown functions to be recovered from the shared measurement.

\renewcommand{\thefixedassumption}{2.1}
\begin{fixedassumption}[Consistency]\label{assumption:2.1}

$S=S(T)$, $M=M(T)$, and $Y=Y(T)$.
\end{fixedassumption}

\indent Assumption 2.1 states that the observed characteristic, measurement, and outcome equal the potential values under the treatment actually received, which links the potential values in the assumptions below to the observed data.

\renewcommand{\thefixedassumption}{2.2}
\begin{fixedassumption}[Overlap and positivity]\label{assumption:2.2}

$F_{X\mid O}\ll F_{X\mid E}$. For $F_{X\mid O}$-almost every $x$, $P(T=t\mid x,E)>0$, and $\pi_t(S(t),x)>0$ holds almost surely in the observational population.
\end{fixedassumption}

\indent Assumption 2.2 requires that both treatment arms of the experiment be available at the covariate values of the target population. Positivity of the selection probability allows us to divide the recovered $\nu_t$ by $\pi_t$ to obtain the outcome mean.

\renewcommand{\thefixedassumption}{2.3}
\begin{fixedassumption}[Randomization in the experiment]\label{assumption:2.3}

For each $t$, $T\perp(S(t),M(t))\mid X,G=E$.
\end{fixedassumption}

\indent Assumption 2.3 states that the treatment is randomized within the experiment, possibly with assignment probabilities that depend on $X$.

\renewcommand{\thefixedassumption}{2.4}
\begin{fixedassumption}[Transportability of the joint distribution]\label{assumption:2.4}

For each $t$ and $F_{X\mid O}$-almost every $x$,
\begin{equation}\tag{2.3}\label{eq:2.3}
F_{S(t),M(t)\mid x,O}=F_{S(t),M(t)\mid x,E}
\end{equation}
\end{fixedassumption}

\indent Under Assumptions 2.1, 2.3, and 2.4, the joint distribution of $(S(t),M(t))$ given $X=x$ in the observational population equals the distribution of $(S,M)$ given $X=x$ in the experimental arm $t$. Randomization equates the distribution of the potential values in the experimental population with their distribution in arm $t$, where they are observed, and equation (2.3) carries this distribution over to the observational population. The resulting joint distribution is the known coefficient of both moment equations in Section 2.3.

\renewcommand{\thefixedassumption}{2.5}
\begin{fixedassumption}[Latent unconfoundedness]\label{assumption:2.5}

For each $t$,
\begin{equation}\tag{2.4}\label{eq:2.4}
T\perp Y(t)\mid S(t),X,G=O
\end{equation}
\end{fixedassumption}

\indent Assumption 2.5 states that, given the characteristic under treatment $t$ and the covariates, selection into treatment in the observational population is independent of the long-term potential outcome. It links the outcome mean $\mu_t$ recovered below to the mean of the potential outcome in the observational population. The integrability conditions used with it are stated in Appendix A.

Assumptions 2.1–2.5 are standard in causal inference and parallel the assumptions of \citet{ref4}, with the characteristic in place of their short-term outcome. The next three assumptions are the key assumptions of this paper.

\renewcommand{\thefixedassumption}{2.6}
\begin{fixedassumption}[Exclusion restriction for selection]\label{assumption:2.6}

For each $t$,
\begin{equation}\tag{2.5}\label{eq:2.5}
P(T=t\mid S(t),M(t),X,O)=\pi_t(S(t),X)
\end{equation}
almost surely.
\end{fixedassumption}

\renewcommand{\thefixedassumption}{2.7}
\begin{fixedassumption}[Exclusion restriction for the outcome mean]\label{assumption:2.7}

For each $t$, $\mathbb E[Y\mid T=t,S,M,X,O]=\mu_t(S,X)$ almost surely.
\end{fixedassumption}

\indent Assumptions 2.6 and 2.7 require that, once the characteristic and the covariates are fixed, the shared measurement carry no additional information about selection into treatment $t$ or about the mean of the long-term outcome. Under Assumption 2.6, the variation of $M(t)$ given the characteristic identifies $\pi_t$, and under Assumption 2.7, the same variation in moments that involve the long-term outcome identifies $\nu_t$. In both examples, these assumptions require that $S$ capture the characteristic that drives selection and the long-term outcome, so that the share of weeks worked or the routine test score adds nothing once the detailed measurement is fixed.

\renewcommand{\thefixedassumption}{2.8}
\begin{fixedassumption}[Completeness]\label{assumption:2.8}

For each $t$ and $F_{X\mid O}$-almost every $x$, every $h\in L^2(F_{S(t)\mid x,O})$ satisfies
\begin{equation}\tag{2.6}\label{eq:2.6}
\mathbb E[h(S(t))\mid M(t),x,O]=0
\quad\Longrightarrow\quad h(S(t))=0\quad\text{almost surely}
\end{equation}
\end{fixedassumption}

\indent Assumption 2.8 requires that a square-integrable function of the characteristic be determined by how the distribution of $S(t)$ varies with the value of $M(t)$. It is the completeness condition of \citet{ref28} for nonparametric instrumental variable models, with the characteristic in the role of the endogenous variable and the shared measurement in the role of the instrument. \citet[Theorems 2.2–2.4]{ref28} give primitive conditions for completeness in terms of the conditional distribution of the endogenous variable given the instrument, here that of $S(t)$ given $M(t)$ and $X=x$, for exponential families, for normal distributions with a mean that is linear in the instrument, and for variables with finitely many values. Collecting the detailed and the coarse measurements together in the experiment provides the information under which completeness can hold, as the following example shows.

\renewcommand{\thefixedexample}{2.1}
\begin{fixedexample}[Measurement error model]\label{example:2.1}

Fix $X=x$ in the observational population and suppress $x$ in the functions below. Let $U_1$, $U_2$, and $U_3$ be mutually independent given $X=x$ and $G=O$, where $U_1$ represents the latent characteristic, $U_2$ is measurement error, and $U_3$ is an outcome disturbance. For each $t\in\{0,1\}$, suppose that
\[
\begin{aligned}
&S(t)=f(t,U_1),\\
&M(t)=g(t,U_1,U_2),\\
&Y(t)=h(t,U_1,U_3),
\end{aligned}
\]
where $f(t,\cdot)$ is one-to-one with a measurable inverse, and that selection into treatment in the observational population depends only on the characteristic in the sense that
\[
P(T=1\mid U_1,U_2,U_3,x,O)=P(T=1\mid U_1,x,O).
\]

This structure implies latent unconfoundedness in Assumption 2.5 and the exclusion restrictions in Assumptions 2.6 and 2.7 when it holds for $F_{X\mid O}$-almost every $x$ and both $t$, together with Assumption 2.1 and the integrability conditions in Appendix A. Because $f(t,\cdot)$ is one-to-one, fixing $S(t)$ given $X=x$ and $G=O$ determines $U_1$, so $T$, $M(t)$, and $Y(t)$ are mutually independent given $S(t)$, $X=x$, and $G=O$. The independence of the measurement error alone does not imply Assumption 2.5, and the selection condition above supplies this implication. Since $M$ is measured after treatment, the measurement function $g$ may depend on $t$.

In addition, if for each $t$ and $F_{X\mid O}$-almost every $x$, $S(t)$ is uniformly distributed on a bounded interval given $X=x$ and $G=O$ (for instance, $U_1$ is uniform on a bounded interval and $f(t,u)=a_t+u$) and $M(t)=S(t)+U_2$ with $U_2\sim\mathcal N(0,\sigma^2)$ for $\sigma^2>0$, then completeness in Assumption 2.8 also holds, as proved in Appendix A. More generally, when $M(t)=S(t)+U_2$ with real-valued $S(t)$ and $U_2$ that are independent given $X=x$ and $G=O$, completeness holds for every distribution of $S(t)$, continuous or discrete, if the characteristic function of $U_2$ is nonzero on a dense subset of the real line. Normal errors satisfy this condition, and so do errors that are uniform on an interval, whose characteristic function vanishes only at isolated points. The designs of the simulation in Section 5, with a uniform or a three-point distribution of the characteristic and a uniform measurement error, are of this kind (Appendix A.2).
\end{fixedexample}
\subsection{Identification result}\label{sec:2.3}

This subsection gives the moments that the two samples identify and obtains the average treatment effect from their unique solutions.

\renewcommand{\thefixedtheorem}{2.1}
\begin{fixedtheorem}[Identification with a shared post-treatment measurement]\label{theorem:2.1}

Suppose that Assumptions 2.1–2.8 and the integrability conditions in Appendix A hold. Then, for each $t$, $F_{X\mid O}$-almost every $x$, and every bounded measurable function $h$,
\begin{equation}\tag{2.7}\label{eq:2.7}
\begin{aligned}
\mathbb E[\mathbf 1\{T=t\}h(M)\mid x,O]
&=\mathbb E[\pi_t(S,x)h(M)\mid T=t,x,E],\\
\mathbb E[\mathbf 1\{T=t\}Yh(M)\mid x,O]
&=\mathbb E[\nu_t(S,x)h(M)\mid T=t,x,E]
\end{aligned}
\end{equation}
Each equation has a unique solution in $L^2(F_{S\mid T=t,x,E})$, and $\mu_t=\nu_t/\pi_t$ is identified. Moreover,
\begin{equation}\tag{2.8}\label{eq:2.8}
\tau=\int\!\left\{\int\mu_1(s,x)\,dF_{S\mid T=1,x,E}(s)-\int\mu_0(s,x)\,dF_{S\mid T=0,x,E}(s)\right\}dF_{X\mid O}(x)
\end{equation}
\end{fixedtheorem}

The left sides of (2.7) are known from the observational data, and the distribution of $(S,M)$ on the right sides is known from the experiment. Because the equations use bounded test functions of $M$, they hold for any distribution of $M$, discrete or continuous. The first equation recovers the selection probability, and the second recovers the product of the selection probability and the outcome mean. Applying completeness to both equations and dividing by the positive selection probability yields the outcome mean as a function of the characteristic. Equation (2.8) averages this function over the distribution of the characteristic in each experimental arm and over the covariate distribution of the observational population. The proof is in Appendix A.

Equation (2.7) yields criteria for estimation once finitely many test functions are chosen. If the functions of the characteristic are approximated by a basis, the right sides become a matrix that the experiment identifies, multiplied by unknown coefficients. Section 3 plugs the resulting estimates of the unknown functions, together with estimates of the functions that represent the weights needed for averaging, into an estimator of the average treatment effect and derives its asymptotic distribution.

\renewcommand{\thefixedremark}{2.1}
\begin{fixedremark}[Identification with a pretreatment proxy]\label{remark:2.1}

The same approach applies when the long-term outcome is the only post-treatment variable in the observational data and both samples instead record a pretreatment proxy $Z$ of the characteristic, such as earnings before random assignment in the job-training example or the score on the routine test of the previous school year in the education example. The experiment then measures the characteristic together with the proxy. Randomization in the experiment and transportability between the two populations are stated given the proxy and the covariates, whose distribution may differ between the populations, and latent unconfoundedness given the characteristic and the covariates is maintained. Once the characteristic and the covariates are fixed, the proxy carries no further information about selection into treatment or about the mean of the long-term outcome, the distribution of the characteristic given the proxy and the covariates is the same in the two populations, and the variation of the proxy moves the distribution of the characteristic enough for completeness. The moment equations (2.7) are then replaced by moments conditional on the proxy and the covariates, and $\tau$ averages the outcome mean over the characteristic in each experimental arm given the proxy and the covariates and then over the distribution of the proxy and the covariates in the observational population. Supplementary Material 1 states these conditions and the identification result, gives the efficient influence function and an estimator, reports a simulation study, and gives conditions for identification when the two samples record both a pretreatment proxy and a shared post-treatment measurement.
\end{fixedremark}

\section{Estimation and inference}\label{sec:3}

This section constructs an estimator of the average treatment effect from the identifying moments of Section 2 and gives its efficiency, asymptotic distribution, and standard errors. Section 3.1 states the sampling assumptions, defines the population functions used in estimation, and gives the efficient influence function, Section 3.2 defines the estimator, and Section 3.3 states the conditions and the main theorem. Section 3.4 treats estimation with cross-fitting.

\subsection{Estimand and efficient influence function}\label{sec:3.1}

This subsection first states the sampling assumptions, then writes the average treatment effect as an average over finitely many strata, expresses the first-order effect of errors in the nuisance functions through functions of the observations, and shows that the resulting influence function is the efficient influence function.
Let $n_E$ and $n_O$ be the sizes of the experimental and observational samples, and let $N=n_E+n_O$. The next three assumptions describe the sampling, restrict the covariates to finitely many values, and bound the selection probability from below.

\renewcommand{\thefixedassumption}{3.1}
\begin{fixedassumption}[Sampling]\label{assumption:3.1}

The two samples are independent, the observations in each sample are independent and identically distributed draws from the respective population, and $n_E/N\to\rho$ for some $\rho\in(0,1)$.
\end{fixedassumption}

\renewcommand{\thefixedassumption}{3.2}
\begin{fixedassumption}[Finitely many covariate values]\label{assumption:3.2}
\leavevmode\par
\begin{enumerate}
\item $X$ takes values in a fixed finite set $\mathcal X$.
\item There is a constant $c>0$ such that $P(X=x\mid O)\ge c$ and $P(T=t,X=x\mid E)\ge c$ for all $t,x$.
\end{enumerate}
\end{fixedassumption}

\renewcommand{\thefixedassumption}{3.3}
\begin{fixedassumption}[Selection probabilities bounded away from zero]\label{assumption:3.3}

There is a constant $a\in(0,1)$ such that, for each $t,x$, $\pi_t\ge a$ holds almost surely on the support of $S\mid T=t,x,E$.
\end{fixedassumption}

Assumption 3.1 governs the variation of averages over the two samples. Assumption 3.2 allows each conditional mean given $X=x$ to be estimated by a sample mean within the stratum $x$, so the estimator below averages within strata, and its asymptotic theory relies on the finite support of $X$. The lower bound on the selection probability in Assumption 3.3 keeps the ratio $\mu_t=\nu_t/\pi_t$ and its estimates stable, and the estimators in Section 3.2 truncate the estimated selection probability at the same constant $a$. Boundedness conditions on the outcome are collected in Appendix B and those on the estimates in Appendix C.

For each $t\in\{0,1\}$ and $x\in\mathcal X$, let
\begin{equation}\tag{3.1}\label{eq:3.1}
\bar\mu_t(x)=\mathbb E[\mu_t(S,x)\mid T=t,x,E]=\mathbb E\!\left[\frac{\nu_t(S,x)}{\pi_t(S,x)}\,\middle|\,T=t,x,E\right]
\end{equation}
denote the outcome mean averaged over the distribution of $S$ in the experimental arm $t$. By (2.8) and (3.1), $\tau=\sum_{x\in\mathcal X}P(X=x\mid O)\{\bar\mu_1(x)-\bar\mu_0(x)\}$. The second expression in (3.1) shows that $\bar\mu_t(x)$ is the mean of the ratio of the two unknown functions identified by (2.7), and this ratio determines how estimation errors in the two functions affect an estimator of $\tau$.

An estimator of $\bar\mu_t(x)$ uses estimates of both $\nu_t$ and $\pi_t$, and their errors move $\bar\mu_t(x)$ to first order. Let $d$ be a bounded function of $S$ with $|d|\le D$, and let $\varepsilon$ be a number with $|\varepsilon|D\le a/2$. Replacing $\nu_t$ by $\nu_t+\varepsilon d$ changes $\bar\mu_t(x)$ by
\[
\mathbb E\!\left[\frac{\nu_t(S,x)+\varepsilon d(S)}{\pi_t(S,x)}\,\middle|\,T=t,x,E\right]-\bar\mu_t(x)
=\varepsilon\,\mathbb E\!\left[\frac{d(S)}{\pi_t(S,x)}\,\middle|\,T=t,x,E\right]
\]
Replacing $\pi_t$ by $\pi_t+\varepsilon d$ instead uses the identity $\nu/(\pi+\varepsilon d)-\nu/\pi=-\varepsilon d\mu/\pi+\varepsilon^2d^2\mu/\{\pi(\pi+\varepsilon d)\}$ with $\mu=\nu/\pi$, which gives
\[
\begin{aligned}
&\mathbb E\!\left[\frac{\nu_t(S,x)}{\pi_t(S,x)+\varepsilon d(S)}\,\middle|\,T=t,x,E\right]-\bar\mu_t(x)\\
&=-\varepsilon\,\mathbb E\!\left[\frac{d(S)\mu_t(S,x)}{\pi_t(S,x)}\,\middle|\,T=t,x,E\right]
+\varepsilon^2\,\mathbb E\!\left[\frac{d(S)^2\mu_t(S,x)}{\pi_t(S,x)\{\pi_t(S,x)+\varepsilon d(S)\}}\,\middle|\,T=t,x,E\right]
\end{aligned}
\]
Because $\pi_t\ge a$ and $\pi_t+\varepsilon d\ge a/2$, the last term is at most $2\varepsilon^2D^2\mathbb E[|\mu_t(S,x)|\mid T=t,x,E]/a^2$ in absolute value. An estimation error $d$ in the estimate of $\nu_t$ or $\pi_t$ therefore shifts the plug-in estimate of $\bar\mu_t(x)$ by the mean of $d$ weighted by $1/\pi_t$ or by $-\mu_t/\pi_t$, up to a term of order $\varepsilon^2$. These shifts are of the order of the estimation error, which for nonparametric estimates of the unknown functions vanishes more slowly than $N^{-1/2}$, so the estimator has to remove them.

To remove these shifts, the estimator subtracts or adds terms that are zero at the true functions and can be computed from the two samples. For a square-integrable function $\omega_t^\nu$ of $M$ and $x$, the second equation of (2.7), which extends from bounded to integrable functions of $M$ (Appendix B), gives with $h=\omega_t^\nu(\cdot,x)$
\[
\mathbb E[\nu_t(S,x)\omega_t^\nu(M,x)\mid T=t,x,E]-\mathbb E[\mathbf 1\{T=t\}Y\omega_t^\nu(M,x)\mid x,O]=0
\]

Replacing $\nu_t$ by $\nu_t+\varepsilon d$ changes the left side by $\varepsilon\mathbb E[d(S)\omega_t^\nu(M,x)\mid T=t,x,E]$, because the second mean does not involve $\nu_t$. Subtracting this term from the plug-in estimate of $\bar\mu_t(x)$ leaves the shift $\varepsilon\mathbb E[d(S)\{1/\pi_t(S,x)-\omega_t^\nu(M,x)\}\mid T=t,x,E]$. Since $d$ depends only on $S$, taking the conditional mean given $S$ inside the expectation writes this shift as
\[
\varepsilon\,\mathbb E\!\left[d(S)\left\{\frac1{\pi_t(S,x)}-\mathbb E[\omega_t^\nu(M,x)\mid S,T=t,x,E]\right\}\,\middle|\,T=t,x,E\right]
\]
The remaining shift therefore vanishes for every bounded $d$ and every $\varepsilon$ exactly when $\omega_t^\nu$ satisfies
\[
\mathbb E[\omega_t^\nu(M,x)\mid S,T=t,x,E]=\frac1{\pi_t(S,x)}
\]
almost surely. The same argument with the first equation of (2.7) applies to errors in $\pi_t$. For a function $\omega_t^\pi$ of $M$ and $x$, replacing $\pi_t$ by $\pi_t+\varepsilon d$ changes $\mathbb E[\pi_t(S,x)\omega_t^\pi(M,x)\mid T=t,x,E]-\mathbb E[\mathbf 1\{T=t\}\omega_t^\pi(M,x)\mid x,O]$ by $\varepsilon\mathbb E[d(S)\omega_t^\pi(M,x)\mid T=t,x,E]$, and adding this term cancels the shift $-\varepsilon\mathbb E[d(S)\mu_t(S,x)/\pi_t(S,x)\mid T=t,x,E]$ for every bounded $d$ exactly when $\mathbb E[\omega_t^\pi(M,x)\mid S,T=t,x,E]=\mu_t(S,x)/\pi_t(S,x)$ almost surely. The next assumption requires such functions to exist.

\renewcommand{\thefixedassumption}{3.4}
\begin{fixedassumption}[Representation of the weights]\label{assumption:3.4}

For each $t\in\{0,1\}$ and $x\in\mathcal X$, there are real-valued functions $\omega_t^\nu,\omega_t^\pi$ of $M$ and $x$ with $\mathbb E[\omega_t^\nu(M,x)^2\mid T=t,x,E]<\infty$ and $\mathbb E[\omega_t^\pi(M,x)^2\mid T=t,x,E]<\infty$ such that
\begin{equation}\tag{3.2}\label{eq:3.2}
\mathbb E[\omega_t^\nu(M,x)\mid S,T=t,x,E]=\frac1{\pi_t(S,x)},\qquad
\mathbb E[\omega_t^\pi(M,x)\mid S,T=t,x,E]=\frac{\mu_t(S,x)}{\pi_t(S,x)}
\end{equation}
almost surely.
\end{fixedassumption}

Assumption 3.4 requires that the weights $1/\pi_t$ and $\mu_t/\pi_t$ be the conditional means given $S$, in the experimental arm $t$, of functions of $M$. We call $\omega_t^\nu$ and $\omega_t^\pi$ the weight functions. Given this representation, iterated expectations give $\mathbb E[d(S)/\pi_t(S,x)\mid T=t,x,E]=\mathbb E[d(S)\omega_t^\nu(M,x)\mid T=t,x,E]$ for bounded $d$, and the right side is linked to a mean in the observational sample through (2.7). Equation (3.2) can have more than one solution, because a function of $M$ with conditional mean zero given $S$ can be added to any solution. The asymptotic distribution in Sections 3.3 and 3.4 holds for the solution to which the weight estimates converge, which we denote by $\omega_t^\nu,\omega_t^\pi$, and the efficiency results use the uniqueness of the solution (Remark 3.1).

We next define two functions that represent the first-order part of the estimation error. For an observation $Z_E=(X,T,M,S)$ of the experimental sample and an observation $Z_O=(X,T,M,Y)$ of the observational sample, let
\begin{equation}\tag{3.3}\label{eq:3.3}
\begin{aligned}
\psi_E(Z_E)={}&\sum_{x\in\mathcal X}\frac{P(X=x\mid O)\mathbf 1\{T=1,X=x\}}{P(T=1,X=x\mid E)}\{\mu_1(S,x)-\bar\mu_1(x)-\nu_1(S,x)\omega_1^\nu(M,x)+\pi_1(S,x)\omega_1^\pi(M,x)\}\\
&-\sum_{x\in\mathcal X}\frac{P(X=x\mid O)\mathbf 1\{T=0,X=x\}}{P(T=0,X=x\mid E)}\{\mu_0(S,x)-\bar\mu_0(x)-\nu_0(S,x)\omega_0^\nu(M,x)+\pi_0(S,x)\omega_0^\pi(M,x)\}
\end{aligned}
\end{equation}
and
\begin{equation}\tag{3.4}\label{eq:3.4}
\begin{aligned}
\psi_O(Z_O)={}&\bar\mu_1(X)-\bar\mu_0(X)-\tau+\mathbf 1\{T=1\}\{Y\omega_1^\nu(M,X)-\omega_1^\pi(M,X)\}\\
&-\mathbf 1\{T=0\}\{Y\omega_0^\nu(M,X)-\omega_0^\pi(M,X)\}
\end{aligned}
\end{equation}
By (3.2), $\mathbb E[\nu_t(S,x)\omega_t^\nu(M,x)\mid T=t,x,E]$ and $\mathbb E[\pi_t(S,x)\omega_t^\pi(M,x)\mid T=t,x,E]$ both equal $\bar\mu_t(x)$, and (2.7) together with these two equalities gives $\mathbb E[\mathbf 1\{T=t\}\{Y\omega_t^\nu(M,x)-\omega_t^\pi(M,x)\}\mid x,O]=0$. Hence both functions have mean zero. The function $\psi_E$ arises from the stratum means and the estimation of the two unknown functions in the experimental sample, and $\psi_O$ from the stratum shares and the moments in the observational sample.

We characterize efficiency by the efficient influence function in the sense of \citet[Chapter 25]{ref32}, adapted to two independent samples. Let the model $\mathcal P$ be the set of pairs of laws of $Z_E$ and $Z_O$ under which $X$ takes values in $\mathcal X$ and Assumptions 2.1–2.8 and the integrability conditions in Appendix A hold. Appendix B defines the one-dimensional submodels with bounded scores along which the functions $\pi_t$ and $\nu_t$ computed from the laws are differentiable in mean square with respect to the parameter of the submodel, the tangent space under the inner product that weights each sample by its share, and the gradients of $\tau$, and it states a tangent space condition formally. The tangent space condition requires that the normalized pair $(\psi_E/\rho,\psi_O/(1-\rho))$ lie in the tangent space.

\renewcommand{\thefixedtheorem}{3.1}
\begin{fixedtheorem}[Efficient influence function and semiparametric efficiency bound]\label{theorem:3.1}

Suppose that Assumptions 2.1–2.8 and 3.1–3.4 hold, together with the integrability conditions in Appendix A and the boundedness and tangent space conditions in Appendix B, and that for each $t\in\{0,1\}$ and $x\in\mathcal X$, (3.2) has a unique square-integrable solution $(\omega_t^\nu,\omega_t^\pi)$. Then the efficient influence function of $\tau$ in the model $\mathcal P$ relative to the smooth submodels of Appendix B is
\[
\Bigl(\frac{\psi_E(Z_E)}{\rho},\frac{\psi_O(Z_O)}{1-\rho}\Bigr)
\]
with $\psi_E$ and $\psi_O$ given by (3.3) and (3.4), and it is the only gradient of $\tau$ that lies in the tangent space. The efficiency bound is
\[
\Sigma^2=\frac1\rho\mathbb E[\psi_E(Z_E)^2\mid E]+\frac1{1-\rho}\mathbb E[\psi_O(Z_O)^2\mid O]
\]
For every sequence of estimators that is regular with respect to the subspace of scores in the tangent space condition, the limiting distribution of $\sqrt N$ times the estimation error is the convolution of $\mathcal N(0,\Sigma^2)$ with some probability distribution, so its variance, when finite, is at least $\Sigma^2$.
\end{fixedtheorem}

Theorem 3.1 gives a lower bound and a way to attain it. For every estimator that is regular with respect to the scores in the tangent space condition, the limiting distribution of $\sqrt N$ times the estimation error has a variance of at least $\Sigma^2$ when this variance is finite, and an estimator whose estimation error equals $n_E^{-1}\sum_{i=1}^{n_E}\psi_E(Z_{E,i})+n_O^{-1}\sum_{j=1}^{n_O}\psi_O(Z_{O,j})$ up to a term of order $o_p(N^{-1/2})$ has asymptotic variance $\Sigma^2$, so the efficient influence function tells us how to construct an efficient estimator. The terms of (3.3) and (3.4) that involve the weight functions are the corrections that remove the first-order effect of the errors in the estimates of $\pi_t$ and $\nu_t$ described above. The estimator of Section 3.2 therefore plugs estimates of the nuisance functions into these corrections and adds them to the plug-in estimate of $\bar\mu_t(x)$, and Theorems 3.2 and 3.3 below show that, under the conditions of Theorem 3.1 and conditions on the estimates of the nuisance functions, the proposed estimators attain the bound.

Along a submodel, $\pi_t$ and $\nu_t$ move with the laws of the two samples through the equations (2.7). Because these functions are differentiable along the smooth submodels of Appendix B, differentiating (2.7) along a submodel gives the derivative of $\tau$ as the sum of the means of the products of $\psi_E$ and $\psi_O$ with the scores of the submodel, for every square-integrable solution of (3.2). As in \citet{ref26}, we study efficiency through derivatives along one-dimensional submodels, and we restrict attention to those along which $\pi_t$ and $\nu_t$ are differentiable. When $S$ and $M$ take finitely many values given $T=t$ and $X=x$, as in the empirical illustration of Section 6, (2.7) is a system of linear equations whose matrix has linearly independent columns by Assumption 2.8, and every regular path of Appendix B is such a submodel. The tangent space condition requires that the normalized influence function be approximated by scores of submodels in $\mathcal P$. The proof is in Appendix B.

Recent work derives efficient influence functions of causal parameters whose identification requires solving such inverse problems. Theorem 4.1 of \citet{ref12} gives the efficient influence function of the average treatment effect in proximal causal inference under conditions on the conditional expectation operators, and our proof of Theorem 3.1 follows their approach.

\renewcommand{\thefixedremark}{3.1}
\begin{fixedremark}[Weight functions that are not unique]\label{remark:3.1}

The influence functions (3.3) and (3.4) depend on the solution $(\omega_t^\nu,\omega_t^\pi)$ of (3.2), and Theorem 3.1 and the efficiency claims of Theorems 3.2 and 3.3 assume that this solution is unique. Solutions fail to be unique when $M$ is richer than $S$, so that some nonzero square-integrable function of $M$ has conditional mean zero given $S$. For example, if $M=S+U_2$ given $T=t$ and $X=x$ in the experiment, with $U_2$ uniform on $[-1,1]$ and independent of $S$, as in the simulation designs of Section 5, then $\mathbb E[\sin(\pi M)\mid S,T=t,x,E]=0$. When $S$ and $M$ take finitely many values given $T=t$ and $X=x$, Assumption 2.8 requires $M$ to take at least as many values as $S$, and uniqueness requires at most as many, so both hold only when they take the same number of values.

When the solutions are not unique, two solutions can still give the same pair of influence functions. Each distinct pair built from a solution is a gradient of $\tau$, and since two gradients in the tangent space coincide, the tangent space condition can hold for at most one of these distinct pairs. In this case, the estimators of Sections 3.2 and 3.4 keep the asymptotic distribution and the consistency of the variance estimator stated in Theorems 3.2 and 3.3, with $\Sigma^2$ computed from the solution to which their weight estimates converge, while the efficiency statements of these theorems require a unique solution.
\end{fixedremark}

\subsection{Estimator}\label{sec:3.2}

This subsection defines the estimator that estimates the nuisance functions and averages within strata on the same full sample, together with notation for sample means. The two independent samples are $Z_{E,i}=(X_i,T_i,M_i,S_i)$, $i=1,\ldots,n_E$, and $Z_{O,j}=(X_j,T_j,M_j,Y_j)$, $j=1,\ldots,n_O$. We write $\widehat{\mathbb E}$ and $\widehat P$ for full-sample analogues of conditional means and shares. For example, $\widehat{\mathbb E}[f(S,M)\mid T=t,x,E]$ is the average of $f(S_i,M_i)$ over the observations with $T_i=t$ and $X_i=x$, $\widehat{\mathbb E}[g(T,M,Y)\mid x,O]$ is the average of $g(T_j,M_j,Y_j)$ over the observations with $X_j=x$, and $\widehat P(T=t,X=x\mid E)$ and $\widehat P(X=x\mid O)$ are the shares of observations satisfying these conditions in the respective samples.

For each $t$, we estimate the nuisance functions $\pi_t,\nu_t,\omega_t^\nu,\omega_t^\pi$ using all observations in the two samples. The estimation method is left unspecified, and the estimates are jointly measurable in the sample and the arguments of the functions. We truncate the estimate of the selection probability to $[a,1]$ with the constant $a$ of Assumption 3.3 and denote the result by $\widehat\pi_t$, write $\widehat\nu_t,\widehat\omega_t^\nu,\widehat\omega_t^\pi$ for the estimates of the product and the two weight functions, and write $\widehat\mu_t=\widehat\nu_t/\widehat\pi_t$ for the estimated ratio. Because $\pi_t$, $\nu_t$, and the weight functions solve linear inverse problems, series, or sieve, estimators are a natural choice, as in nonparametric instrumental variable and conditional moment models \citep{ref28,ref1,ref6,ref8,ref10}. The simulation in Section 5 and the application in Section 6 use series estimates with a Tikhonov penalty. Holding the estimates fixed, we compute stratum means over the same full sample.

For each $t,x$, let
\begin{equation}\tag{3.5}\label{eq:3.5}
\begin{aligned}
\widehat{\bar\mu}_t(x)={}&\widehat{\mathbb E}[\widehat\mu_t(S,x)-\widehat\nu_t(S,x)\widehat\omega_t^\nu(M,x)+\widehat\pi_t(S,x)\widehat\omega_t^\pi(M,x)\mid T=t,x,E]\\
&+\widehat{\mathbb E}[\mathbf 1\{T=t\}\{Y\widehat\omega_t^\nu(M,x)-\widehat\omega_t^\pi(M,x)\}\mid x,O]
\end{aligned}
\end{equation}
and define the estimator of the average treatment effect as
\begin{equation}\tag{3.6}\label{eq:3.6}
\widehat\tau=\sum_{x\in\mathcal X}\widehat P(X=x\mid O)\{\widehat{\bar\mu}_1(x)-\widehat{\bar\mu}_0(x)\}
\end{equation}
The first term of (3.5) adds the two corrections $-\widehat\nu_t\widehat\omega_t^\nu+\widehat\pi_t\widehat\omega_t^\pi$ to the stratum mean of the estimated ratio, and the second term estimates the mean of the same corrections in the observational sample. With the true functions plugged in, the expected correction in the first term and the expectation in the second term are both zero, as shown in Section 3.1. With the estimates plugged in, the corrections in the two terms cancel the first-order part of the error in the estimated ratio, and Appendix C states this property as an identity.

The asymptotic variance of $\widehat\tau$ is $\Sigma^2$ (Theorem 3.2), which equals the efficiency bound of Theorem 3.1 under the conditions of that theorem, and standard errors replace the unknown quantities in (3.3) and (3.4) by estimates. Let $\widehat\psi_E(Z_{E,i})$ be (3.3) evaluated at $Z_{E,i}$ with $\mu_t,\nu_t,\pi_t,\omega_t^\nu,\omega_t^\pi$ replaced by $\widehat\mu_t,\widehat\nu_t,\widehat\pi_t,\widehat\omega_t^\nu,\widehat\omega_t^\pi$, $\bar\mu_t(x)$ by $\widehat{\bar\mu}_t(x)$, and $P(X=x\mid O)$ and $P(T=t,X=x\mid E)$ by $\widehat P(X=x\mid O)$ and $\widehat P(T=t,X=x\mid E)$. Let $\widehat\psi_O(Z_{O,j})$ be (3.4) evaluated at $Z_{O,j}$ with the same replacements and with $\tau$ replaced by $\widehat\tau$. We recall $\Sigma^2$ and define its estimator by
\begin{equation}\tag{3.7}\label{eq:3.7}
\begin{aligned}
\Sigma^2&=\frac1\rho\mathbb E[\psi_E(Z_E)^2\mid E]+\frac1{1-\rho}\mathbb E[\psi_O(Z_O)^2\mid O],\\
\widehat\Sigma^2&=\frac N{n_E}\frac1{n_E}\sum_{i=1}^{n_E}\widehat\psi_E(Z_{E,i})^2+\frac N{n_O}\frac1{n_O}\sum_{j=1}^{n_O}\widehat\psi_O(Z_{O,j})^2
\end{aligned}
\end{equation}
and let $\Sigma$ and $\widehat\Sigma$ be their nonnegative square roots. The standard error is $\widehat\Sigma/\sqrt N$.

\subsection{Asymptotic normality and efficiency}\label{sec:3.3}

This subsection states two conditions on the full-sample estimates and the main theorem. We maintain the conditions of Section 2 for identification and Assumptions 3.1–3.4.

Expectations of functions that involve estimates are taken with the sample held fixed, by integrating with respect to the conditional distribution of the population. For example, $\mathbb E[\widehat\nu_t(S,x)\widehat\omega_t^\nu(M,x)\mid T=t,x,E]=\int\widehat\nu_t(s,x)\widehat\omega_t^\nu(m,x)\,dF_{S,M\mid T=t,x,E}(s,m)$, and $\mathbb E[\,\cdot\mid x,O]$ is the integral with respect to $F_{T,M,Y\mid x,O}$. Conditional expectations given $M$ or $S$ are also taken with the sample fixed, under the same conditional distribution. For a function $f$ of $(S,M)$, the mean-square norm is $\|f\|_{2,t,x}=\mathbb E[f(S,M)^2\mid T=t,x,E]^{1/2}$, and we write $(\widehat{\mathbb E}-\mathbb E)[\,\cdot\mid T=t,x,E]=\widehat{\mathbb E}[\,\cdot\mid T=t,x,E]-\mathbb E[\,\cdot\mid T=t,x,E]$ for the difference between the sample mean and this expectation, and $(\widehat{\mathbb E}-\mathbb E)[\,\cdot\mid x,O]$ for its analogue in the observational sample.

Assumption 3.5 below bounds, by rates of the estimation errors, the deviation from $\bar\mu_t(x)$ that remains when the estimates are held fixed and the two sample means in (3.5) are replaced by expectations. Assumption 3.6 bounds the error from averaging the plugged-in functions over the same sample that was used to estimate them.

\renewcommand{\thefixedassumption}{3.5}
\begin{fixedassumption}[Estimation of the nuisance functions]\label{assumption:3.5}

For each $t,x$, the full-sample estimates $\widehat\pi_t,\widehat\nu_t,\widehat\omega_t^\nu,\widehat\omega_t^\pi$ of Section 3.2 satisfy the following.

\begin{enumerate}
\item $\|\widehat\pi_t-\pi_t\|_{2,t,x}\to_p0$, $\|\widehat\nu_t-\nu_t\|_{2,t,x}\to_p0$, $\|\widehat\omega_t^\nu-\omega_t^\nu\|_{2,t,x}\to_p0$, and $\|\widehat\omega_t^\pi-\omega_t^\pi\|_{2,t,x}\to_p0$.
\item $\|\widehat\pi_t-\pi_t\|_{2,t,x}\times\{\|\widehat\pi_t-\pi_t\|_{2,t,x}+\|\widehat\nu_t-\nu_t\|_{2,t,x}\}=o_p(N^{-1/2})$.
\item The products for $\omega_t^\nu$ satisfy
\[
\begin{aligned}
   \min\Bigl\{&\|\mathbb E[\widehat\nu_t(S,x)-\nu_t(S,x)\mid M,T=t,x,E]\|_{2,t,x}\times\|\widehat\omega_t^\nu-\omega_t^\nu\|_{2,t,x},\\
   &\|\widehat\nu_t-\nu_t\|_{2,t,x}\times\|\mathbb E[\widehat\omega_t^\nu(M,x)-\omega_t^\nu(M,x)\mid S,T=t,x,E]\|_{2,t,x}\Bigr\}=o_p(N^{-1/2})
   \end{aligned}
\]
\item The products for $\omega_t^\pi$ satisfy
\[
\begin{aligned}
   \min\Bigl\{&\|\mathbb E[\widehat\pi_t(S,x)-\pi_t(S,x)\mid M,T=t,x,E]\|_{2,t,x}\times\|\widehat\omega_t^\pi-\omega_t^\pi\|_{2,t,x},\\
   &\|\widehat\pi_t-\pi_t\|_{2,t,x}\times\|\mathbb E[\widehat\omega_t^\pi(M,x)-\omega_t^\pi(M,x)\mid S,T=t,x,E]\|_{2,t,x}\Bigr\}=o_p(N^{-1/2})
   \end{aligned}
\]
\end{enumerate}
\end{fixedassumption}

\renewcommand{\thefixedassumption}{3.6}
\begin{fixedassumption}[Averaging over the estimation sample]\label{assumption:3.6}

For each $t,x$, the full-sample estimates of Section 3.2 and $\widehat\mu_t=\widehat\nu_t/\widehat\pi_t$ satisfy the following.

\begin{enumerate}
\item $(\widehat{\mathbb E}-\mathbb E)[\widehat\mu_t(S,x)-\mu_t(S,x)\mid T=t,x,E]=o_p(N^{-1/2})$.
\item $(\widehat{\mathbb E}-\mathbb E)[\widehat\nu_t(S,x)\widehat\omega_t^\nu(M,x)-\nu_t(S,x)\omega_t^\nu(M,x)\mid T=t,x,E]=o_p(N^{-1/2})$.
\item $(\widehat{\mathbb E}-\mathbb E)[\widehat\pi_t(S,x)\widehat\omega_t^\pi(M,x)-\pi_t(S,x)\omega_t^\pi(M,x)\mid T=t,x,E]=o_p(N^{-1/2})$.
\item $(\widehat{\mathbb E}-\mathbb E)[\mathbf 1\{T=t\}Y\{\widehat\omega_t^\nu(M,x)-\omega_t^\nu(M,x)\}\mid x,O]=o_p(N^{-1/2})$.
\item $(\widehat{\mathbb E}-\mathbb E)[\mathbf 1\{T=t\}\{\widehat\omega_t^\pi(M,x)-\omega_t^\pi(M,x)\}\mid x,O]=o_p(N^{-1/2})$.
\item $\widehat{\mathbb E}[\{\widehat\pi_t(S,x)-\pi_t(S,x)\}^2\mid T=t,x,E]\to_p0$.
\item $\widehat{\mathbb E}[\{\widehat\nu_t(S,x)-\nu_t(S,x)\}^2\mid T=t,x,E]\to_p0$.
\item $\widehat{\mathbb E}[\{\widehat\omega_t^\nu(M,x)-\omega_t^\nu(M,x)\}^2\mid T=t,x,E]\to_p0$.
\item $\widehat{\mathbb E}[\{\widehat\omega_t^\pi(M,x)-\omega_t^\pi(M,x)\}^2\mid T=t,x,E]\to_p0$.
\item $\widehat{\mathbb E}[\mathbf 1\{T=t\}\{\widehat\omega_t^\nu(M,x)-\omega_t^\nu(M,x)\}^2\mid x,O]\to_p0$.
\item $\widehat{\mathbb E}[\mathbf 1\{T=t\}\{\widehat\omega_t^\pi(M,x)-\omega_t^\pi(M,x)\}^2\mid x,O]\to_p0$.
\end{enumerate}
\end{fixedassumption}

Condition (1) of Assumption 3.5 requires consistency of the nuisance function estimates and imposes no particular rate. Conditions (2)–(4) take the product-rate form of \citet{ref11}. They make the products that measure the deviation from $\bar\mu_t(x)$, when the estimates are held fixed and the two sample means in (3.5) are replaced by expectations, vanish faster than $N^{-1/2}$. Because the moment in (3.5) is insensitive to first-order errors in the nuisance functions, only products of errors appear, and each condition holds, for example, when both factors of the product are $o_p(N^{-1/4})$. Nonparametric and machine learning estimators that converge at such slow rates can therefore be used for estimation and inference on $\tau$.

Assumption 3.6 consists of conditions (1)–(5) on the difference between the sample mean of a plugged-in function and its expectation with the estimates held fixed, and conditions (6)–(11) on sample means of squared estimation errors. Conditions (1)–(5) impose directly on the estimates the stochastic equicontinuity of empirical process theory \citep[Chapter 19]{ref32}. They hold if, on events with probability approaching one, each plugged-in function multiplied by the indicator of its stratum lies in a fixed Donsker class for the respective population that contains the corresponding true function, by the results on Donsker classes in Chapter 19 of \citet{ref32} together with the consistency in Assumption 3.5(1).

Conditions (6)–(9) require that the mean squared errors of the four estimates within the experimental strata, and conditions (10) and (11) that those of the two weight estimates among observations with $T=t$ within the observational strata, converge in probability to zero. They are used for the consistency of the variance estimator. They hold if, on events with probability approaching one, each squared error multiplied by the indicator of its stratum lies in a fixed Glivenko–Cantelli class for the respective population, by the uniform laws of large numbers in Chapter 19 of \citet{ref32} together with Assumption 3.5(1).

Series estimates are one example of estimates that can satisfy the product-rate conditions of Assumption 3.5 under additional regularity conditions. The equations (2.7) for $\pi_t$ and $\nu_t$ have the form of nonparametric instrumental variable models with $M$ in the role of the instrument, and the equations (3.2) for the weight functions have the same form with the roles of $S$ and $M$ reversed and with estimated right sides. The $L^2$ convergence rates of series and penalized sieve estimators of such models \citep{ref6,ref10,ref9} are the basis for sufficient conditions. Under sufficient smoothness of the unknown functions and a mild degree of ill-posedness, these rates are $o_p(N^{-1/4})$ for a suitable growth of the numbers of series terms, and applying them here further requires controlling the error in the estimated right sides of (3.2) and fixing the solution of (3.2) that the weight estimates approach. Assumption 3.6 has to be checked separately for the estimator without sample splitting.

\renewcommand{\thefixedtheorem}{3.2}
\begin{fixedtheorem}[Asymptotic distribution and efficiency of the estimator without sample splitting]\label{theorem:3.2}

Suppose that Assumptions 2.1–2.8 and 3.1–3.6 hold, together with the integrability conditions in Appendix A and the boundedness conditions in Appendices B and C. Then
\begin{equation}\tag{3.8}\label{eq:3.8}
\sqrt N(\widehat\tau-\tau)\Rightarrow\mathcal N(0,\Sigma^2),\qquad
\widehat\Sigma^2\to_p\Sigma^2
\end{equation}
If $\Sigma^2>0$, then $\sqrt N(\widehat\tau-\tau)/\widehat\Sigma\Rightarrow\mathcal N(0,1)$, and for $\alpha\in(0,1)$ and the $1-\alpha/2$ quantile $z_{1-\alpha/2}$ of the standard normal distribution, the coverage probability of $\widehat\tau\pm z_{1-\alpha/2}\widehat\Sigma/\sqrt N$ converges to $1-\alpha$. If, in addition, (3.2) has a unique square-integrable solution for each $t$ and $x$ and the tangent space condition in Appendix B holds, then $\widehat\tau$ is regular in the model $\mathcal P$ with respect to the smooth submodels of Appendix B and attains the efficiency bound $\Sigma^2$ of Theorem 3.1.
\end{fixedtheorem}

The estimation error is approximated by the sum of two sample means, $n_E^{-1}\sum_{i=1}^{n_E}\psi_E(Z_{E,i})+n_O^{-1}\sum_{j=1}^{n_O}\psi_O(Z_{O,j})$, and $\Sigma^2$ is the sum of their two contributions. Under the conditions of Theorem 3.1, which include the uniqueness of the solution of (3.2), these summands form the efficient influence function, so the estimator attains the efficiency bound. When the solution is not unique, the asymptotic distribution and the variance estimator remain valid (Remark 3.1). The proof is in Appendix C.

The Donsker and Glivenko–Cantelli conditions behind Assumption 3.6 can fail for classes of nonparametric or machine learning estimators. \citet{ref11} introduce cross-fitting to avoid such conditions, and our estimator can be computed in the same way, as Section 3.4 describes.

\subsection{Estimation with cross-fitting}\label{sec:3.4}

This subsection defines an estimator that splits the two samples into folds and averages within strata over a fold whose observations are left out when the nuisance functions are estimated, and it shows that the conclusions of Theorem 3.2 hold under the consistency and rate conditions of Assumption 3.7. For a fixed integer $K\ge2$, we partition $\{1,\ldots,n_E\}$ into $K$ disjoint sets $\mathcal I_{E,1},\ldots,\mathcal I_{E,K}$ whose sizes differ by at most one, and partition $\{1,\ldots,n_O\}$ into $\mathcal I_{O,1},\ldots,\mathcal I_{O,K}$ in the same way. These index sets are nonrandom and do not depend on the observations. Fold $k$ consists of $Z_{E,i}$, $i\in\mathcal I_{E,k}$, and $Z_{O,j}$, $j\in\mathcal I_{O,k}$. We write $\widehat{\mathbb E}_k$ and $\widehat P_k$ for the analogues of conditional means and shares within fold $k$. For example, $\widehat{\mathbb E}_k[f(S,M)\mid T=t,x,E]$ is the average of $f(S_i,M_i)$ over the observations with $i\in\mathcal I_{E,k}$, $T_i=t$, and $X_i=x$, and $\widehat P_k(T=t,X=x\mid E)$ is the share of observations with $T_i=t$ and $X_i=x$ among $i\in\mathcal I_{E,k}$.

For each $k$, we estimate $\pi_t,\nu_t,\omega_t^\nu,\omega_t^\pi$ using only the observations outside fold $k$ and denote the estimates by $\widehat\pi_t^{(-k)},\widehat\nu_t^{(-k)},\widehat\omega_t^{\nu,(-k)},\widehat\omega_t^{\pi,(-k)}$, where the superscript $(-k)$ indicates that fold $k$ is left out. As in Section 3.2, $\widehat\pi_t^{(-k)}$ is truncated to $[a,1]$, and $\widehat\mu_t^{(-k)}=\widehat\nu_t^{(-k)}/\widehat\pi_t^{(-k)}$. Holding the estimates fixed, we compute stratum means over fold $k$. For each $t,x,k$, let
\begin{equation}\tag{3.9}\label{eq:3.9}
\begin{aligned}
\widehat{\bar\mu}_{t,k}(x)={}&\widehat{\mathbb E}_k[\widehat\mu_t^{(-k)}(S,x)-\widehat\nu_t^{(-k)}(S,x)\widehat\omega_t^{\nu,(-k)}(M,x)+\widehat\pi_t^{(-k)}(S,x)\widehat\omega_t^{\pi,(-k)}(M,x)\mid T=t,x,E]\\
&+\widehat{\mathbb E}_k[\mathbf 1\{T=t\}\{Y\widehat\omega_t^{\nu,(-k)}(M,x)-\widehat\omega_t^{\pi,(-k)}(M,x)\}\mid x,O]
\end{aligned}
\end{equation}
and define the estimator $\widehat\tau$ of the average treatment effect by substituting $\widehat{\bar\mu}_t(x)=K^{-1}\sum_{k=1}^K\widehat{\bar\mu}_{t,k}(x)$ into (3.6). For standard errors, we compute $\widehat\psi_E(Z_{E,i})$ for $i\in\mathcal I_{E,k}$ and $\widehat\psi_O(Z_{O,j})$ for $j\in\mathcal I_{O,k}$ with the replacements of Section 3.2, using $\widehat\mu_t^{(-k)},\widehat\nu_t^{(-k)},\widehat\pi_t^{(-k)},\widehat\omega_t^{\nu,(-k)},\widehat\omega_t^{\pi,(-k)}$ in place of the full-sample estimates, and use $\widehat\Sigma^2$ in (3.7). In this subsection, $\widehat{\bar\mu}_t(x)$, $\widehat\tau$, $\widehat\psi_E$, $\widehat\psi_O$, and $\widehat\Sigma^2$ denote these cross-fitted quantities. Expectations and norms of functions that involve estimates are taken with the sample held fixed, as in Section 3.3.

\renewcommand{\thefixedassumption}{3.7}
\begin{fixedassumption}[Estimates that leave out a fold]\label{assumption:3.7}

For each $k$, $t$, and $x$, the estimates $\widehat\pi_t^{(-k)},\widehat\nu_t^{(-k)},\widehat\omega_t^{\nu,(-k)},\widehat\omega_t^{\pi,(-k)}$ satisfy the following.

\begin{enumerate}
\item $\|\widehat\pi_t^{(-k)}-\pi_t\|_{2,t,x}\to_p0$, $\|\widehat\nu_t^{(-k)}-\nu_t\|_{2,t,x}\to_p0$, $\|\widehat\omega_t^{\nu,(-k)}-\omega_t^\nu\|_{2,t,x}\to_p0$, and $\|\widehat\omega_t^{\pi,(-k)}-\omega_t^\pi\|_{2,t,x}\to_p0$.
\item $\|\widehat\pi_t^{(-k)}-\pi_t\|_{2,t,x}\times\{\|\widehat\pi_t^{(-k)}-\pi_t\|_{2,t,x}+\|\widehat\nu_t^{(-k)}-\nu_t\|_{2,t,x}\}=o_p(N^{-1/2})$.
\item The products for $\omega_t^\nu$ satisfy
\[
\begin{aligned}
   \min\Bigl\{&\|\mathbb E[\widehat\nu_t^{(-k)}(S,x)-\nu_t(S,x)\mid M,T=t,x,E]\|_{2,t,x}\times\|\widehat\omega_t^{\nu,(-k)}-\omega_t^\nu\|_{2,t,x},\\
   &\|\widehat\nu_t^{(-k)}-\nu_t\|_{2,t,x}\times\|\mathbb E[\widehat\omega_t^{\nu,(-k)}(M,x)-\omega_t^\nu(M,x)\mid S,T=t,x,E]\|_{2,t,x}\Bigr\}=o_p(N^{-1/2})
   \end{aligned}
\]
\item The products for $\omega_t^\pi$ satisfy
\[
\begin{aligned}
   \min\Bigl\{&\|\mathbb E[\widehat\pi_t^{(-k)}(S,x)-\pi_t(S,x)\mid M,T=t,x,E]\|_{2,t,x}\times\|\widehat\omega_t^{\pi,(-k)}-\omega_t^\pi\|_{2,t,x},\\
   &\|\widehat\pi_t^{(-k)}-\pi_t\|_{2,t,x}\times\|\mathbb E[\widehat\omega_t^{\pi,(-k)}(M,x)-\omega_t^\pi(M,x)\mid S,T=t,x,E]\|_{2,t,x}\Bigr\}=o_p(N^{-1/2})
   \end{aligned}
\]
\end{enumerate}
\end{fixedassumption}

Conditions (1)–(4) of Assumption 3.7 are conditions (1)–(4) of Assumption 3.5 stated for the estimates that leave out each fold, and the next theorem uses them in place of Assumptions 3.5 and 3.6.

\renewcommand{\thefixedtheorem}{3.3}
\begin{fixedtheorem}[Asymptotic distribution and efficiency of the cross-fitted estimator]\label{theorem:3.3}

Suppose that Assumptions 2.1–2.8, 3.1–3.4, and 3.7 hold, together with the integrability conditions in Appendix A and the boundedness conditions in Appendices B and C. Then (3.8) holds for the estimator $\widehat\tau$ and the variance estimator $\widehat\Sigma^2$ of Section 3.4. If $\Sigma^2>0$, then $\sqrt N(\widehat\tau-\tau)/\widehat\Sigma\Rightarrow\mathcal N(0,1)$, and for $\alpha\in(0,1)$ and the $1-\alpha/2$ quantile $z_{1-\alpha/2}$ of the standard normal distribution, the coverage probability of $\widehat\tau\pm z_{1-\alpha/2}\widehat\Sigma/\sqrt N$ converges to $1-\alpha$. If, in addition, (3.2) has a unique square-integrable solution for each $t$ and $x$ and the tangent space condition in Appendix B holds, then $\widehat\tau$ is regular in the model $\mathcal P$ with respect to the smooth submodels of Appendix B and attains the efficiency bound $\Sigma^2$ of Theorem 3.1.
\end{fixedtheorem}

In Theorem 3.3, the independence between each fold and the estimates computed without it takes the place of Assumption 3.6. With the estimates held fixed, each quantity in Assumption 3.6 computed over fold $k$ is the difference between an average of independent observations and its expectation, and the consistency in Assumption 3.7(1) makes this difference small enough. Plugging nuisance functions estimated on other folds into a moment that is insensitive to first-order errors in them follows the double/debiased machine learning of \citet{ref11}, who call such moments Neyman orthogonal and combine them with cross-fitting. The proof is in Appendix C.

\section{Partial identification without the exclusion restriction for the outcome mean}\label{sec:4}

This section bounds the average treatment effect when the exclusion restriction for the outcome mean in Assumption 2.7 may fail and the other assumptions of Section 2 are maintained. Section 4.1 derives sharp bounds from an optimal transport problem that places no restriction on how the outcome mean depends on the shared measurement and shows that the bounds remain sharp and are computed by linear programs when linear restrictions on the outcome mean are added, and Section 4.2 describes how the bounds are estimated and how confidence intervals for $\tau$ are formed.

\subsection{Bounds from optimal transport}\label{sec:4.1}

This subsection derives sharp bounds on $\tau$ from Assumptions 2.1–2.6 and 2.8 without any restriction in place of Assumption 2.7. Without Assumption 2.7, the selection probability remains identified, and it turns the mean of each potential outcome into a weighted mean of the observed outcome. The first equation of (2.7) is derived from Assumptions 2.1, 2.3, 2.4, and 2.6 alone (Appendix A.1), so by Assumption 2.8 it identifies $\pi_t$. By Assumption 2.5, $\mathbf 1\{T=t\}$ and $Y(t)$ are independent given $S(t)$ and $X=x$ in the observational population, so $\mathbb E[\mathbf 1\{T=t\}Y(t)\mid S(t),x,O]=\pi_t(S(t),x)\mathbb E[Y(t)\mid S(t),x,O]$. Dividing by $\pi_t(S(t),x)>0$ (Assumption 2.2), taking expectations, and using $Y=Y(t)$ on $\{T=t\}$ (Assumption 2.1) give the inverse probability weighting representation
\begin{equation}\tag{4.1}\label{eq:4.1}
\mathbb E[Y(t)\mid x,O]=\mathbb E[\mathbf 1\{T=t\}YW(t)\mid x,O],\qquad W(t)=\frac1{\pi_t(S(t),x)}
\end{equation}
The observational data record $Y$ but not $S$, so the right side of (4.1) is not identified in general. Under Assumption 2.7, the moment equations of Section 2 pin it down, and without Assumption 2.7, we bound it over the values that are consistent with the two samples. We assume that the measurements take finitely many values.

\renewcommand{\thefixedassumption}{4.1}
\begin{fixedassumption}[Finite support of the characteristic and the shared measurement]\label{assumption:4.1}

For each $t$ and $F_{X\mid O}$-almost every $x$, the conditional distribution of $(S(t),M(t))$ given $X=x$ and $G=O$ has finite support.
\end{fixedassumption}

Under this assumption, (4.1) becomes a sum over the cells $\{T=t,M=m\}$ of the observational data, $\mathbb E[Y(t)\mid x,O]=\sum_mP(T=t,M=m\mid x,O)\,\mathbb E[YW(t)\mid T=t,M=m,x,O]$. In this sum, the probabilities $P(T=t,M=m\mid x,O)$ are identified from the observational data, whereas the means $\mathbb E[YW(t)\mid T=t,M=m,x,O]$ are not identified in general, because they depend on the joint distribution of $Y$ and $W(t)$. 

The marginal distributions of $Y$ and of $W(t)$ within each cell are nonetheless identified. The conditional distribution of $Y$ given $T=t$, $M=m$, $X=x$, and $G=O$ is observed directly in the observational data, whereas that of $W(t)$ is identified once the selection probability, which the two samples identify, is known. Indeed, by Assumption 2.6, the treatment depends on $(S(t),M(t))$ only through $\pi_t(S(t),x)$, and by Assumptions 2.1, 2.3, and 2.4, the distribution of $(S(t),M(t))$ in the observational population is that of $(S,M)$ in the experimental arm $t$. Hence, as Appendix A.3 shows step by step, for each value $m$ with $P(T=t,M=m\mid x,O)>0$,
\begin{equation}\tag{4.2}\label{eq:4.2}
\begin{aligned}
P(W(t)\le w\mid T=t,M=m,x,O)&=\frac{\mathbb E[\pi_t(S,x)\mathbf 1\{1/\pi_t(S,x)\le w\}\mid M=m,T=t,x,E]}
{\mathbb E[\pi_t(S,x)\mid M=m,T=t,x,E]}.
\end{aligned}
\end{equation}

The largest and smallest values of $\mathbb E[YW(t)\mid T=t,M=m,x,O]$ that are consistent with the two identified marginal distributions are the values of optimal transport problems, which maximize and minimize the mean of $YW(t)$ over the joint distributions with these marginal distributions. The next proposition bounds $\mathbb E[Y(t)\mid x,O]$ by these values. To state it, let $F_{t,m,x}$ and $G_{t,m,x}$ denote the conditional distribution functions of $Y$ and of $W(t)$ given $T=t$, $M=m$, $X=x$, and $G=O$, let $\Pi(F,G)$ denote the set of joint distributions on $\mathbb R^2$ with marginal distribution functions $F$ and $G$, and write $H^{-1}(u)=\inf\{z:H(z)\ge u\}$, $0<u<1$, for the quantile function of a distribution function $H$.

\renewcommand{\thefixedproposition}{4.1}
\begin{fixedproposition}[Sharp bounds from optimal transport]\label{proposition:4.1}

Suppose that Assumptions 2.1–2.6, 2.8, and 4.1 and the integrability conditions in Appendix A hold. For each $t$ and $x$, let
\begin{equation}\tag{4.3}\label{eq:4.3}
\begin{aligned}
&L_t(x)=\sum_mP(T=t,M=m\mid x,O)\inf_{\gamma\in\Pi(F_{t,m,x},G_{t,m,x})}\int yw\,d\gamma(y,w),\\
&U_t(x)=\sum_mP(T=t,M=m\mid x,O)\sup_{\gamma\in\Pi(F_{t,m,x},G_{t,m,x})}\int yw\,d\gamma(y,w),
\end{aligned}
\end{equation}
where the sums run over the values $m$ with $P(T=t,M=m\mid x,O)>0$. Then $L_t(x)\le\mathbb E[Y(t)\mid x,O]\le U_t(x)$ for $F_{X\mid O}$-almost every $x$, and
\begin{equation}\tag{4.4}\label{eq:4.4}
\int\{L_1(x)-U_0(x)\}\,dF_{X\mid O}(x)\le\tau\le\int\{U_1(x)-L_0(x)\}\,dF_{X\mid O}(x)
\end{equation}
The bounds in (4.4) are sharp. Moreover, because $Y$ and $W(t)$ are real valued and the surplus $yw$ is supermodular, the infimum and the supremum in (4.3) are attained by the couplings that pair the quantiles of the two distributions in the opposite and in the same order, so that
\[
\begin{aligned}
&L_t(x)=\sum_mP(T=t,M=m\mid x,O)\int_0^1F_{t,m,x}^{-1}(u)\,G_{t,m,x}^{-1}(1-u)\,du,\\
&U_t(x)=\sum_mP(T=t,M=m\mid x,O)\int_0^1F_{t,m,x}^{-1}(u)\,G_{t,m,x}^{-1}(u)\,du
\end{aligned}
\]
\end{fixedproposition}

In each cell, the infimum and the supremum in (4.3) are the values of Kantorovich optimal transport problems with surplus $yw$, and optimal transport has been used to characterize identified sets in many problems in econometrics \citep{ref19}. The quantile form in the last part of the proposition is the monotone rearrangement \citep[see, e.g., Theorem 11 of][]{ref19}. Because the two marginal distributions fix $\mathbb E[Y^2]$ and $\mathbb E[W(t)^2]$, maximizing $\mathbb E[YW(t)]$ is the same as minimizing $\mathbb E[(Y-W(t))^2]$, and the same theorem covers this transport problem, whose cost is a convex function of $y-w$ \citep[Section 1.5.1]{ref19}. Each problem in (4.3) is a linear program when $Y$ and $W(t)$ take finitely many values in the cell. If $Y$ takes the values $y_j$ and $W(t)$ the values $w_k$, a coupling in $\Pi(F_{t,m,x},G_{t,m,x})$ is a set of nonnegative masses $\gamma_{jk}$ on the pairs $(y_j,w_k)$ whose sums over $k$ and over $j$ are the masses of the two marginal distributions, and $\int yw\,d\gamma=\sum_{j,k}y_jw_k\gamma_{jk}$ is linear in these masses. Restrictions that are linear in the joint distribution of $Y$ and $W(t)$, or of $Y$ and $S(t)$, enter this program as additional linear constraints, so the bounds under such restrictions are computed just as easily. Section 6.3 imposes two of them that are plausible in the Job Corps illustration.

The bounds are sharp because the data and Assumptions 2.1–2.6 and 2.8 determine the two marginal distributions in each cell but leave their coupling free. Any coupling can be realized by drawing $Y(t)$ given $(S(t),M(t))$ from it, in the same way for treated and untreated units. This leaves the distribution of $(T,S(t),M(t))$ and the observed data unchanged, and because the treatment depends on $M(t)$ only through $S(t)$ by Assumption 2.6, it also keeps Assumption 2.5. The extreme couplings therefore attain the ends, and their mixtures attain every value in between. The same holds under linear restrictions, because the couplings that satisfy them form a convex set and the construction keeps them.

\subsection{Estimation of the bounds and inference on the average treatment effect}\label{sec:4.2}

This subsection describes how the ends are estimated and how confidence intervals for $\tau$ are formed. We replace the functions in Proposition 4.1 by sample analogues within each cell $x$ and arm $t$ and average over the cells with the shares $\widehat P(X=x\mid O)$. The selection probability is estimated in the same way as the point identification. The ends integrate the empirical quantile functions of $Y$ in the observational cells against those of the inverse of the estimated selection probability, whose distribution (4.2) is estimated from the experimental units of the same arm, cell, and value of $M$.

The confidence intervals cover the true value of $\tau$ rather than the whole interval of bounds, as in \citet{ref22}. For estimated lower and upper ends $\widehat L$ and $\widehat U$ with standard errors $s_L$ and $s_U$, the interval is $[\widehat L-cs_L,\widehat U+cs_U]$, where $c$ solves $\Phi(c+\max(\widehat U-\widehat L,0)/\max(s_L,s_U))-\Phi(-c)=1-\alpha$ and $\Phi$ is the standard normal distribution function. When the estimated ends are close, $c$ approaches the two-sided critical value, and when they are far apart, it approaches the one-sided critical value. The standard errors come from estimated influence functions of the ends, which include the effects of estimating the selection probability, the distributions within the cells, and the cell shares in the two samples, with the penalty held fixed. Supplementary Material 2 gives the details of the computation and these influence functions.

\section{Simulation study}\label{sec:5}

This section compares three versions of the proposed estimator with three estimators that use the shared measurement in place of the characteristic or the observational sample alone, in designs in which selection into treatment in the observational population depends on the characteristic and the shared measurement records the characteristic with error. Section 5.1 describes the designs, Section 5.2 the estimators, Section 5.3 the results, and Section 5.4 the bounds of Section 4. Supplementary Material 2 gives the details of the computation.

\subsection{Design}\label{sec:5.1}

This subsection describes the two data-generating designs. In each replication, we draw an experimental sample of size $n_E$ and an observational sample of size $n_O$ independently, with $n_E=n_O\in\{1{,}000,2{,}000,4{,}000,8{,}000\}$, and each sample size has 1,000 replications. The covariate $X$ takes the values $0,1,2,3$ with probability $0.25$ each in the experimental population and with probabilities $0.3,0.3,0.2,0.2$ in the observational population. In both populations, $X$, $U_1$, $U_2$, and $U_3$ are mutually independent, and $U_2$ and $U_3$ are uniform on $[-1,1]$. The potential values in both populations and the treatment in the observational population are generated by
\begin{equation}\tag{5.1}\label{eq:5.1}
\begin{aligned}
&S(t)=U_1+(1+0.5X)t,\\
&M(t)=S(t)+U_2,\\
&Y(t)=2S(t)+3U_1+5U_3,\\
&P(T=1\mid U_1,U_2,U_3,X,O)=0.5 + 0.25U_1
\end{aligned}
\end{equation}
and in the experimental population, $T$ is independent of $(X,U_1,U_2,U_3)$ with $P(T=1\mid E)=0.5$. The experimental sample contains $(X,T,M,S)$ and the observational sample $(X,T,M,Y)$.

The two designs differ only in the distribution of $U_1$. In the continuous design, $U_1$ is uniform on $[-1,1]$, and in the discrete design, it is uniform on $\{-1,0,1\}$, so the characteristic takes three values within each cell, that is, each value of $X$, and arm. The effect of the treatment on the long-term outcome is $Y(1)-Y(0)=2\{S(1)-S(0)\}=2+X$ for every unit, so the target (2.1) is $\tau=2+\mathbb E[X\mid O]=2+(0.3+0.4+0.6)=3.30$ in both designs.

Both designs satisfy Assumptions 2.1–2.4 by construction, and as special cases of the measurement error model in Example 2.1, they satisfy Assumptions 2.5–2.7. Completeness in Assumption 2.8 holds because the characteristic function of $U_2$ vanishes only at isolated points (Appendix A.2). The selection probability lies in $[0.25,0.75]$, so Assumption 3.3 holds with the truncation constant $a=0.10$. In both designs, bounded weight functions satisfying (3.2) exist but are not unique (Section S2.1.1 of Supplementary Material 2).

\subsection{Estimators}\label{sec:5.2}

We report three versions of the proposed estimator, all computed with series estimates of the nuisance functions. Proposed (Tikhonov) is the estimator of Section 3.2 with series estimates that include a Tikhonov penalty. Proposed (Tikhonov, cross-fitted) is the estimator of Section 3.4, in which the same series estimates are computed from the observations outside each fold, and we report it with $K=2$ and $K=4$ folds. Proposed (unpenalized) computes the estimator of Section 3.2 with the same series estimates and the penalty set to zero, as defined below. The basis $b_J$ in $S$ consists of the monomials of degree at most 2 in both designs, so $J=3$, and in the discrete design it spans all functions of the three values of the characteristic. The test functions $\varphi_L$ are the monomials of degree at most 5 in $M$, so $L=6$, and both variables are standardized in the experimental sample. In all three versions and in the bounds of Section 5.4, the estimated selection probability is truncated to $[0.10,1]$.

We estimate the selection probability $\pi_t$ and the product $\nu_t$ in the same way by series methods, using the components of $\varphi_L$ as test functions in (2.7). For $\pi_t(s,x)=b_J(s)^\top\beta$ and $\nu_t(s,x)=b_J(s)^\top\gamma$, the two equations of (2.7) become $\mathbb E[\mathbf 1\{T=t\}\varphi_L(M)\mid x,O]=Q_t(x)\beta$ and $\mathbb E[\mathbf 1\{T=t\}Y\varphi_L(M)\mid x,O]=Q_t(x)\gamma$ with $Q_t(x)=\mathbb E[\varphi_L(M)b_J(S)^\top\mid T=t,x,E]$, two linear systems with the same matrix. With the sample means $\widehat Q_t(x)$ and $\widehat V(x)=\widehat{\mathbb E}[\varphi_L(M)\varphi_L(M)^\top\mid x,O]$, the Tikhonov versions estimate both coefficient vectors with the regularized matrix
\begin{equation}\tag{5.2}\label{eq:5.2}
\widehat A_{t,\lambda}(x)=\{\widehat Q_t(x)^\top\widehat V(x)^{-1}\widehat Q_t(x)+\lambda I\}^{-1}\widehat Q_t(x)^\top\widehat V(x)^{-1}
\end{equation}
where $I$ is the $J\times J$ identity matrix and $\lambda>0$ is a penalty. The estimates of $\beta$ and $\gamma$ are $\widehat A_{t,\lambda}(x)$ times the sample means of $\mathbf 1\{T=t\}\varphi_L(M)$ and $\mathbf 1\{T=t\}Y\varphi_L(M)$ over the observational units in cell $x$, and the same matrix also gives the estimates of the two weight functions. Each of the two estimates minimizes the squared residuals of its moment equations weighted by $\widehat V(x)^{-1}$ plus $\lambda$ times the squared norm of its coefficients. The penalty is chosen from $10^{-6},10^{-5},\ldots,10^{0}$ by two-fold validation of the moment conditions, with one value common to all cells and arms, and the cross-fitted version chooses it from the observations outside each fold. Proposed (unpenalized) uses the matrix (5.2) with $\lambda=0$, which requires $\widehat Q_t(x)^\top\widehat V(x)^{-1}\widehat Q_t(x)$ to be invertible. The standard errors of the three versions are those of (3.7), computed for the cross-fitted version as in Section 3.4.

The three comparison estimators use the shared measurement in place of the characteristic or the observational sample alone, and we compute them with the cells of the proposed estimator. ESC is the experimental selection correction of \citet{ref4}, a control function approach, with $M$ in place of their short-term outcome. Its control variable is the empirical distribution function of $M$ in the same cell and arm of the experimental sample, evaluated at the observed $M$. Within each cell and arm of the observational sample, ESC regresses $Y$ on the monomials of degree at most 5 in this control variable. ESC Imputation regresses $Y$ on $\varphi_L(M)$ within each cell and arm of the observational sample and averages the fitted values over the same cell and arm of the experimental sample. Obs Regression is the least squares coefficient on $T$ in the regression of $Y$ on an intercept, $T$, and cell indicators in the observational sample. Their standard errors are computed, like those of the proposed estimator, from estimates of their influence functions.

\subsection{Results}\label{sec:5.3}

Tables 2 and 3 report, for the continuous and discrete designs, the errors of the estimators and the coverage of their 95\% intervals, and the notes to the tables define the columns. In both designs, the Tikhonov version has the smallest absolute bias of the seven estimators at every sample size, and its bias is small compared with the standard deviation of its estimates. The coverage of its 95\% intervals rises toward the nominal level as the samples grow, from 81.0\% at $n_E=n_O=1{,}000$ to 92.4\% at 8,000 in the discrete design (Table 3) and from 68.6\% to 89.3\% in the continuous design (Table 2). The biases of the comparison estimators do not shrink as the samples grow. The biases of ESC and ESC Imputation stay at about a quarter of $\tau$ in the continuous design and a third of it in the discrete design, and the bias of Obs Regression stays at about half of $\tau$ in the continuous design and about as large as $\tau$ in the discrete design. As the samples grow, the coverage of ESC and ESC Imputation falls toward zero, and that of Obs Regression is zero at every sample size.

The cross-fitted versions with two and four folds have a larger absolute bias and lower coverage than the version without sample splitting at every sample size in both designs, and at 8,000 their coverage is between 72.1\% and 85.7\%. The unpenalized version is computed in every replication, but its estimates are more dispersed than those of the Tikhonov version, its absolute bias exceeds that of the Tikhonov version at every sample size in both designs, and its absolute bias is smaller than those of all three comparison estimators only at 4,000 and 8,000 in the discrete design. The Tikhonov version has a smaller absolute bias than all three comparison estimators in both designs at every sample size, and it is therefore the version we recommend.

\begin{table}[tbp]
\centering
\def\MainTextTableBody{
\begin{tabular}{rlrrrrr}
\toprule
$n_E=n_O$ & Estimator & Bias & SD & RMSE & SE & Coverage \\
\midrule
1,000 & Proposed (Tikhonov) & \textminus{}0.239 & 2.792 & 2.800 & 0.752 & 68.6 \\
1,000 & Proposed (Tikhonov, cross-fitted, 2 folds) & 1.683 & 4.858 & 5.139 & 1.349 & 39.4 \\
1,000 & Proposed (Tikhonov, cross-fitted, 4 folds) & 1.162 & 2.515 & 2.769 & 1.227 & 57.4 \\
1,000 & Proposed (unpenalized) & \textminus{}12.860 & 22.653 & 26.039 & 12.285 & 97.9 \\
1,000 & ESC & 0.858 & 0.268 & 0.898 & 0.272 & 12.8 \\
1,000 & ESC Imputation & 0.869 & 0.274 & 0.911 & 0.280 & 12.9 \\
1,000 & Obs Regression & 1.680 & 0.247 & 1.698 & 0.255 & 0.0 \\
2,000 & Proposed (Tikhonov) & \textminus{}0.235 & 1.551 & 1.568 & 0.518 & 76.3 \\
2,000 & Proposed (Tikhonov, cross-fitted, 2 folds) & 0.931 & 4.164 & 4.265 & 0.714 & 53.6 \\
2,000 & Proposed (Tikhonov, cross-fitted, 4 folds) & 0.630 & 1.301 & 1.445 & 0.725 & 72.4 \\
2,000 & Proposed (unpenalized) & \textminus{}4.449 & 3.941 & 5.942 & 4.477 & 98.8 \\
2,000 & ESC & 0.851 & 0.191 & 0.872 & 0.189 & 0.5 \\
2,000 & ESC Imputation & 0.854 & 0.192 & 0.875 & 0.189 & 0.6 \\
2,000 & Obs Regression & 1.673 & 0.180 & 1.682 & 0.180 & 0.0 \\
4,000 & Proposed (Tikhonov) & \textminus{}0.206 & 0.986 & 1.006 & 0.378 & 79.9 \\
4,000 & Proposed (Tikhonov, cross-fitted, 2 folds) & 0.490 & 0.977 & 1.092 & 0.415 & 63.3 \\
4,000 & Proposed (Tikhonov, cross-fitted, 4 folds) & 0.352 & 0.691 & 0.775 & 0.453 & 78.5 \\
4,000 & Proposed (unpenalized) & \textminus{}1.853 & 1.635 & 2.470 & 1.976 & 98.9 \\
4,000 & ESC & 0.853 & 0.134 & 0.863 & 0.133 & 0.0 \\
4,000 & ESC Imputation & 0.854 & 0.133 & 0.864 & 0.132 & 0.0 \\
4,000 & Obs Regression & 1.667 & 0.128 & 1.672 & 0.128 & 0.0 \\
8,000 & Proposed (Tikhonov) & \textminus{}0.094 & 0.420 & 0.430 & 0.233 & 89.3 \\
8,000 & Proposed (Tikhonov, cross-fitted, 2 folds) & 0.296 & 0.495 & 0.577 & 0.271 & 73.1 \\
8,000 & Proposed (Tikhonov, cross-fitted, 4 folds) & 0.206 & 0.375 & 0.428 & 0.284 & 85.7 \\
8,000 & Proposed (unpenalized) & \textminus{}0.853 & 0.746 & 1.133 & 0.975 & 99.7 \\
8,000 & ESC & 0.848 & 0.093 & 0.854 & 0.094 & 0.0 \\
8,000 & ESC Imputation & 0.848 & 0.092 & 0.853 & 0.093 & 0.0 \\
8,000 & Obs Regression & 1.666 & 0.090 & 1.669 & 0.090 & 0.0 \\
\bottomrule
\end{tabular}
}
\setlength{\tabcolsep}{4pt}\small
\sbox{\MainTextTableBox}{\MainTextTableBody}
\def\MainTextTableFont{small}
\ifdim\wd\MainTextTableBox>\textwidth
  \footnotesize\sbox{\MainTextTableBox}{\MainTextTableBody}
  \def\MainTextTableFont{footnotesize}
\fi
\typeout{TABLE-WIDTH 2 font=\MainTextTableFont; width=\the\wd\MainTextTableBox; limit=\the\textwidth}
\ifdim\wd\MainTextTableBox>\textwidth
  \PackageError{main-text}{Table 2 remains too wide}{Stop and report; no scaling allowed}
\fi
\usebox{\MainTextTableBox}
\setcounter{table}{1}
\caption{Errors and coverage of the estimators in the continuous design}
\label{table:2}
\par\smallskip{\small\NoteLeft Note: The true effect is $\tau=3.30$, and each sample size has 1,000 replications. Bias is the mean of the estimates minus $\tau$, SD is the standard deviation of the estimates, and RMSE is the square root of the mean of the squared differences between the estimates and $\tau$. SE is the mean of the standard errors, and Coverage is the percentage of replications in which the estimate plus or minus 1.96 standard errors contains $\tau$. The standard errors of the three versions of the proposed estimator are those of (3.7), and those of the comparison estimators are computed from estimates of their influence functions. All 1,000 replications are used for every estimator and sample size.\par}
\end{table}

\begin{table}[tbp]
\centering
\def\MainTextTableBody{
\begin{tabular}{rlrrrrr}
\toprule
$n_E=n_O$ & Estimator & Bias & SD & RMSE & SE & Coverage \\
\midrule
1,000 & Proposed (Tikhonov) & 0.035 & 1.670 & 1.670 & 0.685 & 81.0 \\
1,000 & Proposed (Tikhonov, cross-fitted, 2 folds) & 1.608 & 2.298 & 2.804 & 0.886 & 43.2 \\
1,000 & Proposed (Tikhonov, cross-fitted, 4 folds) & 1.028 & 1.459 & 1.784 & 0.862 & 63.8 \\
1,000 & Proposed (unpenalized) & \textminus{}5.111 & 5.222 & 7.305 & 5.358 & 98.2 \\
1,000 & ESC & 1.110 & 0.355 & 1.166 & 0.351 & 13.1 \\
1,000 & ESC Imputation & 1.116 & 0.369 & 1.176 & 0.342 & 10.5 \\
1,000 & Obs Regression & 3.340 & 0.300 & 3.353 & 0.300 & 0.0 \\
2,000 & Proposed (Tikhonov) & 0.067 & 0.742 & 0.744 & 0.427 & 82.9 \\
2,000 & Proposed (Tikhonov, cross-fitted, 2 folds) & 0.799 & 1.073 & 1.337 & 0.566 & 63.4 \\
2,000 & Proposed (Tikhonov, cross-fitted, 4 folds) & 0.582 & 0.759 & 0.956 & 0.523 & 74.6 \\
2,000 & Proposed (unpenalized) & \textminus{}1.764 & 1.938 & 2.620 & 1.985 & 98.5 \\
2,000 & ESC & 1.115 & 0.246 & 1.142 & 0.238 & 0.2 \\
2,000 & ESC Imputation & 1.112 & 0.242 & 1.138 & 0.233 & 0.4 \\
2,000 & Obs Regression & 3.348 & 0.216 & 3.355 & 0.212 & 0.0 \\
4,000 & Proposed (Tikhonov) & 0.030 & 0.433 & 0.434 & 0.287 & 84.3 \\
4,000 & Proposed (Tikhonov, cross-fitted, 2 folds) & 0.474 & 0.584 & 0.752 & 0.330 & 64.6 \\
4,000 & Proposed (Tikhonov, cross-fitted, 4 folds) & 0.335 & 0.410 & 0.530 & 0.312 & 77.6 \\
4,000 & Proposed (unpenalized) & \textminus{}0.688 & 0.763 & 1.027 & 0.796 & 98.7 \\
4,000 & ESC & 1.111 & 0.164 & 1.123 & 0.167 & 0.0 \\
4,000 & ESC Imputation & 1.104 & 0.162 & 1.116 & 0.164 & 0.0 \\
4,000 & Obs Regression & 3.339 & 0.144 & 3.342 & 0.150 & 0.0 \\
8,000 & Proposed (Tikhonov) & 0.008 & 0.232 & 0.232 & 0.197 & 92.4 \\
8,000 & Proposed (Tikhonov, cross-fitted, 2 folds) & 0.249 & 0.252 & 0.354 & 0.205 & 72.1 \\
8,000 & Proposed (Tikhonov, cross-fitted, 4 folds) & 0.163 & 0.238 & 0.288 & 0.206 & 84.8 \\
8,000 & Proposed (unpenalized) & \textminus{}0.265 & 0.333 & 0.425 & 0.366 & 98.1 \\
8,000 & ESC & 1.109 & 0.116 & 1.115 & 0.118 & 0.0 \\
8,000 & ESC Imputation & 1.102 & 0.114 & 1.108 & 0.115 & 0.0 \\
8,000 & Obs Regression & 3.329 & 0.104 & 3.331 & 0.106 & 0.0 \\
\bottomrule
\end{tabular}
}
\setlength{\tabcolsep}{4pt}\small
\sbox{\MainTextTableBox}{\MainTextTableBody}
\def\MainTextTableFont{small}
\ifdim\wd\MainTextTableBox>\textwidth
  \footnotesize\sbox{\MainTextTableBox}{\MainTextTableBody}
  \def\MainTextTableFont{footnotesize}
\fi
\typeout{TABLE-WIDTH 3 font=\MainTextTableFont; width=\the\wd\MainTextTableBox; limit=\the\textwidth}
\ifdim\wd\MainTextTableBox>\textwidth
  \PackageError{main-text}{Table 3 remains too wide}{Stop and report; no scaling allowed}
\fi
\usebox{\MainTextTableBox}
\setcounter{table}{2}
\caption{Errors and coverage of the estimators in the discrete design}
\label{table:3}
\par\smallskip{\small\NoteLeft Note: The entries are defined as in Table 2. All 1,000 replications are used for every estimator and sample size.\par}
\end{table}

\subsection{Bounds without the exclusion restriction for the outcome mean}\label{sec:5.4}

This subsection computes the optimal transport bounds of Section 4 in the designs of Section 5.1 and reports the coverage of their 95\% confidence intervals for $\tau$. In both designs, $M$ is continuous, so Assumption 4.1 does not hold, and we pair the quantiles as in Proposition 4.1 within ten groups of $M$ formed by the deciles of $M$ in each cell and arm of the experimental sample. Because the joint distribution of $Y$ and $W(t)$ within each group has the two marginal distributions used in this pairing, the bounds computed in this way contain $\tau$. The estimators and the confidence intervals are those of Section 4.2 with the series of Section 5.2, and Supplementary Material 2 gives the details.

Table 4 shows that the confidence intervals contain $\tau$ at least at the nominal level in the continuous design at every sample size and in the discrete design from 4,000 on. In the continuous design, they contain $\tau$ in at least 98.8\% of the replications. In the discrete design, their coverage rises from 85.8\% at 1,000 to 98.1\% at 8,000, above the 92.4\% of the Tikhonov version at 8,000 (Table 3), which uses Assumption 2.7. In both designs, the confidence intervals become shorter as the samples grow.

\begin{table}[tbp]
\centering
\def\MainTextTableBody{
\begin{tabular}{lrrrrrr}
\toprule
Design & $n_E=n_O$ & Lower & Upper & CI lower & CI upper & Coverage \\
\midrule
Continuous & 1,000 & 3.077 & 6.124 & 2.285 & 6.737 & 98.8 \\
Continuous & 2,000 & 2.826 & 5.827 & 2.249 & 6.242 & 100.0 \\
Continuous & 4,000 & 2.789 & 5.699 & 2.363 & 6.001 & 99.9 \\
Continuous & 8,000 & 2.798 & 5.612 & 2.486 & 5.826 & 100.0 \\
Discrete & 1,000 & 3.606 & 5.942 & 2.770 & 6.601 & 85.8 \\
Discrete & 2,000 & 3.360 & 5.767 & 2.795 & 6.184 & 92.7 \\
Discrete & 4,000 & 3.259 & 5.661 & 2.866 & 5.942 & 96.3 \\
Discrete & 8,000 & 3.232 & 5.573 & 2.958 & 5.763 & 98.1 \\
\bottomrule
\end{tabular}
}
\setlength{\tabcolsep}{4pt}\small
\sbox{\MainTextTableBox}{\MainTextTableBody}
\def\MainTextTableFont{small}
\ifdim\wd\MainTextTableBox>\textwidth
  \footnotesize\sbox{\MainTextTableBox}{\MainTextTableBody}
  \def\MainTextTableFont{footnotesize}
\fi
\typeout{TABLE-WIDTH 4 font=\MainTextTableFont; width=\the\wd\MainTextTableBox; limit=\the\textwidth}
\ifdim\wd\MainTextTableBox>\textwidth
  \PackageError{main-text}{Table 4 remains too wide}{Stop and report; no scaling allowed}
\fi
\usebox{\MainTextTableBox}
\setcounter{table}{3}
\caption{Estimated bounds and coverage of the confidence intervals for $\tau$ in the designs with a shared measurement}
\label{table:4}
\par\smallskip{\small\NoteLeft Note: The true effect is $\tau=3.30$, and each sample size has 1,000 replications. Lower and Upper are the means of the estimated ends of the interval (4.4), computed by pairing the quantiles as in Proposition 4.1 within the decile groups of $M$ in the experimental sample. A decile group with no observations is merged with an adjacent group, which happens in 46 and 128 of the 1,000 replications at $n_E=n_O=1{,}000$ in the continuous and discrete designs and in 1 replication at 2,000 in the discrete design. CI lower and CI upper are the means of the ends of the 95\% confidence intervals of Section 4.2, and Coverage is the percentage of replications in which these confidence intervals contain $\tau$. The standard errors are computed from the influence functions of Section 4.2.\par}
\end{table}

\section{Empirical illustration with the Job Corps data}\label{sec:6}

This section illustrates the estimator of Section 3 under the design of Section 2 with data from the National Job Corps Study. Section 6.1 describes the data and the construction of the experimental and observational samples, Section 6.2 describes the estimators and compares their errors with a pseudo-true value computed from the experiment, and Section 6.3 reports the bounds of Section 4 and narrows them under monotonicity restrictions on the outcome mean that are plausible in this example. Supplementary Material 2 gives the details of the sample construction and the computation.

\subsection{Data and sample construction}\label{sec:6.1}

This subsection describes the data, the variables, and how the two samples are drawn. Our data are the public data from the randomized evaluation of Job Corps, a training program for disadvantaged youth in the United States in which the offer to enroll was randomly assigned \citep{ref30}. The data are distributed with the R package causalweight. Of the 9,240 persons in the public sample, we use the 9,084 with nonmissing years of education, which is one of the covariates of the comparison regression in Section 6.2, and of these persons 5,483 were assigned to treatment and 3,601 to control.

We use year-4 weekly earnings as the outcome and obtain the characteristic and the shared measurement by discretizing two variables recorded in the data. The target is the average effect of the offer $T\in\{0,1\}$ on year-4 weekly earnings $Y$. The characteristic $S\in\{1,2,3\}$ is the category of the recorded year-3 weekly earnings of a person, with $S=1$ for no earnings in year 3, $S=2$ for positive weekly earnings at or below the median of positive year-3 weekly earnings among the 9,084 persons (about 188 dollars), and $S=3$ for weekly earnings above that median. The shared measurement $M$ is the level of the recorded share of weeks worked in year 3, which records whether a person worked in no week of year 3, worked a positive share of weeks at or below the median of the positive shares among the 9,084 persons (about 73 percent), or worked a larger share. Both variables are thus the recorded values of each person coarsened into three categories. The pretreatment variable $X$ takes six values, the cells formed by gender and three groups of age at random assignment, 16–17, 18–19, and 20–24 years, which are the age groups for which \citet{ref30} report separate impact estimates. The three earnings categories contain 905, 2,260, and 2,318 of the 5,483 treated persons and 667, 1,496, and 1,438 of the 3,601 control persons, and the three levels of $M$ contain 898, 2,284, and 2,301 of the treated and 663, 1,515, and 1,423 of the control persons.

We construct an observational sample in which selection into treatment depends on the characteristic, because the offer was assigned independently of the potential values within each cell and a sample drawn with equal probabilities would be another experiment. We treat the 9,084 persons as the population and, in each of 1,000 replications, draw from it with replacement two independent samples of 4,500 persons each. The experimental sample is drawn with equal probabilities and records $(X,T,S,M)$. The observational sample records $(X,T,M,Y)$, and its draws keep the share of each cell in the population but distort, within each cell, the probabilities of drawing treated and control persons in each earnings category. In the resulting observational population, the selection probability $\pi_1(k,x)=P(T=1\mid S(1)=k,x,O)$ of a person in cell $x$ whose earnings category under the offer is $k$ depends on $k$, and comparisons between treated and control persons given $X$ and $M$ are biased.

Nonexperimental comparisons of participants with nonparticipants can differ substantially from experimental estimates \citep{ref23}, and participants in training programs often experience a decline in earnings before they enter \citep{ref3,ref20}. To evaluate the estimators under negative selection, we build into the observational sample a selection rule under which persons with lower year-3 earnings under the offer are more likely to receive it. Within each cell, a person is drawn with probability proportional to a weight that equals $8/13$, $1/2$, and $5/13$ for treated persons with no year-3 earnings, with positive earnings at or below the median, and with earnings above the median, and $5/13$, $1/2$, and $8/13$ for control persons in the same categories, so the treated and control weights of each category sum to one. Within each cell, $\pi_1(k,x)$ is then proportional to $8/13$, $1/2$, and $5/13$ for $k=1,2,3$. Under this selection rule, about three quarters of the persons with no year-3 earnings under the offer receive the offer in the observational population, compared with about half of the persons above the median. Section S2.2.1 of Supplementary Material 2 gives the weights, and Section S2.2.8 reports the results under two other selection rules, under which the selection probability rises with the earnings category or is lowest in the middle category.

Figure 2 plots the selection probability $\pi_1(k,x)$ of each cell against the earnings category $k$ under the offer, together with the same probability in the experimental population, $P(T=1\mid S(1)=k,x,E)$, which equals the share of the offer in cell $x$ for every $k$. In the experimental population, the selection probability does not depend on the earnings category under the offer, and it lies between 0.56 and 0.66 across the cells. In the observational population, it falls with the earnings category in equal steps within each cell.

\begin{figure}[htbp]
\centering
\includegraphics[max width=\textwidth]{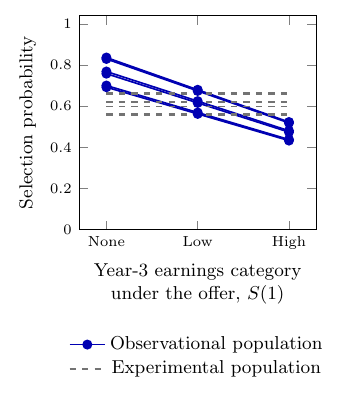}
\setcounter{figure}{1}
\caption{Selection probability by year-3 earnings category in the observational and experimental populations}
\label{figure:2}
\par\smallskip{\small\NoteLeft Note: Each solid line connects $\pi_1(k,x)=P(T=1\mid S(1)=k,x,O)$, $k=1,2,3$, for one of the six cells $x$, and each dashed line gives $P(T=1\mid S(1)=k,x,E)$, which equals the share of the offer among the persons in cell $x$ for every $k$. The ratios between the values of $\pi_1(k,x)$ for different $k$ are the same in every cell, so the solid lines have the same shape and differ across cells by a common factor. Lines of cells with nearly equal values overlap. None, Low, and High are the earnings categories $k=1,2,3$ of Section 6.1 under the offer.\par}
\end{figure}

We compare the estimates with the pseudo-true value
\[
\tilde\tau=\sum_{x}\bar p(x)\{\bar Y_1(x)-\bar Y_0(x)\}=19.50
\]
where $\bar p(x)$ is the share of cell $x$ among the 9,084 persons and $\bar Y_t(x)$ is the mean of $Y$ among the persons with $T=t$ and $X=x$. It is common to all estimators.

In this construction, Assumptions 2.1–2.6, 3.2, and 3.3 hold by design for the populations from which the samples are drawn, extending the independent sampling scheme with equal sample sizes yields Assumption 3.1, and the pseudo-true value $\tilde\tau$ equals the average treatment effect of the observational population, so the coverage reported below refers to repeated draws from the 9,084 persons. Because $S$ and $M$ each take three values, completeness in Assumption 2.8 and the existence of the weight functions in Assumption 3.4 follow from a rank condition on the joint probabilities of the earnings categories and the levels of $M$, which holds in every cell and arm, and the weight functions are unique.

In this construction, Assumption 2.7 is violated, although the violation is small relative to the variation in year-4 earnings across the earnings categories. The assumption requires that, among the persons with the same cell, arm, and earnings category, the mean of year-4 earnings be the same at every level of $M$. The least-squares regression of $Y_i$ on the indicators of the combinations of cell and arm, of the two higher earnings categories, and of the two higher levels of $M$, computed from the 9,084 persons, is
\begin{equation}\tag{6.1}\label{eq:6.1}
\widehat Y_i=\sum_{x}\sum_{t=0}^{1}\widehat\alpha_{x,t}\,\mathbf 1\{X_i=x,T_i=t\}+\underset{(23.25)}{121.12}\,\mathbf 1\{S_i=2\}+\underset{(23.90)}{268.71}\,\mathbf 1\{S_i=3\}\underset{(23.27)}{-35.78}\,\mathbf 1\{M_i=2\}\underset{(23.93)}{-5.12}\,\mathbf 1\{M_i=3\}
\end{equation}
where $x$ runs over the six cells, $S_i=2$ and $S_i=3$ indicate positive year-3 earnings below and at or above the median, $M_i=2$ and $M_i=3$ indicate a positive share of weeks worked in year 3 at or below and above the median, and heteroskedasticity-robust (HC3) standard errors are in parentheses. The test that both coefficients on the levels of $M$ are zero rejects with a p-value below 0.0001, but these coefficients are much smaller than those on the earnings categories. Section 6.3 therefore drops Assumption 2.7 and computes the bounds of Section 4, together with monotonicity restrictions that narrow them.

\subsection{Estimation and results}\label{sec:6.2}

This subsection describes the estimators and reports their errors and coverage. We compute the three versions of the proposed estimator of Section 5.2, with the estimated selection probability truncated to $[0.05,1]$. The basis $b_J$ in $S$ consists of the indicators of the three earnings categories and the test functions $\varphi_L$ are the indicators of the three levels of $M$, so $J=L=3$. As in Section 5.2, the cross-fitted version uses two and four folds. We also compute the three comparison estimators of Section 5.2, which use $M$ or pretreatment covariates in place of $S$. ESC uses a quadratic function of its single control variable, ESC Imputation uses the indicator test functions $\varphi_L(M)$, and Obs Regression adds 26 pretreatment covariates of the public data to the regression of Section 5.2.

Table 5 reports the errors of the estimators and the coverage of their 95\% intervals over the 1,000 replications. Proposed (Tikhonov) has a smaller absolute bias and a higher coverage than the three comparison estimators. Its bias is less than half of the standard deviation of its estimates, and its coverage is 92.4\%. The cross-fitted version with four folds has an even smaller absolute bias and coverage of 94.9\%, whereas with two folds its estimates are about three times as dispersed and its coverage is 90.4\%. The unpenalized version is close to the Tikhonov version. The two comparison estimators that use $M$ in place of the earnings category are biased downward by more than twice the absolute bias of the Tikhonov version, and their coverage is 69.4\% and 77.4\%. Obs Regression is biased downward by more than the pseudo-true value itself, so its mean estimate is negative although the pseudo-true value is positive, and its coverage is 0.2\%.

\begin{table}[tbp]
\centering
\def\MainTextTableBody{
\begin{tabular}{lrrrrr}
\toprule
Estimator & Bias & SD & RMSE & SE & Coverage \\
\midrule
Proposed (Tikhonov) & 3.33 & 7.84 & 8.51 & 7.52 & 92.4 \\
Proposed (Tikhonov, cross-fitted, 2 folds) & \textminus{}0.51 & 22.99 & 22.99 & 8.85 & 90.4 \\
Proposed (Tikhonov, cross-fitted, 4 folds) & 1.97 & 7.80 & 8.04 & 7.90 & 94.9 \\
Proposed (unpenalized) & 3.54 & 7.90 & 8.66 & 8.03 & 93.8 \\
ESC & \textminus{}8.21 & 5.58 & 9.92 & 5.53 & 69.4 \\
ESC Imputation & \textminus{}7.09 & 5.81 & 9.16 & 5.77 & 77.4 \\
Obs Regression & \textminus{}28.76 & 5.71 & 29.32 & 5.59 & 0.2 \\
\bottomrule
\end{tabular}
}
\setlength{\tabcolsep}{4pt}\small
\sbox{\MainTextTableBox}{\MainTextTableBody}
\def\MainTextTableFont{small}
\ifdim\wd\MainTextTableBox>\textwidth
  \footnotesize\sbox{\MainTextTableBox}{\MainTextTableBody}
  \def\MainTextTableFont{footnotesize}
\fi
\typeout{TABLE-WIDTH 5 font=\MainTextTableFont; width=\the\wd\MainTextTableBox; limit=\the\textwidth}
\ifdim\wd\MainTextTableBox>\textwidth
  \PackageError{main-text}{Table 5 remains too wide}{Stop and report; no scaling allowed}
\fi
\usebox{\MainTextTableBox}
\setcounter{table}{4}
\caption{Errors and coverage of the estimators}
\label{table:5}
\par\smallskip{\small\NoteLeft Note: The pseudo-true value is $\tilde\tau=19.50$. There are 1,000 replications, in all of which every estimate and standard error was computed. The cross-fitted version splits each sample into two and four folds. Bias is the mean of the estimates minus $\tilde\tau$, SD is the standard deviation of the estimates, and RMSE is the square root of the mean of the squared differences between the estimates and $\tilde\tau$. SE is the mean of the standard errors, and Coverage is the percentage of replications in which the estimate plus or minus 1.96 standard errors contains $\tilde\tau$. Estimates are in dollars per week.\par}
\end{table}

\subsection{Bounds without the exclusion restriction for the outcome mean and with monotonicity restrictions}\label{sec:6.3}

This subsection computes the bounds of Section 4, because Assumption 2.7 fails in this construction (Section 6.1), and narrows them with two monotonicity restrictions on the outcome mean that are plausible in this example. Since $S$ and $M$ take three values each, Assumption 4.1 holds. The estimators and the confidence intervals are those of Section 4.2, with the indicator bases of Section 6.2, and Supplementary Material 2 gives the details of the optimal transport bounds.

The bounds of Proposition 4.1 place no restriction on how the mean of the outcome depends on the characteristic and the shared measurement, but in applications the direction of this dependence is often known. To impose it, we write
\[
\mu_t(s,m,x)=\mathbb E[Y\mid T=t,S=s,M=m,X=x,G=O]
\]
for the conditional mean in Assumption 2.7, which states that $\mu_t(s,m,x)=\mu_t(s,x)$ does not depend on $m$, and we restrict the direction in which it moves with $s$ and with $m$. Condition (1) of Assumption 6.1 adds a new linear restriction, and condition (2) relaxes Assumption 2.7.

\renewcommand{\thefixedassumption}{6.1}
\begin{fixedassumption}[Monotone outcome means]\label{assumption:6.1}

For each $t$ and $F_{X\mid O}$-almost every $x$, over the pairs $(s,m)$ with $P(S(t)=s,M(t)=m\mid T=t,x,O)>0$,

\begin{enumerate}
\item $\mu_t(s,m,x)$ is nondecreasing in $s$ for each $m$, and
\item $\mu_t(s,m,x)$ is nondecreasing in $m$ for each $s$.
\end{enumerate}
\end{fixedassumption}

Assumption 6.1 restricts the direction in which the outcome mean moves with the characteristic and with the shared measurement. Condition (1) requires that, among units with the same treatment, covariates, and shared measurement, those with a higher characteristic have a weakly higher mean outcome. Condition (2) requires that, among units with the same treatment, covariates, and characteristic, those with a higher shared measurement have a weakly higher mean outcome. Condition (2) weakens Assumption 2.7, which requires this mean to be the same at every value of $m$, to the requirement that the mean not decrease in $m$. It therefore allows the conditional mean to increase with the shared measurement, as when workers who worked more weeks have accumulated more work experience and earn more later, in which case Assumption 2.7 fails.

Condition (1) is plausible in this example and holds in the data. Among persons with the same arm, gender, age group, and level of the share of weeks worked in year 3, a higher category of year-3 weekly earnings can reflect higher pay per week worked or more hours, which depend on skills and job characteristics that carry over to the next year, so these persons can be expected to earn more in year 4 as well. In this construction, $\mu_t(s,m,x)$ is the mean of $Y$ among the 9,084 persons with $T_i=t$, $S_i=s$, $M_i=m$, and $X_i=x$, because the selection probability depends on the potential values only through $S(t)$ (Section S2.2.3 of Supplementary Material 2), so both conditions can be checked in these persons. Figure 3 compares these means within every pair of adjacent earnings categories and every pair of adjacent levels of $M$. In the left table, the mean of $Y$ is higher in the higher category in all 31 pairs of adjacent earnings categories that share an arm, cell, and level of $M$ and both contain persons, so condition (1) holds in every pair.

Condition (2) has support but fails in part of the data. Among persons with the same arm, gender, age group, and earnings category, those who worked more weeks in year 3 may have accumulated more work experience and be more attached to their jobs, which suggests higher year-4 earnings. In the right table of Figure 3, however, the mean of $Y$ is lower at the higher level of $M$ in 12 of the 31 pairs of adjacent levels that share an arm, cell, and earnings category and both contain persons. Each of these decreases is modest or rests on very few persons. In five of them, at least one of the two levels contains five or fewer persons, so the observed order of the two means is unreliable, and in the other seven, the decrease is less than a quarter of the average difference between adjacent earnings categories in the left table.

\begin{figure}[htbp]
\centering
\includegraphics[max width=\textwidth,max height=0.78\textheight]{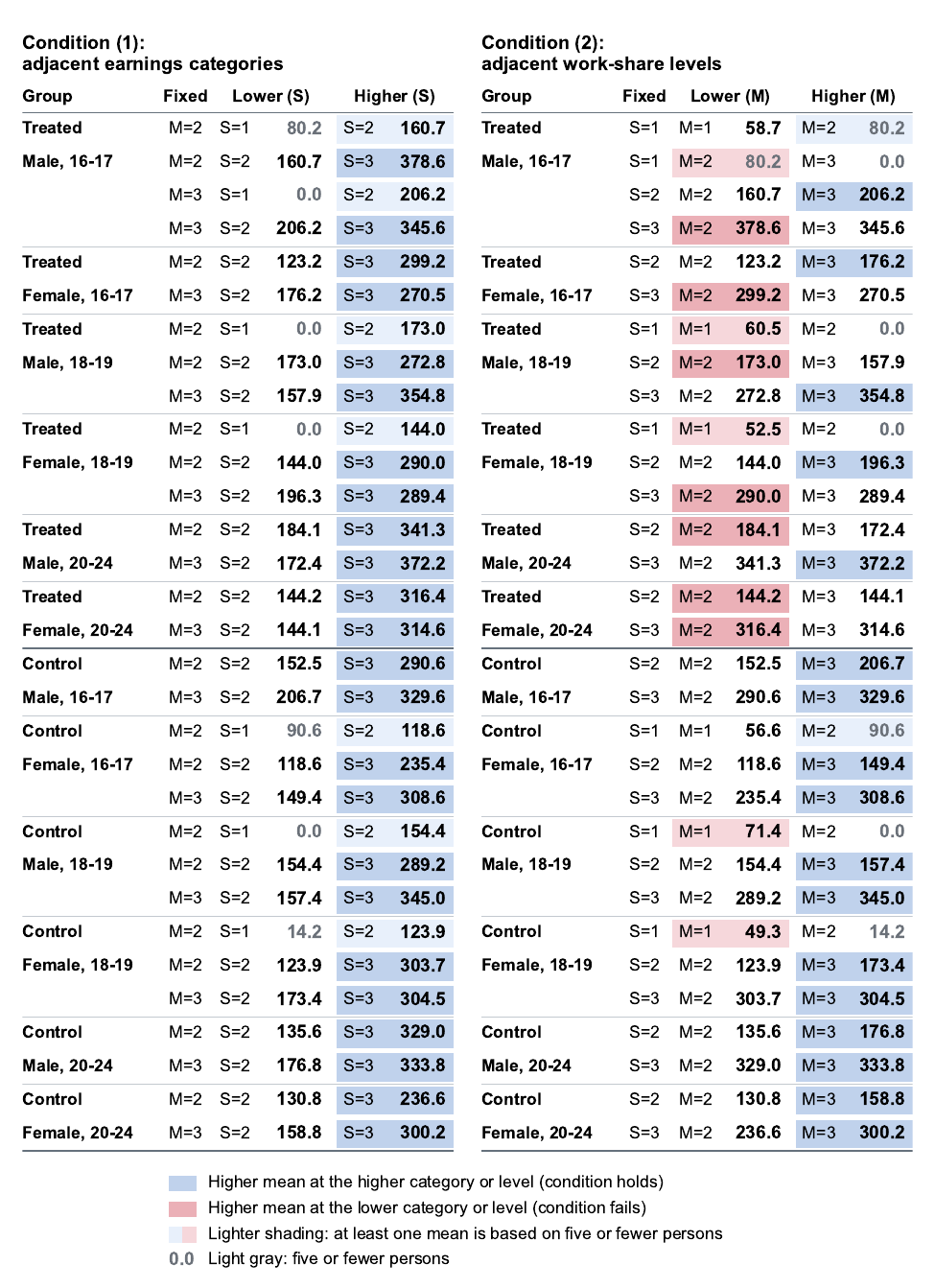}
\setcounter{figure}{2}
\caption{Mean year-4 weekly earnings by year-3 earnings category and level of the share of weeks worked}
\label{figure:3}
\par\smallskip{\footnotesize\NoteLeft Note: Each row compares the mean of $Y$ at two adjacent categories or levels that share an arm, a cell (gender and age group), and the value of the other variable, which is given in the column Fixed. The left table compares adjacent earnings categories $S$ (condition (1) of Assumption 6.1) and the right table compares adjacent levels of $M$ (condition (2)). In each row, the columns Lower ($S$) and Higher ($S$) in the left table and Lower ($M$) and Higher ($M$) in the right table give the mean of $Y$ in dollars per week at the lower and at the higher category or level, each labeled by its value of $S$ or $M$. The larger mean is shaded, blue if it is at the higher category or level and red if it is at the lower one, and the shading is lighter when at least one of the two means is based on five or fewer persons. A mean based on five or fewer persons is printed in light gray. Pairs in which either category or level contains no persons are omitted.\par}
\end{figure}

Under Assumption 4.1, both conditions are linear restrictions on the couplings within the cells. For each $m$, let $\gamma_m$ denote the joint distribution of $(S(t),Y)$ given $T=t$, $M=m$, $X=x$, and $G=O$. Its marginal distributions are $\lambda_{t,m,x}(s)=P(S(t)=s\mid T=t,M(t)=m,x,O)$ and $F_{t,m,x}$, where $\lambda_{t,m,x}(s)$ by Assumptions 2.1, 2.3, 2.4, and 2.6 equals $\pi_t(s,x)P(S=s,M=m\mid T=t,x,E)\big/\sum_{s'}\pi_t(s',x)P(S=s',M=m\mid T=t,x,E)$ and is identified because Assumption 2.8 identifies $\pi_t$. Each coupling determines the conditional means through
\[
\mu_t(s,m,x)=\frac{\int y\,\gamma_m(\{s\},dy)}{\lambda_{t,m,x}(s)},
\]
so each inequality in Assumption 6.1 is a linear inequality in the collection of couplings $(\gamma_m)_m$. Because $W(t)$ is a function of $S(t)$, the objective of (4.3) can also be written as a linear function of these couplings, so the bounds under one or both conditions of Assumption 6.1 are the minimum and the maximum of this objective over the couplings $(\gamma_m)_m$ that satisfy these inequalities. We compute three intervals, the optimal transport bounds of Proposition 4.1 and the bounds under condition (1) of Assumption 6.1 alone and under both of its conditions.

Table 6 reports the three intervals. The monotonicity restrictions shorten the interval considerably. Under condition (1) alone, the distance between the means of the estimated ends is about half of that for the optimal transport bounds, and adding condition (2) shortens the interval further. Computed from the 9,084 persons, the lower end of the optimal transport bounds is negative, whereas under condition (1) alone it is positive, so condition (1), which holds in these data, determines the sign of the effect. The means of the estimated lower ends follow the same pattern. All three confidence intervals cover $\tilde\tau$ in at least 98.3\% of the replications. The confidence interval excludes zero in almost no replication for the optimal transport bounds (0.4\%) and in about half of the replications under the monotonicity restrictions (48.1\% under condition (1) alone and 52.8\% under both conditions). Condition (2) fails in part of these data (Figure 3), so the interval under both conditions rests on a restriction that the 9,084 persons do not satisfy.

Under condition (1) alone, the dual solutions of the linear programs computed from the 9,084 persons are not unique for some ends, so these ends need not be differentiable in the underlying distributions, and we interpret the standard errors as those of the linear approximation at the sample solution.

\begin{table}[tbp]
\centering
\def\MainTextTableBody{
\begin{tabular}{lrrrrr}
\toprule
Bounds & Lower & Upper & CI lower & CI upper & Coverage \\
\midrule
OT bounds & \textminus{}5.91 & 30.27 & \textminus{}17.85 & 43.51 & 100.0 \\
OT bounds with monotonicity in $S$ & 10.23 & 28.54 & \textminus{}0.38 & 41.55 & 99.5 \\
OT bounds with monotonicity in $S$ and $M$ & 10.69 & 23.30 & 0.36 & 35.27 & 98.3 \\
\bottomrule
\end{tabular}
}
\setlength{\tabcolsep}{4pt}\small
\sbox{\MainTextTableBox}{\MainTextTableBody}
\def\MainTextTableFont{small}
\ifdim\wd\MainTextTableBox>\textwidth
  \footnotesize\sbox{\MainTextTableBox}{\MainTextTableBody}
  \def\MainTextTableFont{footnotesize}
\fi
\typeout{TABLE-WIDTH 6 font=\MainTextTableFont; width=\the\wd\MainTextTableBox; limit=\the\textwidth}
\ifdim\wd\MainTextTableBox>\textwidth
  \PackageError{main-text}{Table 6 remains too wide}{Stop and report; no scaling allowed}
\fi
\usebox{\MainTextTableBox}
\setcounter{table}{5}
\caption{Bounds on the effect and coverage of the confidence intervals for $\tilde\tau$}
\label{table:6}
\par\smallskip{\small\NoteLeft Note: The pseudo-true value is $\tilde\tau=19.50$, and there are 1,000 replications. OT denotes optimal transport. The first row reports the interval (4.4) of Proposition 4.1, the second the interval under condition (1) of Assumption 6.1 alone, and the third the interval under both of its conditions. Lower and Upper are the means of the estimated ends, CI lower and CI upper are the means of the ends of the 95\% confidence intervals of Section 4.2, and Coverage is the percentage of replications in which these confidence intervals contain $\tilde\tau$. Estimates are in dollars per week.\par}
\end{table}

Figure 4 places the estimates of Sections 6.2 and 6.3 side by side. Under the monotonicity restrictions, the means of the estimated lower ends are positive, and the intervals between the means of the estimated ends contain the mean estimates of all point estimators except Obs Regression, whose mean estimate is negative. The mean of the estimated lower ends of the optimal transport bounds is negative.

\begin{figure}[htbp]
\centering
\includegraphics[max width=\textwidth]{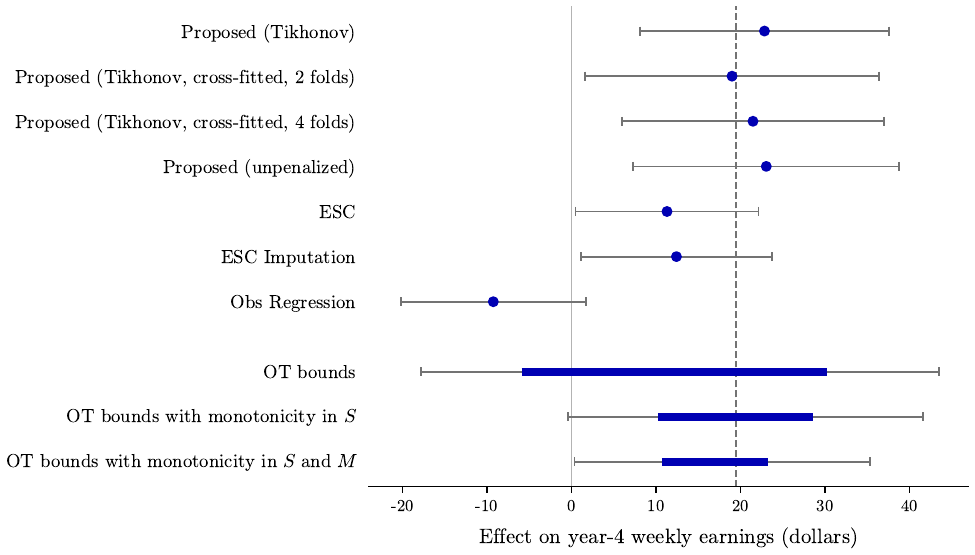}
\setcounter{figure}{3}
\caption{Point estimates, estimated bounds, and 95\% confidence intervals}
\label{figure:4}
\par\smallskip{\small\NoteLeft Note: Each point is the mean of the estimates of a point estimator over the 1,000 replications, and each thick bar connects the means of the estimated ends of an interval. Each thin line connects the means of the ends of the 95\% confidence intervals. The dashed line marks the pseudo-true value $\tilde\tau=19.50$, and the solid line marks zero. OT denotes optimal transport.\par}
\end{figure}

\section{Conclusion}\label{sec:7}

This paper asks what a short-term experiment should measure so that, combined with observational data, it identifies the average long-term treatment effect in the observational population. When the observational data record a coarse post-treatment measurement of the characteristics that drive selection, the experiment should collect the same measurement and measure the characteristics in enough detail to remove confounding, while distinguishing no more levels than the coarse measurement can separate. Under exclusion restrictions and completeness, such an experiment and the observational data identify the effect, and the proposed estimator, which is based on the efficient influence function, is asymptotically normal and attains the semiparametric efficiency bound under the conditions that we give. In simulations and in the Job Corps illustration, it has a smaller absolute bias than estimators that use the coarse measurement in place of the characteristics. When the exclusion restriction for the outcome mean is not plausible, an optimal transport problem gives sharp bounds on the effect, and monotonicity restrictions on the outcome mean narrow them through linear programs. Partial identification without the exclusion restriction for selection is left for future work.

\section*{Declaration of generative AI and AI-assisted technologies in the manuscript preparation process}

During the preparation of this work the author used Claude (Anthropic), Codex (OpenAI), and Gemini (Google) to draft and revise text, translate parts of the manuscript into English, write and check code for the simulations and the empirical illustration, and check derivations and proofs. After using these tools, the author reviewed and edited the content as needed and takes full responsibility for the content of the published article.

\clearpage
\begin{center}
{\LARGE\bfseries Appendix\par}
\vspace{1em}
{\large\bfseries What Should an Experiment Measure? Combining a Short-Term Experiment with Confounded Observational Data to Identify Long-Term Treatment Effects\par}
\vspace{1em}
Kentaro Kawato\par
Graduate School of Economics, The University of Tokyo\par
\end{center}
\section*{Contents}
\AppendixSectionEntry{A}{Proofs for Sections 2 and 4}
\AppendixSubsectionEntry{A.1}{Proof of Theorem 2.1}
\AppendixSubsectionEntry{A.2}{Completeness under additive measurement error}
\AppendixSubsectionEntry{A.3}{Derivation of equation (4.2)}
\AppendixSectionEntry{B}{Proof of Theorem 3.1}
\AppendixSubsectionEntry{B.1}{Model, smooth submodels, and tangent space}
\AppendixSubsectionEntry{B.2}{Lemmas and proof of Theorem 3.1}
\AppendixSubsectionEntry{B.3}{Proofs of Lemmas B.1–B.3}
\AppendixSubsectionEntry{B.4}{Proof of Lemma B.4}
\AppendixSectionEntry{C}{Proofs of Theorems 3.2 and 3.3}
\AppendixSubsectionEntry{C.1}{Proof of Theorem 3.2}
\AppendixSubsectionEntry{C.2}{Lemmas and proof of Theorem 3.3}
\AppendixSectionEntry{D}{Supplementary materials}
\appendix
\AppendixNumbering
\setcounter{secnumdepth}{3}

\section{Proofs for Sections 2 and 4}\label{sec:A}

Appendix A.1 proves Theorem 2.1, the identification result with a shared post-treatment measurement in Section 2, by deriving the two moment equations, showing that their solutions are unique, and relating the solutions to the means of the potential outcomes. Appendix A.2 gives a general sufficient condition for completeness under additive measurement error and verifies, as special cases, completeness in the second part of Example 2.1 in Section 2.2 and in the simulation designs of Section 5. Appendix A.3 derives equation (4.2) of Section 4.1, which gives the distribution of the inverse of the selection probability within the cells of the observational data.

\subsection{Proof of Theorem 2.1}\label{sec:A.1}

\subsubsection{Integrability conditions}\label{sec:A.1.1}

Throughout, we assume that $\mathbb E[|Y(t)|\mid O]<\infty$ and that $\mathbb E[Y^2\mid T=t,x,O]<\infty$ for $F_{X\mid O}$-almost every $x$. These are the integrability conditions in Appendix A to which the main text refers. The first condition makes the estimand and the conditional means of the potential outcomes well defined, and the second ensures that the solution of the second equation of (2.7) lies in the class on which completeness is imposed. Under Assumption B.1 in Appendix B, these conditions hold for the moments used in Section 3. Proposition 4.1 uses, in addition, that $L_t$ and $U_t$ in (4.3) are integrable with respect to $F_{X\mid O}$, which holds when $X$ takes finitely many values, and the ends of the bounds under the linear restrictions of Section 4.1 are then integrable because they lie between $L_t$ and $U_t$.

\subsubsection{Proof of Theorem 2.1}\label{sec:A.1.2}

The proof shows that the solutions of (2.7) are $\pi_t$ and $\nu_t$ and that averaging their ratio as in (2.8) gives $\tau$. Assumptions 2.6 and 2.7 first turn the two observational moments into averages of functions of $S(t)$ in the observational population, and Assumptions 2.3 and 2.4 carry these averages to the experiment. Completeness then applies to the difference between two solutions, whose conditional expectation is zero. Finally, Assumption 2.5 links the solutions to the means of the potential outcomes.

We recall equation (2.7). For each $t$, for $F_{X\mid O}$-almost every $x$, and for every bounded measurable function $h$,
\[
\begin{aligned}
\mathbb E[\mathbf 1\{T=t\}h(M)\mid x,O]
&=\mathbb E[\pi_t(S,x)h(M)\mid T=t,x,E],\\
\mathbb E[\mathbf 1\{T=t\}Yh(M)\mid x,O]
&=\mathbb E[\nu_t(S,x)h(M)\mid T=t,x,E]
\end{aligned}
\]

Fix a bounded measurable function $h$ and fix $t$ and $x$. Consistency, iterated expectations, and the exclusion restriction for selection in (2.5) give
\begin{equation}\tag{A.1}\label{eq:A.1}
\begin{aligned}
\mathbb E[\mathbf1\{T=t\}h(M)\mid x,O]
&=\mathbb E[\mathbf1\{T=t\}h(M(t))\mid x,O]\\
&=\mathbb E[\pi_t(S(t),x)h(M(t))\mid x,O]\\
&=\mathbb E[\pi_t(S,x)h(M)\mid T=t,x,E]
\end{aligned}
\end{equation}
where the last equality follows from transportability of the joint distribution and randomization in the experiment. By Assumptions 2.1 and 2.2, the distribution $F_{S\mid T=t,x,O}$ of $S$ among the units with $T=t$ in the observational population has density $\pi_t(\cdot,x)/P(T=t\mid x,O)$ with respect to $F_{S(t)\mid x,O}$, and this density is positive $F_{S(t)\mid x,O}$-almost everywhere, so the two distributions have the same null sets. Hence $\mu_t(\cdot,x)$, which is defined $F_{S\mid T=t,x,O}$-almost everywhere, is defined $F_{S(t)\mid x,O}$-almost everywhere, and $\mu_t(S(t),x)$ below is defined almost surely in the observational population. For the equation with the outcome, Assumption 2.7 first replaces the observed outcome by its conditional mean, and the same calculation gives
\begin{equation}\tag{A.2}\label{eq:A.2}
\begin{aligned}
\mathbb E[\mathbf1\{T=t\}Yh(M)\mid x,O]
&=\mathbb E[\mathbf1\{T=t\}\mu_t(S,x)h(M)\mid x,O]\\
&=\mathbb E[\pi_t(S(t),x)\mu_t(S(t),x)h(M(t))\mid x,O]\\
&=\mathbb E[\nu_t(S,x)h(M)\mid T=t,x,E]
\end{aligned}
\end{equation}
Each integral is well defined under the integrability conditions above and the boundedness of $h$.

Because the selection probability takes values in $[0,1]$, $\pi_t\in L^2(F_{S\mid T=t,x,E})$. Conditional Jensen's inequality and $\pi_t^2\le\pi_t$ give
\[
\begin{aligned}
\mathbb E[\nu_t(S(t),x)^2\mid x,O]
&\le\mathbb E[\pi_t(S(t),x)\mu_t(S(t),x)^2\mid x,O]\\
&=\mathbb E[\mathbf1\{T=t\}\mu_t(S,x)^2\mid x,O]\\
&\le\mathbb E[\mathbf1\{T=t\}Y^2\mid x,O]<\infty
\end{aligned}
\]
and, by transportability, $\nu_t$ also lies in $L^2(F_{S\mid T=t,x,E})$.

Let $d$ be the difference between two $L^2$ solutions of the equation in (A.1). Since $\mathbb E[d(S)h(M)\mid T=t,x,E]=0$ for every bounded measurable $h(M)$, the defining property of conditional expectation gives $\mathbb E[d(S)\mid M,T=t,x,E]=0$. By Assumptions 2.3 and 2.4, this is equivalent to $\mathbb E[d(S(t))\mid M(t),x,O]=0$, and (2.6) in Assumption 2.8 gives $d=0$. The same argument applies to the difference between two solutions of the equation in (A.2). Hence $\pi_t$ and $\nu_t$ are uniquely determined, and positivity in Assumption 2.2 identifies $\mu_t=\nu_t/\pi_t$ on the support of $F_{S(t)\mid x,O}$.

Consistency and the equality of conditional means that follows from the conditional independence in Assumption 2.5 and $\mathbb E[|Y(t)|\mid O]<\infty$ in Appendix A.1.1 give, on the support of $F_{S\mid T=t,x,O}$,
\[
\mu_t(S(t),x)=\mathbb E[Y(t)\mid T=t,S(t),x,O]
=\mathbb E[Y(t)\mid S(t),x,O]
\]
and positivity extends this equality to the support of $F_{S(t)\mid x,O}$. Iterated expectations, transportability, and randomization in the experiment, applied in this order, give
\[
\mathbb E[Y(t)\mid x,O]
=\int\mu_t(s,x)\,dF_{S(t)\mid x,O}(s)
=\int\mu_t(s,x)\,dF_{S\mid T=t,x,E}(s)
\]
Finally, averaging over $F_{X\mid O}$ and taking the difference between $t=1$ and $t=0$ gives
\[
\tau=\int\!\left\{\int\mu_1(s,x)\,dF_{S\mid T=1,x,E}(s)-\int\mu_0(s,x)\,dF_{S\mid T=0,x,E}(s)\right\}dF_{X\mid O}(x)
\]
which is (2.8). $\square$

\subsection{Completeness under additive measurement error}\label{sec:A.2}

This subsection gives a sufficient condition for completeness, stated in terms of the distribution of the error $U_2$, when the measurement in Example 2.1 takes the form $M(t)=S(t)+U_2$. It derives, as special cases, completeness in the second part of Example 2.1 and in the designs of Section 5. Fix $(t,x)$ and suppress the conditioning on $X=x$ in the observational population. Suppose that $S(t)$ and $U_2$ are real-valued, that $U_2$ is independent of $S(t)$, and that the set of points at which the characteristic function $u\mapsto\mathbb E[e^{iuU_2}]$ of $U_2$ is nonzero is dense in the real line. No condition is placed on the distribution of $S(t)$. Then, if $h\in L^2(F_{S(t)\mid x,O})$ satisfies $\mathbb E[h(S(t))\mid M(t)]=0$ almost surely, $h(S(t))=0$ almost surely.

This result applies Proposition 2.4 of \citet{ref15} to the real-valued model $M(t)=S(t)+U_2$ and weakens its condition on the error. That proposition establishes completeness for integrable functions in a model in which a function of the measurement equals a function of the latent variable plus an error, when the error is independent of the latent variable given the common variables, the function that transforms the latent variable is one-to-one, and the zeros of the characteristic function of the error are isolated. In $M(t)=S(t)+U_2$, both functions are the identity, so the one-to-one condition holds. If the zeros are isolated, the set of points at which the characteristic function is nonzero is dense in the real line, so the condition on the error above is weaker. Our conclusion is restricted to square-integrable functions, and the proof follows the same argument. Completeness for square-integrable functions is $L^2$-completeness in the terminology of \citet{ref2}, which lies between completeness for integrable functions and completeness for bounded functions. Assumption 2.8 imposes completeness in this form. We prove the result here to keep the paper self-contained.

The Cauchy–Schwarz inequality gives $\mathbb E[|h(S(t))|]\le\mathbb E[h(S(t))^2]^{1/2}<\infty$. Fix a real number $u$. Because $e^{iuM(t)}$ is a bounded function of $M(t)$, iterated expectations and the independence of $U_2$ and $S(t)$ give
\[
0=\mathbb E\bigl[\mathbb E[h(S(t))\mid M(t)]\,e^{iuM(t)}\bigr]
=\mathbb E[h(S(t))e^{iuM(t)}]
=\mathbb E[h(S(t))e^{iuS(t)}]\,\mathbb E[e^{iuU_2}]
\]
Hence $\mathbb E[h(S(t))e^{iuS(t)}]=0$ for every $u$ with $\mathbb E[e^{iuU_2}]\ne0$. Since $\mathbb E[|h(S(t))|]<\infty$, the dominated convergence theorem shows that $u\mapsto\mathbb E[h(S(t))e^{iuS(t)}]$ is continuous, and because it vanishes on a dense set, it vanishes for every $u$. Thus the Fourier transform of the finite signed measure $B\mapsto\mathbb E[h(S(t))\mathbf 1\{S(t)\in B\}]$ is identically zero, and the uniqueness of the Fourier transform implies that this measure is zero. Taking $B=\{s:h(s)>0\}$ and $B=\{s:h(s)<0\}$ shows that the positive and negative parts of $h(S(t))$ both have expectation zero, so $h(S(t))=0$ almost surely. $\square$

In the second part of Example 2.1, $U_2$ is independent of $S(t)$ and normally distributed with mean zero and variance $\sigma^2$. Its characteristic function $\exp(-\sigma^2u^2/2)$ is nonzero for every $u$, so the result above gives Assumption 2.8 for each $t$ and $x$.

In the two designs of Section 5, given $X=x$, $S(t)=U_1+(1+0.5x)t$ and $M(t)=S(t)+U_2$, where $U_2$ is uniformly distributed on $[-1,1]$ and independent of $(U_1,X)$. The characteristic function of $U_2$ equals $\sin(u)/u$ for $u\ne0$ and 1 at $u=0$, so its zeros are the points $u=k\pi$ with $k$ a nonzero integer, and the set of points at which it is nonzero is dense. The result above therefore gives Assumption 2.8 for each $t$ and $x$, both in the continuous design, in which $U_1$ is uniformly distributed, and in the discrete design, in which $U_1$ takes three values.

\subsection{Derivation of equation (4.2)}\label{sec:A.3}

This subsection derives equation (4.2) of Section 4.1, which gives the distribution of $W(t)=1/\pi_t(S(t),x)$ among the units with $T=t$ and $M=m$ in the observational population. The distribution of $S(t)$ among these units can differ from the distribution of $S$ among the units of the experimental arm $t$ with $M=m$, because selection into treatment can depend on $S(t)$, and weighting the experimental distribution by the selection probability $\pi_t$ recovers it. We recall equation (4.2). For each $t$, $F_{X\mid O}$-almost every $x$, each value $m$ with $P(T=t,M=m\mid x,O)>0$, and every real number $w$,
\[
\begin{aligned}
P(W(t)\le w\mid T=t,M=m,x,O)&=\frac{\mathbb E[\pi_t(S,x)\mathbf 1\{1/\pi_t(S,x)\le w\}\mid M=m,T=t,x,E]}
{\mathbb E[\pi_t(S,x)\mid M=m,T=t,x,E]}.
\end{aligned}
\]

Fix $t$, a value $x$ outside the $F_{X\mid O}$-null set on which one of Assumptions 2.2–2.4 and 2.6 may fail, a value $m$ with $P(T=t,M=m\mid x,O)>0$, and a real number $w$. By Assumption 2.2, $\pi_t(S(t),x)>0$ almost surely in the observational population, so $W(t)$ is a function of $S(t)$. By consistency in Assumption 2.1, $M=M(t)$ on $\{T=t\}$, so $\{T=t,M=m\}=\{T=t,M(t)=m\}$ and $P(M(t)=m\mid x,O)\ge P(T=t,M=m\mid x,O)>0$. The definition of conditional probability then gives
\begin{equation}\tag{A.3}\label{eq:A.3}
\begin{aligned}
P(W(t)\le w\mid T=t,M=m,x,O)
&=P(W(t)\le w\mid T=t,M(t)=m,x,O)\\
&=\frac{P(W(t)\le w,T=t\mid M(t)=m,x,O)}{P(T=t\mid M(t)=m,x,O)}
\end{aligned}
\end{equation}

For the numerator, we write the probability as the conditional mean of a product of indicators, condition on $(S(t),M(t))$, and use (2.5) in Assumption 2.6:
\begin{equation}\tag{A.4}\label{eq:A.4}
\begin{aligned}
P(W(t)\le w,T=t\mid M(t)=m,x,O)
&=\mathbb E[\mathbf 1\{W(t)\le w\}\mathbf 1\{T=t\}\mid M(t)=m,x,O]\\
&=\mathbb E\bigl[\mathbf 1\{W(t)\le w\}P(T=t\mid S(t),M(t),x,O)\bigm|M(t)=m,x,O\bigr]\\
&=\mathbb E[\pi_t(S(t),x)\mathbf 1\{1/\pi_t(S(t),x)\le w\}\mid M(t)=m,x,O]
\end{aligned}
\end{equation}
The second equality is the law of iterated expectations, in which $\mathbf 1\{W(t)\le w\}$ is taken outside the inner conditional expectation because it is a function of $S(t)$. The third equality uses $P(T=t\mid S(t),M(t),x,O)=\pi_t(S(t),x)$, which is (2.5), and $W(t)=1/\pi_t(S(t),x)$.

The same calculation without the indicator of $\{W(t)\le w\}$ gives the denominator,
\begin{equation}\tag{A.5}\label{eq:A.5}
\begin{aligned}
P(T=t\mid M(t)=m,x,O)
&=\mathbb E[P(T=t\mid S(t),M(t),x,O)\mid M(t)=m,x,O]\\
&=\mathbb E[\pi_t(S(t),x)\mid M(t)=m,x,O]
\end{aligned}
\end{equation}
which is positive because $P(T=t,M(t)=m\mid x,O)=P(T=t,M=m\mid x,O)>0$.

The right sides of (A.4) and (A.5) are conditional means of bounded functions of $S(t)$ given $M(t)=m$ in the observational population, and the last step replaces them by conditional means in the experimental arm $t$. By Assumptions 2.1, 2.3, and 2.4, the joint distribution of $(S(t),M(t))$ given $X=x$ in the observational population equals the distribution of $(S,M)$ given $T=t$ and $X=x$ in the experimental population. Hence $P(M=m\mid T=t,x,E)=P(M(t)=m\mid x,O)>0$, and for every bounded measurable function $g$,
\[
\begin{aligned}
\mathbb E[g(S(t))\mid M(t)=m,x,O]
&=\frac{\mathbb E[g(S(t))\mathbf 1\{M(t)=m\}\mid x,O]}{P(M(t)=m\mid x,O)}\\
&=\frac{\mathbb E[g(S)\mathbf 1\{M=m\}\mid T=t,x,E]}{P(M=m\mid T=t,x,E)}\\
&=\mathbb E[g(S)\mid M=m,T=t,x,E]
\end{aligned}
\]

We apply this equality with $g(s)=\pi_t(s,x)\mathbf 1\{1/\pi_t(s,x)\le w\}$ to the right side of (A.4) and with $g(s)=\pi_t(s,x)$ to the right side of (A.5). Both functions take values in $[0,1]$, with the first set to zero where $\pi_t(s,x)=0$. Substituting the results into (A.3) gives (4.2). The calculation uses $\mathbf 1\{W(t)\le w\}$ only as a bounded function of $S(t)$, so replacing it by $\mathbf 1\{S(t)\in B\}$ for a Borel set $B$ gives, in the same way, the conditional distribution of $S(t)$ given $T=t$, $M=m$, $X=x$, and $G=O$. $\square$

\section{Proof of Theorem 3.1}\label{sec:B}

This appendix proves Theorem 3.1 of Section 3, which gives the efficient influence function and the semiparametric efficiency bound. Table B.1 lists the contents of the subsections, and Figure B.1 shows how the results used in the proof depend on one another. Theorems 3.2 and 3.3 on the estimators of Section 3 are proved in Appendix C.

\begin{table}[tbp]
\centering
\def\MainTextTableBody{
\begin{tabular}{lp{.79\textwidth}}
\toprule
Subsection & Contents \\
\midrule
B.1 & Defines the model, smooth submodels, the tangent space, and gradients, and states Assumption B.2 \\
B.2 & States Lemmas B.1–B.4 and uses them to prove Theorem 3.1 \\
B.3 & Proves Lemmas B.1–B.3 \\
B.4 & Proves Lemma B.4 \\
\bottomrule
\end{tabular}
}
\setlength{\tabcolsep}{4pt}\small
\sbox{\MainTextTableBox}{\MainTextTableBody}
\def\MainTextTableFont{small}
\typeout{TABLE-WIDTH B.1 font=\MainTextTableFont; width=\the\wd\MainTextTableBox; limit=\the\textwidth}
\ifdim\wd\MainTextTableBox>\textwidth
  \PackageError{main-text}{Table B.1 remains too wide}{Stop and report; no scaling allowed}
\fi
\usebox{\MainTextTableBox}
\renewcommand{\thetable}{B.1}
\caption{Subsections of Appendix B and their contents}
\label{table:B.1}
\end{table}
Figure B.1 shows how the results used to prove Theorem 3.1 depend on one another, where an arrow from one result to another means that the first is used to prove the second. Theorem 3.1 follows from Lemmas B.2, B.3, and B.4. Lemma B.2 states that the influence functions $\psi_E,\psi_O$ have mean zero and finite second moments, Lemma B.3 that the two-sample log likelihood ratio is locally asymptotically normal, and Lemma B.4 that the right derivative of $\tau$ along a smooth submodel equals the sum of the expected products of the influence functions and the scores. By Lemmas B.2 and B.4, the normalized influence function $(\psi_E/\rho,\psi_O/(1-\rho))$ is a gradient of $\tau$, and by Assumption B.2 this gradient lies in the tangent space. Lemma B.3 enters through the convolution theorem, which makes the efficiency bound a lower bound on the asymptotic variance of regular estimators. The proofs of Lemmas B.2 and B.4 use Lemma B.1, which extends the observational moment equations (2.7) to unbounded functions.

\begin{figure}[htbp]
\centering
\includegraphics[max width=\textwidth]{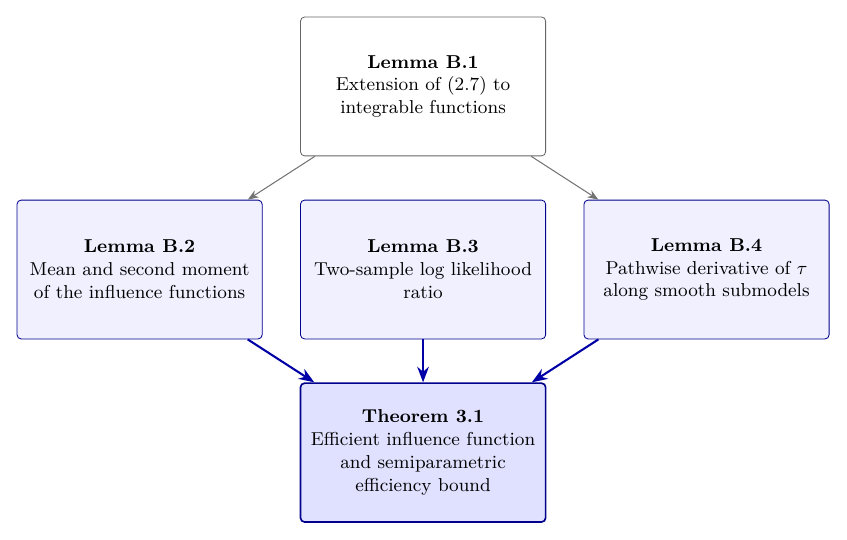}
\renewcommand{\thefigure}{B.1}
\caption{Dependencies among the results used to prove Theorem 3.1}
\label{figure:B.1}
\par\smallskip{\small\NoteLeft Note: Thick arrows start from the lemmas that directly yield Theorem 3.1.\par}
\end{figure}

\renewcommand{\thefixedassumption}{B.1}
\begin{fixedassumption}[Boundedness]\label{assumption:B.1}

There is a constant $C>0$ such that, for each $t\in\{0,1\}$ and $x\in\mathcal X$, the following hold.
\begin{enumerate}
\item $|\mu_t|\le C$ almost surely on the support of $S\mid T=t,x,E$.
\item $|Y|\le C$ almost surely in the observational population.
\end{enumerate}
\end{fixedassumption}

Assumption B.1 bounds the true functions and the outcome, and together with $\pi_t\le1$ it gives $|\nu_t|\le C$. The boundedness of the estimates is stated in Assumption C.1 of Appendix C. Below, we take versions of $\pi_t(\cdot,x)$ and $\mu_t(\cdot,x)$ with $a\le\pi_t(s,x)\le1$ and $|\mu_t(s,x)|\le C$ for every $s$, where $a$ is the constant in Assumption 3.3. Since versions differ only on null sets of $S\mid T=t,x,E$, this choice leaves (2.7), (3.2)–(3.4), and the expectations below unchanged, and $|\nu_t(s,x)|\le C$ for every $s$.

We recall four equations of the main text that the proofs in Appendices B and C use repeatedly. Equation (2.7) expresses the two observational moments as experimental averages. For each $t$, for $F_{X\mid O}$-almost every $x$, and for every bounded measurable function $h$,
\begin{NoHyper}\begin{equation}\tag{2.7}
\begin{aligned}
\mathbb E[\mathbf 1\{T=t\}h(M)\mid x,O]
&=\mathbb E[\pi_t(S,x)h(M)\mid T=t,x,E],\\
\mathbb E[\mathbf 1\{T=t\}Yh(M)\mid x,O]
&=\mathbb E[\nu_t(S,x)h(M)\mid T=t,x,E]
\end{aligned}
\end{equation}\end{NoHyper}
Equation (3.2) states the conditions that Assumption 3.4 imposes on the weight functions $\omega_t^\nu,\omega_t^\pi$,
\begin{NoHyper}\begin{equation}\tag{3.2}
\mathbb E[\omega_t^\nu(M,x)\mid S,T=t,x,E]=\frac1{\pi_t(S,x)},\qquad
\mathbb E[\omega_t^\pi(M,x)\mid S,T=t,x,E]=\frac{\mu_t(S,x)}{\pi_t(S,x)}
\end{equation}\end{NoHyper}
equation (3.6) defines the estimator from the stratum means $\widehat{\bar\mu}_t(x)$ in (3.5),
\begin{NoHyper}\begin{equation}\tag{3.6}
\widehat\tau=\sum_{x\in\mathcal X}\widehat P(X=x\mid O)\{\widehat{\bar\mu}_1(x)-\widehat{\bar\mu}_0(x)\}
\end{equation}\end{NoHyper}
and equation (3.7) defines the asymptotic variance $\Sigma^2$ and its estimator $\widehat\Sigma^2$,
\begin{NoHyper}\begin{equation}\tag{3.7}
\begin{aligned}
\Sigma^2&=\frac1\rho\mathbb E[\psi_E(Z_E)^2\mid E]+\frac1{1-\rho}\mathbb E[\psi_O(Z_O)^2\mid O],\\
\widehat\Sigma^2&=\frac N{n_E}\frac1{n_E}\sum_{i=1}^{n_E}\widehat\psi_E(Z_{E,i})^2+\frac N{n_O}\frac1{n_O}\sum_{j=1}^{n_O}\widehat\psi_O(Z_{O,j})^2
\end{aligned}
\end{equation}\end{NoHyper}

\subsection{Model, smooth submodels, and tangent space}\label{sec:B.1}

This subsection defines, for two independent samples, the smooth submodels, tangent space, efficient influence function, and regular estimators that Theorem 3.1 uses, and it states the tangent space condition, Assumption B.2. The definitions adapt the one-sample definitions in Chapter 25 of \citet{ref32} to a two-sample inner product. We write $P_E,P_O$ for the observed laws and $P=(P_E,P_O)$ for their pair. The model $\mathcal P$ of the main text is the set of all pairs of observed images of full-data laws under which the range of $X$ is contained in $\mathcal X$ and Assumptions 2.1–2.8 and the integrability conditions in Appendix A hold. By Theorem 2.1, $\tau$ is a function on $\mathcal P$.

We write $L_2^0(P_E),L_2^0(P_O)$ for the sets of real-valued functions of the observations that have mean zero and are square-integrable. With $\rho$ of Assumption 3.1, the two-sample inner product is
\begin{equation}\tag{B.1}\label{eq:B.1}
\langle(a_E,a_O),(b_E,b_O)\rangle_\rho
=\rho\mathbb E[a_Eb_E\mid E]+(1-\rho)\mathbb E[a_Ob_O\mid O]
\end{equation}
and $\|\cdot\|_\rho$ denotes its norm. A path $\{P_\varepsilon=(P_{E,\varepsilon},P_{O,\varepsilon}):0\le\varepsilon<\varepsilon_0\}$ in $\mathcal P$, with $\varepsilon_0>0$, is a family in which every $P_\varepsilon$ belongs to $\mathcal P$ and $P_0=P$. A path is regular if its likelihood ratios have first-order expansions with bounded scores, that is, if for each $G\in\{E,O\}$, $P_{G,\varepsilon}$ is absolutely continuous with respect to $P_G$ and there exist $a_G\in L_2^0(P_G)$, bounded under $P_G$, and functions $r_{G,\varepsilon}$ whose essential suprema $\|r_{G,\varepsilon}\|_\infty$ under $P_G$ tend to zero as $\varepsilon\downarrow0$, such that the density $dP_{G,\varepsilon}/dP_G$ equals $1+\varepsilon a_G+\varepsilon r_{G,\varepsilon}$. We call $(a_E,a_O)$ the score of the path. Like the paths with bounded scores in Chapter 25 of \citet{ref32}, a regular path is differentiable in quadratic mean,
\begin{equation}\tag{B.2}\label{eq:B.2}
\int\left\{dP_{G,\varepsilon}^{1/2}-dP_G^{1/2}-\frac\varepsilon2a_G\,dP_G^{1/2}\right\}^2=o(\varepsilon^2)\quad(\varepsilon\downarrow0),\qquad G\in\{E,O\},
\end{equation}
The integral in (B.2) is written with the square roots of densities with respect to a sigma-finite measure that dominates both $P_G$ and $P_{G,\varepsilon}$, and it does not depend on the choice of this measure. Along a regular path, $P_{G,\varepsilon}$ is absolutely continuous with respect to $P_G$, so the density $dP_{G,\varepsilon}/dP_G$ is defined with probability one for observations from $P_G$.

The selection probability and the product are also uniquely determined under the law along the path, as the solutions of (2.7). Take a regular path in $\mathcal P$. For each $\varepsilon$, we write $\pi_{t,\varepsilon},\mu_{t,\varepsilon}$ for the functions in (2.2) of a full-data law whose observed image is $P_\varepsilon$, and we set $\nu_{t,\varepsilon}=\pi_{t,\varepsilon}\mu_{t,\varepsilon}$. For small $\varepsilon$, the densities $dP_{G,\varepsilon}/dP_G=1+\varepsilon a_G+\varepsilon r_{G,\varepsilon}$ of both samples lie between $1/2$ and $3/2$, so $P_\varepsilon(X=x\mid O)\ge c/2$ and $P_\varepsilon(T=t,X=x\mid E)\ge c/2$ for all $t,x$. Since $P_\varepsilon$ belongs to $\mathcal P$, Theorem 2.1 applied to $P_\varepsilon$ shows that, for all $t,x$, $\pi_{t,\varepsilon}(\cdot,x)$ and $\nu_{t,\varepsilon}(\cdot,x)$ are the unique solutions of (2.7) under $P_\varepsilon$ that are square-integrable with respect to the distribution of $S\mid T=t,x,E$ under $P_\varepsilon$, and in particular they do not depend on the choice of the full-data law. The bounds on the densities imply that the distributions of $S\mid T=t,x,E$ under $P_\varepsilon$ and under $P$ are mutually absolutely continuous and have the same space of square-integrable functions. Hence $\pi_{t,\varepsilon}(\cdot,x)$ and $\nu_{t,\varepsilon}(\cdot,x)$ are elements of $L^2(F_{S\mid T=t,x,E})$ and can be compared with $\pi_t(\cdot,x)$ and $\nu_t(\cdot,x)$ under the true law. As in Section 3.3, we measure the size of a difference by $\|f\|_{2,t,x}=\mathbb E[f(S,M)^2\mid T=t,x,E]^{1/2}$ for a function $f$ of $(S,M)$.

We restrict attention to regular paths along which the selection probability and the product are differentiable. A regular path in $\mathcal P$ is a smooth submodel if, for each $t\in\{0,1\}$ and $x\in\mathcal X$, there are functions $\dot\pi_t(\cdot,x)$ and $\dot\nu_t(\cdot,x)$ of $S$ with $\|\dot\pi_t\|_{2,t,x}<\infty$ and $\|\dot\nu_t\|_{2,t,x}<\infty$ such that $\|(\pi_{t,\varepsilon}-\pi_t)/\varepsilon-\dot\pi_t\|_{2,t,x}\to0$ and $\|(\nu_{t,\varepsilon}-\nu_t)/\varepsilon-\dot\nu_t\|_{2,t,x}\to0$ as $\varepsilon\downarrow0$. Because $\pi_t$ and $\nu_t$ along the path solve (2.7) under the law along the path, their changes cannot be read directly from the changes in the observed laws, and, following \citet{ref26} in deriving efficiency bounds from derivatives along submodels, we restrict attention to submodels on which these functions are differentiable. The efficiency statements of Section 3 are relative to the smooth submodels. Lemma B.4 in Appendix B.2.1 shows that, for every smooth submodel with score $(a_E,a_O)$, $\tau(P_\varepsilon)$ is right differentiable at $\varepsilon=0$ and
\begin{equation}\tag{B.3}\label{eq:B.3}
\left.\frac{d\tau(P_\varepsilon)}{d\varepsilon}\right|_{\varepsilon=0+}
=\mathbb E[\psi_E(Z_E)a_E(Z_E)\mid E]+\mathbb E[\psi_O(Z_O)a_O(Z_O)\mid O]
\end{equation}
for the functions in (3.3) and (3.4) constructed from any square-integrable solution of (3.2).

When $S$ and $M$ take finitely many values given $T=t$ and $X=x$, every regular path in $\mathcal P$ is a smooth submodel. In this case, (2.7) with the indicators of the values of $M$ as $h$ is, under each law along the path, a system of linear equations for the values of $\pi_{t,\varepsilon}(\cdot,x)$ and another for the values of $\nu_{t,\varepsilon}(\cdot,x)$, with a common matrix whose columns are linearly independent by Assumption 2.8. The entries of the matrix and of the right-hand sides are conditional probabilities and means under $P_\varepsilon$, which are right differentiable at $\varepsilon=0$ because the densities of the path are. Hence the solutions, obtained by applying the left inverse $(A^\top A)^{-1}A^\top$ of the matrix $A$ to the right-hand sides, are right differentiable as well.

The tangent space is the closure under $\|\cdot\|_\rho$ of the set of all linear combinations of the scores of smooth submodels. A pair $(b_E,b_O)\in L_2^0(P_E)\times L_2^0(P_O)$ is a gradient of $\tau$ if, for every smooth submodel, $\tau(P_\varepsilon)$ is right differentiable at $\varepsilon=0$ and its right derivative equals the inner product $\langle(b_E,b_O),(a_E,a_O)\rangle_\rho$ with the score. A gradient in the tangent space is called the efficient influence function (EIF) of $\tau$. When the coefficients of the two sample averages are $(\psi_E,\psi_O)$, the corresponding element under this inner product is $(\psi_E/\rho,\psi_O/(1-\rho))$. The squared norm of the EIF is called the efficiency bound. To define the regularity of an estimator, take a smooth submodel and, for large $N$, let the two samples be drawn from $P_{1/\sqrt N}=(P_{E,1/\sqrt N},P_{O,1/\sqrt N})$, and fix a linear subspace of $L_2^0(P_E)\times L_2^0(P_O)$. A sequence of estimators is regular with respect to that subspace if, for every smooth submodel whose score lies in the subspace, $\sqrt N$ times the difference between the estimator and $\tau(P_{1/\sqrt N})$ converges in distribution under $P_{1/\sqrt N}$ to a limit that does not depend on the submodel. If this holds for every smooth submodel, the sequence of estimators is called regular in the model $\mathcal P$. Chapter 25 of \citet{ref32} likewise defines these notions relative to a given collection of paths.

\renewcommand{\thefixedassumption}{B.2}
\begin{fixedassumption}[Tangent space condition]\label{assumption:B.2}

There exists a linear subspace of $L_2^0(P_E)\times L_2^0(P_O)$ that satisfies the following two conditions.
\begin{enumerate}
\item Every element is the score of some smooth submodel.
\item Its closure under $\|\cdot\|_\rho$ contains $(\psi_E/\rho,\psi_O/(1-\rho))$.
\end{enumerate}
\end{fixedassumption}

Assumption B.2 requires that the normalized influence function can be approximated by scores of smooth submodels. Condition (1) asks for a linear space of scores, which the convolution theorem needs to turn the efficiency bound into a lower bound. Condition (2) requires that the normalized pair built from (3.3) and (3.4) be approximated by these scores. Since the elements of the subspace in condition (1) are scores, its closure is contained in the tangent space, and condition (2) therefore implies that $(\psi_E/\rho,\psi_O/(1-\rho))$ lies in the tangent space. If the scores of all smooth submodels form a linear space, that space can serve as the subspace in condition (1), and conditions (1) and (2) are then equivalent to $(\psi_E/\rho,\psi_O/(1-\rho))$ lying in the tangent space. The functions in (3.3) and (3.4) depend on the fixed solution $\omega_t^\nu,\omega_t^\pi$ of (3.2), and Assumption B.2 is imposed on the pair constructed from that solution (see Remark 3.1). Below, we fix one linear subspace that satisfies Assumption B.2 and call it the subspace in Assumption B.2.

\subsection{Lemmas and proof of Theorem 3.1}\label{sec:B.2}

This subsection proves Theorem 3.1 of Section 3.1, whose boundedness and tangent space conditions in Appendix B are Assumptions B.1 and B.2. Appendix B.2.1 states Lemmas B.1–B.4, which the proof uses, and Appendix B.2.2 proves the theorem. Lemmas B.1–B.3 are proved in Appendix B.3, and Lemma B.4 in Appendix B.4.

\subsubsection{Lemmas used in the proof}\label{sec:B.2.1}

This part states the four lemmas used to prove Theorem 3.1. Lemma B.1 extends (2.7) to unbounded functions, Lemma B.2 concerns the means and second moments of the influence functions, Lemma B.3 the two-sample log likelihood ratio, and Lemma B.4 the derivative of $\tau$ along a smooth submodel.

\renewcommand{\thefixedlemma}{B.1}
\begin{fixedlemma}[Extension of (2.7) to integrable functions]\label{lemma:B.1}

Suppose that Assumptions 2.1–2.8, the integrability conditions in Appendix A, Assumptions 3.2 and 3.3, and Assumption B.1 hold, and fix $t\in\{0,1\}$ and $x\in\mathcal X$.

\begin{enumerate}
\item For every nonnegative measurable function $g$, $\mathbb E[\mathbf 1\{T=t\}g(M)\mid x,O]=\mathbb E[\pi_t(S,x)g(M)\mid T=t,x,E]$, where both sides may be infinite.
\item If a measurable function $g$ satisfies $\mathbb E[|g(M)|\mid T=t,x,E]<\infty$, then $\mathbb E[\mathbf 1\{T=t\}g(M)\mid x,O]=\mathbb E[\pi_t(S,x)g(M)\mid T=t,x,E]$ and $\mathbb E[\mathbf 1\{T=t\}Yg(M)\mid x,O]=\mathbb E[\nu_t(S,x)g(M)\mid T=t,x,E]$, and all these expectations are finite.
\item If Assumption 3.4 also holds, then $\mathbb E[\nu_t(S,x)\omega_t^\nu(M,x)\mid T=t,x,E]$ and $\mathbb E[\pi_t(S,x)\omega_t^\pi(M,x)\mid T=t,x,E]$ both equal $\bar\mu_t(x)$, and $\mathbb E[\mathbf 1\{T=t\}\{Y\omega_t^\nu(M,x)-\omega_t^\pi(M,x)\}\mid x,O]=0$. Hence $\mathbb E[\psi_E(Z_E)\mid E]=\mathbb E[\psi_O(Z_O)\mid O]=0$.
\end{enumerate}
\end{fixedlemma}

The main conclusion of Lemma B.1 is claim (3), which states that the functions in (3.3) and (3.4) have mean zero. Because the weight functions need not be bounded, claims (1) and (2) first extend (2.7), which is stated for bounded functions, to integrable functions by the standard argument with truncation and convergence theorems.

\renewcommand{\thefixedlemma}{B.2}
\begin{fixedlemma}[Mean and second moment of the influence functions]\label{lemma:B.2}

Suppose that Assumptions 2.1–2.8, the integrability conditions in Appendix A, Assumptions 3.1–3.4, and Assumption B.1 hold. Then $\mathbb E[\psi_E(Z_E)\mid E]=\mathbb E[\psi_O(Z_O)\mid O]=0$, $\mathbb E[\psi_E(Z_E)^2\mid E]<\infty$, and $\mathbb E[\psi_O(Z_O)^2\mid O]<\infty$.
\end{fixedlemma}

Lemma B.2 ensures that $\psi_E/\rho$ and $\psi_O/(1-\rho)$ belong to $L_2^0(P_E)$ and $L_2^0(P_O)$, respectively, so that they are candidates for a gradient under the inner product of Appendix B.1.

\renewcommand{\thefixedlemma}{B.3}
\begin{fixedlemma}[Two-sample log likelihood ratio]\label{lemma:B.3}

Suppose that Assumption 3.1 holds, and take a regular path in $\mathcal P$ with score $(a_E,a_O)$. Then, as $N\to\infty$, under the true law $P$,
\begin{equation}\tag{B.4}\label{eq:B.4}
\begin{aligned}
&\sum_{i=1}^{n_E}\log\frac{dP_{E,1/\sqrt N}}{dP_E}(Z_{E,i})+\sum_{j=1}^{n_O}\log\frac{dP_{O,1/\sqrt N}}{dP_O}(Z_{O,j})\\
&=\frac1{\sqrt N}\left\{\sum_{i=1}^{n_E}a_E(Z_{E,i})+\sum_{j=1}^{n_O}a_O(Z_{O,j})\right\}-\frac12\|(a_E,a_O)\|_\rho^2+o_p(1)
\end{aligned}
\end{equation}
and the term in braces on the right-hand side, divided by $\sqrt N$, converges in distribution to a normal distribution with mean zero and variance $\|(a_E,a_O)\|_\rho^2$.
\end{fixedlemma}

Lemma B.3 approximates the two-sample log likelihood ratio between the law shifted by $1/\sqrt N$ along the path and the true law by the normalized sum of the scores minus half the squared norm of Appendix B.1. The proof of Theorem 3.1 uses it in the convolution theorem, and Appendix C uses it with Le Cam's third lemma to show that the estimator is regular.

\renewcommand{\thefixedlemma}{B.4}
\begin{fixedlemma}[Pathwise derivative of $\tau$ along smooth submodels]\label{lemma:B.4}

Suppose that Assumptions 2.1–2.8, the integrability conditions in Appendix A, Assumptions 3.2–3.4, and Assumption B.1 hold, and take a smooth submodel with score $(a_E,a_O)$. Then $\tau(P_\varepsilon)$ is right differentiable at $\varepsilon=0$, and, for the functions in (3.3) and (3.4) constructed from any square-integrable solution of (3.2),
\[
\left.\frac{d\tau(P_\varepsilon)}{d\varepsilon}\right|_{\varepsilon=0+}
=\mathbb E[\psi_E(Z_E)a_E(Z_E)\mid E]+\mathbb E[\psi_O(Z_O)a_O(Z_O)\mid O]
\]
which is (B.3) of Appendix B.1.
\end{fixedlemma}

Lemma B.4 gives (B.3), stated in Appendix B.1, along every smooth submodel. Appendix B.4 proves it by writing $\tau(P_\varepsilon)$ in terms of stratum shares and stratum means and differentiating with respect to $\varepsilon$.

\subsubsection{Proof of Theorem 3.1}\label{sec:B.2.2}

This part proves the four claims of Theorem 3.1. Claim (1) states that the pair $(\psi_E/\rho,\psi_O/(1-\rho))$ is a gradient of $\tau$, claim (2) that it is the unique gradient in the tangent space, claim (3) that the efficiency bound equals $\Sigma^2$, and claim (4) that the limiting distribution of a regular estimator is a convolution with $\mathcal N(0,\Sigma^2)$. Lemmas B.2 and B.4 give claim (1), Assumption B.2 and the properties of the inner product give claim (2), the definition of the inner product gives claim (3), and Lemma B.3 and the convolution theorem give claim (4).

\textbf{Proof of Theorem 3.1.} We prove claim (1). Lemma B.2 gives $\psi_E\in L_2^0(P_E)$ and $\psi_O\in L_2^0(P_O)$, and since $\rho$ and $1-\rho$ are positive by Assumption 3.1, $\psi_E/\rho\in L_2^0(P_E)$ and $\psi_O/(1-\rho)\in L_2^0(P_O)$. Take any smooth submodel with score $(a_E,a_O)$. The definition (B.1) gives
\begin{equation}\tag{B.5}\label{eq:B.5}
\begin{aligned}
\left\langle\left(\frac{\psi_E}{\rho},\frac{\psi_O}{1-\rho}\right),(a_E,a_O)\right\rangle_\rho
&=\rho\mathbb E\left[\frac{\psi_E(Z_E)}{\rho}a_E(Z_E)\mid E\right]+(1-\rho)\mathbb E\left[\frac{\psi_O(Z_O)}{1-\rho}a_O(Z_O)\mid O\right]\\
&=\mathbb E[\psi_E(Z_E)a_E(Z_E)\mid E]+\mathbb E[\psi_O(Z_O)a_O(Z_O)\mid O]
\end{aligned}
\end{equation}
Lemma B.4 shows that $\tau(P_\varepsilon)$ is right differentiable at $\varepsilon=0$ and that its right derivative equals the right-hand side of (B.3), which is the right-hand side of (B.5). The right derivative therefore equals the inner product on the left-hand side of (B.5). Since the path is arbitrary, $(\psi_E/\rho,\psi_O/(1-\rho))$ is a gradient of $\tau$.

We prove claim (2). Since the elements of the subspace in Assumption B.2 are scores, they belong to the set of all linear combinations of scores, so their closure is contained in the tangent space. By Assumption B.2(2), $(\psi_E/\rho,\psi_O/(1-\rho))$ belongs to this closure and hence to the tangent space, and together with claim (1) this makes it the EIF by definition. For uniqueness, take a gradient $(b_E,b_O)$ in the tangent space. For every smooth submodel, the inner products of the two gradients with the score equal the same right derivative, so for the score $(a_E,a_O)$,
\begin{equation}\tag{B.6}\label{eq:B.6}
\left\langle\left(b_E-\frac{\psi_E}{\rho},b_O-\frac{\psi_O}{1-\rho}\right),(a_E,a_O)\right\rangle_\rho=0
\end{equation}
By linearity of the inner product, (B.6) also holds for every linear combination of scores, and by continuity of the inner product under $\|\cdot\|_\rho$, it holds for every element of their closure, which is the tangent space. Both $(b_E,b_O)$ and $(\psi_E/\rho,\psi_O/(1-\rho))$ lie in the tangent space, so their difference does too, and substituting the difference for $(a_E,a_O)$ in (B.6) gives
\[
\begin{aligned}
0
&=\left\langle\left(b_E-\frac{\psi_E}{\rho},b_O-\frac{\psi_O}{1-\rho}\right),\left(b_E-\frac{\psi_E}{\rho},b_O-\frac{\psi_O}{1-\rho}\right)\right\rangle_\rho\\
&=\rho\mathbb E\left[\left(b_E-\frac{\psi_E}{\rho}\right)^2\mid E\right]+(1-\rho)\mathbb E\left[\left(b_O-\frac{\psi_O}{1-\rho}\right)^2\mid O\right]
\end{aligned}
\]
Both coefficients are positive and both expectations are nonnegative, so both expectations are zero. Hence $b_E=\psi_E/\rho$ almost surely under $P_E$ and $b_O=\psi_O/(1-\rho)$ almost surely under $P_O$.

We prove claim (3). By claim (2), the EIF is $(\psi_E/\rho,\psi_O/(1-\rho))$, and by the definition (B.1) its squared norm is
\begin{equation}\tag{B.7}\label{eq:B.7}
\begin{aligned}
\left\|\left(\frac{\psi_E}{\rho},\frac{\psi_O}{1-\rho}\right)\right\|_\rho^2
&=\rho\mathbb E\left[\left(\frac{\psi_E(Z_E)}{\rho}\right)^2\mid E\right]+(1-\rho)\mathbb E\left[\left(\frac{\psi_O(Z_O)}{1-\rho}\right)^2\mid O\right]\\
&=\frac1\rho\mathbb E[\psi_E(Z_E)^2\mid E]+\frac1{1-\rho}\mathbb E[\psi_O(Z_O)^2\mid O]=\Sigma^2
\end{aligned}
\end{equation}
where the last equality is the definition of $\Sigma^2$ in (3.7). By Lemma B.2, $\Sigma^2$ is finite.

We prove claim (4). We use the standard convolution theorem for regular estimation \citep[Chapter 25]{ref32} in its form for locally asymptotically normal sequences of experiments. Take finitely many linearly independent elements $(a_{E,1},a_{O,1}),\ldots,(a_{E,k},a_{O,k})$ of the subspace in Assumption B.2. For $h=(h_1,\ldots,h_k)\in\mathbb R^k$, the linearity of the subspace implies that $\sum_{l=1}^kh_l(a_{E,l},a_{O,l})$ also belongs to it, so by Assumption B.2(1) there is a smooth submodel with this score. Choose one such path for each $h$, with the constant path $P_\varepsilon=P$ at $h=0$, and let the laws at $\varepsilon=1/\sqrt N$ define an experiment indexed by $h$. This sequence of experiments has three properties. First, Lemma B.3 applied to the score $\sum_lh_l(a_{E,l},a_{O,l})$ shows that the log likelihood ratio equals, up to $o_p(1)$, the inner product of $h$ with a random vector that converges to a normal distribution, minus $\frac12\sum_{l,l'}h_lh_{l'}\langle(a_{E,l},a_{O,l}),(a_{E,l'},a_{O,l'})\rangle_\rho$, and the $(l,l')$ entry of the covariance matrix of the limit is $\langle(a_{E,l},a_{O,l}),(a_{E,l'},a_{O,l'})\rangle_\rho$. The sequence of experiments is therefore locally asymptotically normal, with information given by the inner product (B.1). Second, the right derivative in claim (1), evaluated along $\varepsilon=1/\sqrt N$, shows that $\sqrt N\{\tau(P_{1/\sqrt N})-\tau(P)\}$ converges to $\sum_lh_l\langle(\psi_E/\rho,\psi_O/(1-\rho)),(a_{E,l},a_{O,l})\rangle_\rho$. Third, if the sequence of estimators is regular with respect to the subspace in Assumption B.2, the limiting distribution of $\sqrt N$ times the difference between the estimator and $\tau(P_{1/\sqrt N})$ under $P_{1/\sqrt N}$ does not depend on $h$. Given these three properties, the convolution theorem shows that the limiting distribution is the convolution of some probability distribution with a normal distribution whose variance is the squared norm of the orthogonal projection of $(\psi_E/\rho,\psi_O/(1-\rho))$ onto the span of $(a_{E,1},a_{O,1}),\ldots,(a_{E,k},a_{O,k})$. Writing $\sigma_k^2$ for this squared norm, the characteristic function of the limiting distribution at every real $u$ is the product of $\exp(-\sigma_k^2u^2/2)$ and the characteristic function of the other factor of the convolution.

We extend this conclusion to $\Sigma^2$ by enlarging the finite collection of elements. By Assumption B.2(2), there is a sequence of elements of the subspace that converges to $(\psi_E/\rho,\psi_O/(1-\rho))$ under $\|\cdot\|_\rho$, and we take as the finite collection a basis of the span of its first $k$ elements (if this span contains only zero, then $\sigma_k^2=0$ and the conclusion above holds directly). By the properties of orthogonal projection, the squared norm of $(\psi_E/\rho,\psi_O/(1-\rho))$ minus the squared norm $\sigma_k^2$ of its projection equals the squared norm of their difference. Because the projection is the closest element of the span, this difference is at most the squared distance to the $k$-th element of the sequence, which lies in the span, and this tends to zero as $k\to\infty$. By (B.7), $\sigma_k^2\to\Sigma^2$. The characteristic function of the other factor of the convolution equals the characteristic function of the limiting distribution multiplied by $\exp(\sigma_k^2u^2/2)$, so as $k\to\infty$ it converges for each $u$ to the characteristic function of the limiting distribution multiplied by $\exp(\Sigma^2u^2/2)$. This function equals one at $u=0$ and is continuous, so by Lévy's continuity theorem it is the characteristic function of a probability distribution. Hence the characteristic function of the limiting distribution is the product of $\exp(-\Sigma^2u^2/2)$ and the characteristic function of that probability distribution, and the limiting distribution is the convolution of a normal distribution with mean zero and variance $\Sigma^2$ with that probability distribution. If the limiting distribution has a finite variance, this variance is at least $\Sigma^2$, because the variance of the sum of two independent random variables is the sum of their variances. This proves the four claims of Theorem 3.1. $\square$

The convolution theorem used in this part and (B.9) in Appendix B.3.3 are standard results for regular estimation in Chapter 25 of \citet{ref32}.

\subsection{Proofs of Lemmas B.1–B.3}\label{sec:B.3}

This subsection proves Lemmas B.1, B.2, and B.3 of Appendix B.2.1 in turn.

\subsubsection{Proof of Lemma B.1}\label{sec:B.3.1}

\textbf{Proof of Lemma B.1.} Equation (2.7) holds for $F_{X\mid O}$-almost every $x$, and since every point of $\mathcal X$ has positive probability by Assumption 3.2, it holds for every $x\in\mathcal X$. Claim (1) follows by applying the first equation of (2.7) to the bounded truncation $\min\{g,n\}$ and letting $n\to\infty$ with the monotone convergence theorem. Claim (2) follows by applying both equations of (2.7) to the two-sided truncation $\max\{-n,\min\{g,n\}\}$ and letting $n\to\infty$ with the dominated convergence theorem. The dominating functions are $\max\{C,1\}|g(M)|$ on the experimental side and $\max\{C,1\}\mathbf 1\{T=t\}|g(M)|$ on the observational side, where the constant comes from $|Y|\le C$ in Assumption B.1 and $|\nu_t|\le C$. The observational one is integrable by the equation that claim (1) gives for $|g|$ and by $\pi_t\le1$.

We next prove claim (3). Iterated expectations and (3.2) give
\begin{equation}\tag{B.8}\label{eq:B.8}
\begin{aligned}
&\mathbb E[\nu_t(S,x)\omega_t^\nu(M,x)\mid T=t,x,E]\\
&=\mathbb E[\nu_t(S,x)\mathbb E[\omega_t^\nu(M,x)\mid S,T=t,x,E]\mid T=t,x,E]=\mathbb E[\mu_t(S,x)\mid T=t,x,E],\\
&\mathbb E[\pi_t(S,x)\omega_t^\pi(M,x)\mid T=t,x,E]\\
&=\mathbb E[\pi_t(S,x)\mathbb E[\omega_t^\pi(M,x)\mid S,T=t,x,E]\mid T=t,x,E]=\mathbb E[\mu_t(S,x)\mid T=t,x,E]
\end{aligned}
\end{equation}
and the right-hand sides equal $\bar\mu_t(x)$, where the last equality in the first equation uses $\nu_t/\pi_t=\mu_t$. Each expectation is finite by $|\nu_t|\le C$, $\pi_t\le1$, and the square integrability of $\omega_t^\nu,\omega_t^\pi$. Since square-integrable functions are integrable, claim (2) applied to $g=\omega_t^\nu(\cdot,x)$ and $g=\omega_t^\pi(\cdot,x)$, together with (B.8), gives
\[
\begin{aligned}
\mathbb E[\mathbf 1\{T=t\}\{Y\omega_t^\nu(M,x)-\omega_t^\pi(M,x)\}\mid x,O]
&=\mathbb E[\nu_t(S,x)\omega_t^\nu(M,x)-\pi_t(S,x)\omega_t^\pi(M,x)\mid T=t,x,E]\\
&=\bar\mu_t(x)-\bar\mu_t(x)=0
\end{aligned}
\]
Finally, we compute the means of the two functions. Since the probability that the observed $(T,X)$ equals $(t,x)$ is $P(T=t,X=x\mid E)$, the expectation of each term in (3.3) is $P(X=x\mid O)\mathbb E[\mu_t(S,x)-\bar\mu_t(x)-\nu_t(S,x)\omega_t^\nu(M,x)+\pi_t(S,x)\omega_t^\pi(M,x)\mid T=t,x,E]=P(X=x\mid O)\{\bar\mu_t(x)-\bar\mu_t(x)-\bar\mu_t(x)+\bar\mu_t(x)\}=0$, so $\mathbb E[\psi_E(Z_E)\mid E]=0$. In (3.4), $\mathbb E[\bar\mu_1(X)-\bar\mu_0(X)\mid O]=\tau$ and the equalities above given $X=x$ yield $\mathbb E[\psi_O(Z_O)\mid O]=0$. This proves claim (3). $\square$

\subsubsection{Proof of Lemma B.2}\label{sec:B.3.2}

\textbf{Proof of Lemma B.2.} The mean-zero property is Lemma B.1(3), so we show that the second moments are finite. When the observed $(T,X)$ equals $(t,x)$, only one term of the sum in (3.3) remains, and its weight $P(X=x\mid O)/P(T=t,X=x\mid E)$ is at most $c^{-1}$ by Assumption 3.2(2). Assumption B.1 gives $|\mu_t|,|\bar\mu_t|,|\nu_t|\le C$ and $\pi_t\le1$, and Assumption 3.4 makes $\omega_t^\nu,\omega_t^\pi$ square-integrable, so the remaining term is square-integrable and $\mathbb E[\psi_E(Z_E)^2\mid E]<\infty$. For (3.4), Lemma B.1(1) gives $\mathbb E[\mathbf 1\{T=t\}\omega_t^\nu(M,x)^2\mid x,O]=\mathbb E[\pi_t(S,x)\omega_t^\nu(M,x)^2\mid T=t,x,E]<\infty$, and the same holds for $\omega_t^\pi$, which together with $|Y|\le C$ and $|\bar\mu_t|\le C$ gives $\mathbb E[\psi_O(Z_O)^2\mid O]<\infty$. $\square$

\subsubsection{Proof of Lemma B.3}\label{sec:B.3.3}

\textbf{Proof of Lemma B.3.} We apply the local asymptotic normality of paths that are differentiable in quadratic mean to each sample and add the two expansions. For a regular path of Appendix B.1, $P_{G,\varepsilon}$ is absolutely continuous with respect to $P_G$, so the likelihood ratio $dP_{G,\varepsilon}/dP_G$ is the density of $P_{G,\varepsilon}$ with respect to $P_G$ and has no singular part. The joint law of the two samples under $P_{1/\sqrt N}$ is likewise absolutely continuous with respect to the joint law under $P$, and its likelihood ratio is the product of the likelihood ratios of the individual observations. The log likelihood ratio on the left-hand side of (B.4) is therefore the log likelihood ratio of the joint laws of the two samples. This fact is used when Le Cam's third lemma is applied in Appendix C.1.6.

We use the following standard result for paths that are differentiable in quadratic mean \citep[Chapter 25]{ref32}. Let $W_1,\ldots,W_n$ be an independent and identically distributed sample of size $n$ from a law $Q$, and let $Q_\varepsilon$ be a path of the form (B.2) with score $b\in L_2^0(Q)$. For a real sequence $s_n$ with $s_n\ge0$ for large $n$ that converges to $s\ge0$, under $Q$,
\begin{equation}\tag{B.9}\label{eq:B.9}
\sum_{k=1}^n\log\frac{dQ_{s_n/\sqrt n}}{dQ}(W_k)=\frac{s_n}{\sqrt n}\sum_{k=1}^nb(W_k)-\frac{s^2}2\mathbb E_Q[b^2]+o_p(1)
\end{equation}
We apply this to the experimental sample with $n=n_E$, $Q=P_E$, $b=a_E$, and $s_{n_E}=\sqrt{n_E/N}$. Then $s_{n_E}/\sqrt{n_E}=1/\sqrt N$, and Assumption 3.1 gives $s_{n_E}\to\sqrt\rho$. Since the experimental sample is independent and identically distributed by Assumption 3.1, (B.9) gives
\begin{equation}\tag{B.10}\label{eq:B.10}
\sum_{i=1}^{n_E}\log\frac{dP_{E,1/\sqrt N}}{dP_E}(Z_{E,i})=\frac1{\sqrt N}\sum_{i=1}^{n_E}a_E(Z_{E,i})-\frac\rho2\mathbb E[a_E^2\mid E]+o_p(1)
\end{equation}
For the observational sample, $s_{n_O}=\sqrt{n_O/N}\to\sqrt{1-\rho}$, so in the same way
\begin{equation}\tag{B.11}\label{eq:B.11}
\sum_{j=1}^{n_O}\log\frac{dP_{O,1/\sqrt N}}{dP_O}(Z_{O,j})=\frac1{\sqrt N}\sum_{j=1}^{n_O}a_O(Z_{O,j})-\frac{1-\rho}2\mathbb E[a_O^2\mid O]+o_p(1)
\end{equation}
Adding (B.10) and (B.11), the two quadratic terms sum to $-\rho\mathbb E[a_E^2\mid E]/2-(1-\rho)\mathbb E[a_O^2\mid O]/2=-\|(a_E,a_O)\|_\rho^2/2$, which yields (B.4). For the convergence in distribution, the term in braces divided by $\sqrt N$ can be written as
\begin{equation}\tag{B.12}\label{eq:B.12}
\frac1{\sqrt N}\left\{\sum_{i=1}^{n_E}a_E(Z_{E,i})+\sum_{j=1}^{n_O}a_O(Z_{O,j})\right\}
=\sqrt{\frac{n_E}N}\,\frac1{\sqrt{n_E}}\sum_{i=1}^{n_E}a_E(Z_{E,i})+\sqrt{\frac{n_O}N}\,\frac1{\sqrt{n_O}}\sum_{j=1}^{n_O}a_O(Z_{O,j})
\end{equation}
By the central limit theorem, the two normalized sums converge in distribution to normal distributions with mean zero and variances $\mathbb E[a_E^2\mid E]$ and $\mathbb E[a_O^2\mid O]$, respectively, and they are independent by Assumption 3.1. By $\sqrt{n_E/N}\to\sqrt\rho$, $\sqrt{n_O/N}\to\sqrt{1-\rho}$, and Slutsky's lemma, the expression converges in distribution to a normal distribution with mean zero and variance $\rho\mathbb E[a_E^2\mid E]+(1-\rho)\mathbb E[a_O^2\mid O]=\|(a_E,a_O)\|_\rho^2$. $\square$

\subsection{Proof of Lemma B.4}\label{sec:B.4}

\textbf{Proof of Lemma B.4.} We differentiate the stratum shares and the stratum means along the smooth submodel and compare the result with the expected products of the influence functions and the scores. We assume Assumptions 2.1–2.8, the integrability conditions in Appendix A, Assumptions 3.2–3.4, and Assumption B.1, and we fix a smooth submodel $\{P_\varepsilon\}$ with score $(a_E,a_O)$ and remainders $r_{E,\varepsilon},r_{O,\varepsilon}$ as in Appendix B.1. We take $\varepsilon$ small enough that the densities $dP_{G,\varepsilon}/dP_G$ lie between $1/2$ and $3/2$ in both samples, write $\mathbb E_{P_\varepsilon}$ for expectations under $P_\varepsilon$, and suppress the arguments $(S,x)$ and $(M,x)$ of the functions when $t,x$ are fixed.

Along the path, $\tau$ is again a combination of stratum shares and stratum means. With $\bar\mu_{t,\varepsilon}(x)=\mathbb E_{P_\varepsilon}[\mu_{t,\varepsilon}(S,x)\mid T=t,x,E]$, (2.8) of Theorem 2.1 applied to $P_\varepsilon$ gives $\tau(P_\varepsilon)=\sum_{x\in\mathcal X}P_\varepsilon(X=x\mid O)\{\bar\mu_{1,\varepsilon}(x)-\bar\mu_{0,\varepsilon}(x)\}$. We express derivatives through the conditional covariance $\operatorname{Cov}(f,a_E\mid T=t,x,E)=\mathbb E[fa_E\mid T=t,x,E]-\mathbb E[f\mid T=t,x,E]\,\mathbb E[a_E\mid T=t,x,E]$, which is finite for every function $f$ of $(S,M)$ with $\mathbb E[|f|\mid T=t,x,E]<\infty$ because $a_E$ is bounded, and we define $\operatorname{Cov}(g,a_O\mid x,O)$ in the same way for a function $g$ of $(T,M,Y)$. Given $T=t,X=x$, the density of the experimental law under $P_\varepsilon$ with respect to that under $P$ is the ratio of $1+\varepsilon a_E+\varepsilon r_{E,\varepsilon}$ to its conditional mean, which equals $1+\varepsilon\{a_E-\mathbb E[a_E\mid T=t,x,E]\}$ up to a function whose essential supremum is $o(\varepsilon)$. Hence, for every such $f$ and every family $f_\varepsilon$ with $\mathbb E[|(f_\varepsilon-f)/\varepsilon-\dot f|\mid T=t,x,E]\to0$ for an integrable $\dot f$,
\begin{equation}\tag{B.13}\label{eq:B.13}
\begin{aligned}
&\lim_{\varepsilon\downarrow0}\frac{\mathbb E_{P_\varepsilon}[f\mid T=t,x,E]-\mathbb E[f\mid T=t,x,E]}\varepsilon=\operatorname{Cov}(f,a_E\mid T=t,x,E),\\
&\lim_{\varepsilon\downarrow0}\frac{\mathbb E_{P_\varepsilon}[f_\varepsilon\mid T=t,x,E]-\mathbb E[f\mid T=t,x,E]}\varepsilon=\mathbb E[\dot f\mid T=t,x,E]+\operatorname{Cov}(f,a_E\mid T=t,x,E)
\end{aligned}
\end{equation}
where the second limit follows from the first and the boundedness of the conditional density. The same limits hold in the observational sample given $X=x$, with $a_O$ in place of $a_E$.

We first differentiate (2.7) along the path with the weight functions as test functions. Because $|Y|\le C$ also holds under $P_\varepsilon$, which is absolutely continuous with respect to $P$, and $\mu_{t,\varepsilon}$ is a conditional mean of $Y$, the bounds $|\mu_{t,\varepsilon}|\le C$ and $|\nu_{t,\varepsilon}|\le C$ hold almost surely under the true law. Equation (2.7) holds under $P_\varepsilon$ because $P_\varepsilon$ belongs to $\mathcal P$, and the truncation argument in the proof of Lemma B.1, applied under $P_\varepsilon$ with these bounds, extends it to $h=\omega_t^\nu$ and $h=\omega_t^\pi$, which are square-integrable under $P$ and hence under $P_\varepsilon$. We differentiate both sides of the extended equations with (B.13). On the left-hand sides, both the functions $\nu_{t,\varepsilon}\omega_t^\nu,\pi_{t,\varepsilon}\omega_t^\pi$ and the experimental law move, and by the Cauchy–Schwarz inequality the difference quotients of the functions converge in mean absolute value to $\dot\nu_t\omega_t^\nu,\dot\pi_t\omega_t^\pi$. On the right-hand sides, only the observational law moves. Equating the derivatives gives
\begin{equation}\tag{B.14}\label{eq:B.14}
\begin{aligned}
\mathbb E[\dot\nu_t\omega_t^\nu\mid T=t,x,E]&=\operatorname{Cov}(\mathbf 1\{T=t\}Y\omega_t^\nu,a_O\mid x,O)-\operatorname{Cov}(\nu_t\omega_t^\nu,a_E\mid T=t,x,E),\\
\mathbb E[\dot\pi_t\omega_t^\pi\mid T=t,x,E]&=\operatorname{Cov}(\mathbf 1\{T=t\}\omega_t^\pi,a_O\mid x,O)-\operatorname{Cov}(\pi_t\omega_t^\pi,a_E\mid T=t,x,E)
\end{aligned}
\end{equation}
where the observational covariances are finite by Lemma B.1(1).

We next differentiate the stratum means. The identities $\nu_{t,\varepsilon}=\pi_{t,\varepsilon}\mu_{t,\varepsilon}$ and $\nu_t=\pi_t\mu_t$ give $\pi_t(\mu_{t,\varepsilon}-\mu_t)=(\nu_{t,\varepsilon}-\nu_t)-\mu_{t,\varepsilon}(\pi_{t,\varepsilon}-\pi_t)$. Since $\pi_t\ge a$, $|\mu_{t,\varepsilon}|\le C$, and the difference quotients of $\pi_t$ and $\nu_t$ converge in $\|\cdot\|_{2,t,x}$, this identity gives $\|\mu_{t,\varepsilon}-\mu_t\|_{2,t,x}\le C\varepsilon$. Writing $\mu_{t,\varepsilon}=\mu_t+(\mu_{t,\varepsilon}-\mu_t)$ on its right-hand side and dividing by $\varepsilon\pi_t$ gives
\[
\frac{\mu_{t,\varepsilon}-\mu_t}{\varepsilon}=\frac1{\pi_t}\cdot\frac{\nu_{t,\varepsilon}-\nu_t}{\varepsilon}-\frac{\mu_t}{\pi_t}\cdot\frac{\pi_{t,\varepsilon}-\pi_t}{\varepsilon}-\frac{(\mu_{t,\varepsilon}-\mu_t)(\pi_{t,\varepsilon}-\pi_t)}{\varepsilon\pi_t}
\]
and the Cauchy–Schwarz inequality bounds the mean absolute value of the last term by $C\varepsilon$. The difference quotient of $\mu_t$ therefore converges in mean absolute value to $(\dot\nu_t-\mu_t\dot\pi_t)/\pi_t$, and the second limit in (B.13) gives
\[
\left.\frac{d\bar\mu_{t,\varepsilon}(x)}{d\varepsilon}\right|_{\varepsilon=0+}=\operatorname{Cov}(\mu_t,a_E\mid T=t,x,E)+\mathbb E\!\left[\frac{\dot\nu_t}{\pi_t}-\frac{\mu_t\dot\pi_t}{\pi_t}\,\middle|\,T=t,x,E\right]
\]
Since $\dot\nu_t$ and $\dot\pi_t$ are functions of $S$, iterated expectations and (3.2) replace $1/\pi_t$ and $\mu_t/\pi_t$ by $\omega_t^\nu$ and $\omega_t^\pi$, so the last expectation equals $\mathbb E[\dot\nu_t\omega_t^\nu-\dot\pi_t\omega_t^\pi\mid T=t,x,E]$, and (B.14) gives
\begin{equation}\tag{B.15}\label{eq:B.15}
\begin{aligned}
\left.\frac{d\bar\mu_{t,\varepsilon}(x)}{d\varepsilon}\right|_{\varepsilon=0+}
={}&\operatorname{Cov}(\mu_t-\nu_t\omega_t^\nu+\pi_t\omega_t^\pi,a_E\mid T=t,x,E)\\
&+\operatorname{Cov}(\mathbf 1\{T=t\}(Y\omega_t^\nu-\omega_t^\pi),a_O\mid x,O)
\end{aligned}
\end{equation}

Finally, we differentiate $\tau(P_\varepsilon)$ and compare the result with the expected products. The stratum shares satisfy $\{P_\varepsilon(X=x\mid O)-P(X=x\mid O)\}/\varepsilon=\mathbb E[\mathbf 1\{X=x\}(a_O+r_{O,\varepsilon})\mid O]\to P(X=x\mid O)\mathbb E[a_O\mid x,O]$. Since $\mathcal X$ is finite, the product rule and (B.15) give
\[
\begin{aligned}
\left.\frac{d\tau(P_\varepsilon)}{d\varepsilon}\right|_{\varepsilon=0+}
={}&\sum_{x\in\mathcal X}P(X=x\mid O)\{\bar\mu_1(x)-\bar\mu_0(x)\}\,\mathbb E[a_O\mid x,O]\\
&+\sum_{x\in\mathcal X}P(X=x\mid O)\sum_{t\in\{0,1\}}(2t-1)\operatorname{Cov}(\mu_t-\nu_t\omega_t^\nu+\pi_t\omega_t^\pi,a_E\mid T=t,x,E)\\
&+\sum_{x\in\mathcal X}P(X=x\mid O)\sum_{t\in\{0,1\}}(2t-1)\operatorname{Cov}(\mathbf 1\{T=t\}(Y\omega_t^\nu-\omega_t^\pi),a_O\mid x,O)
\end{aligned}
\]
The second sum equals $\mathbb E[\psi_E(Z_E)a_E(Z_E)\mid E]$, and the first and third sums add up to $\mathbb E[\psi_O(Z_O)a_O(Z_O)\mid O]$, which gives (B.3). Indeed, in (3.3), the term for $(t,x)$, taken with its sign, is nonzero only when $(T,X)=(t,x)$, and the expression in braces has conditional mean zero given $T=t,X=x$ by Lemma B.1(3), so the expected product of this term and $a_E$ is $(2t-1)P(X=x\mid O)$ times the conditional covariance in the second sum, from which the constant $\bar\mu_t(x)$ drops out. In (3.4), $\mathbb E[\tau a_O(Z_O)\mid O]=0$ because the score has mean zero, the term $\bar\mu_1(X)-\bar\mu_0(X)$ gives the first sum, and the terms $\mathbf 1\{T=t\}\{Y\omega_t^\nu(M,X)-\omega_t^\pi(M,X)\}$ have conditional mean zero given $X=x$ by Lemma B.1(3), so their expected products with $a_O$ given $X=x$ are the conditional covariances in the third sum. $\square$

\section{Proofs of Theorems 3.2 and 3.3}\label{sec:C}

This appendix proves Theorems 3.2 and 3.3 of Section 3, which give the asymptotic distribution and efficiency of the estimator without sample splitting and of the cross-fitted estimator. Table C.1 lists the contents of the subsections, and Figure C.1 shows how the results used in the proofs depend on one another.

\begin{table}[tbp]
\centering
\def\MainTextTableBody{
\begin{tabular}{lp{.79\textwidth}}
\toprule
Subsection & Contents \\
\midrule
C.1 & Proves Theorem 3.2 through events with probability approaching one (C.1.1), Lemma C.1 (C.1.2), Lemma C.2 (C.1.3), the asymptotic linear representation (C.1.4), Lemma C.3 (C.1.5), and studentization and efficiency (C.1.6) \\
C.2 & Defines events with probability approaching one (C.2.1), states and proves Lemma C.4 (C.2.2) and Lemma C.5 (C.2.3), and then proves Theorem 3.3 (C.2.4) \\
\bottomrule
\end{tabular}
}
\setlength{\tabcolsep}{4pt}\small
\sbox{\MainTextTableBox}{\MainTextTableBody}
\def\MainTextTableFont{small}
\typeout{TABLE-WIDTH C.1 font=\MainTextTableFont; width=\the\wd\MainTextTableBox; limit=\the\textwidth}
\ifdim\wd\MainTextTableBox>\textwidth
  \PackageError{main-text}{Table C.1 remains too wide}{Stop and report; no scaling allowed}
\fi
\usebox{\MainTextTableBox}
\renewcommand{\thetable}{C.1}
\caption{Subsections of Appendix C and their contents}
\label{table:C.1}
\end{table}
The arrows in Figure C.1 have the same meaning as in Figure B.1. The core of the proof of Theorem 3.2 is Lemma C.2, which splits the error in a stratum mean into sample averages evaluated at the true functions and a remainder that tends to zero faster than $N^{-1/2}$. Among the terms of this remainder, the discrepancy that arises when the estimates are held fixed and the sample averages are replaced by expectations is bounded with Lemma C.1, which expresses it through products of estimation errors. Lemma C.2 gives the asymptotic linear representation of the estimator (Appendix C.1.4), and together with the consistency of the variance estimator in Lemma C.3 it gives the limit of the studentized estimator. Efficiency is shown in Appendix C.1.6 with Lemmas B.2–B.4 and Theorem 3.1, which is proved in Appendix B.

The proof of Theorem 3.3 establishes the conclusions of Lemmas C.2 and C.3 again for the cross-fitted estimator and applies the arguments of Appendices C.1.4 and C.1.6. The terms due to plugging in the estimates that leave out a fold and the mean squared errors of the variance estimator are bounded by Lemma C.4, which conditions on the observations outside a fold, and the averages over the folds are turned into sums over the full samples by Lemma C.5 (Appendix C.2.4).

\begin{figure}[htbp]
\centering
\includegraphics[max width=\textwidth]{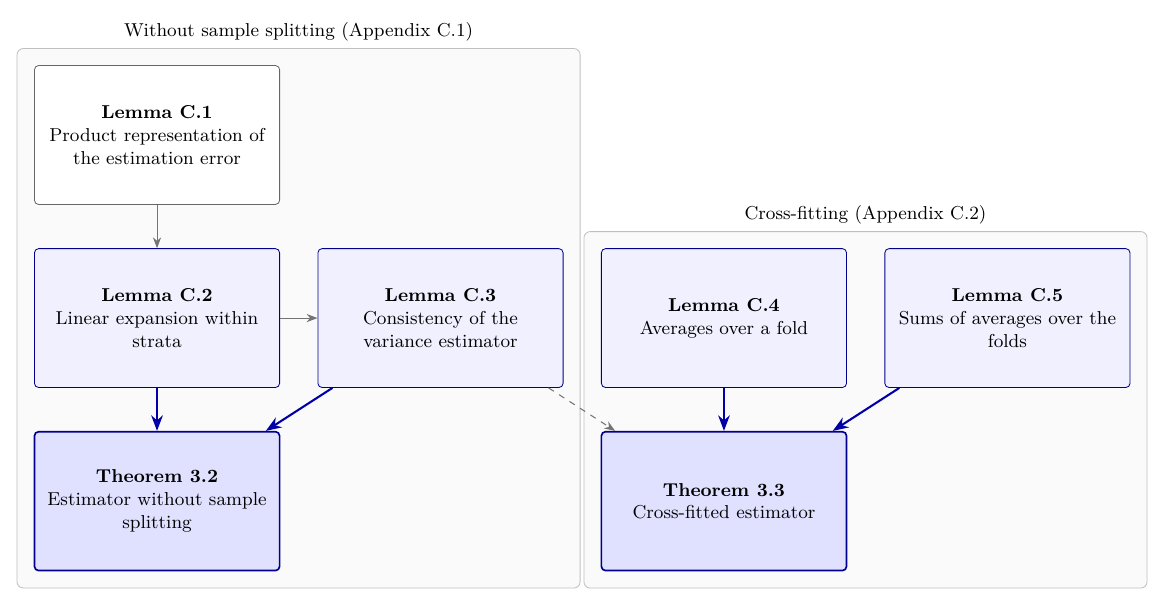}
\renewcommand{\thefigure}{C.1}
\caption{Dependencies among the results used to prove Theorems 3.2 and 3.3}
\label{figure:C.1}
\par\smallskip{\small\NoteLeft Note: Thick arrows start from the lemmas that directly yield the theorems. The dashed arrow indicates that the proof of Theorem 3.3 applies the arguments for Lemmas C.2 and C.3 fold by fold. The events with probability approaching one (Appendices C.1.1 and C.2.1) are used throughout the proofs of the respective theorems.\par}
\end{figure}

\renewcommand{\thefixedassumption}{C.1}
\begin{fixedassumption}[Boundedness of the estimates]\label{assumption:C.1}

For the constant $C$ in Assumption B.1, the following hold for each $t\in\{0,1\}$ and $x\in\mathcal X$.
\begin{enumerate}
\item On events with probability approaching one, the full-sample estimates of Section 3.2 satisfy $|\widehat\nu_t(s,x)|\le C$ for all $s$.
\item For each $k$, on events with probability approaching one, the estimates of Section 3.4 that leave out fold $k$ satisfy $|\widehat\nu_t^{(-k)}(s,x)|\le C$ for all $s$.
\end{enumerate}
\end{fixedassumption}

Condition (1) of Assumption C.1 bounds the full-sample estimates, and condition (2) bounds the estimates that leave out a fold. Theorem 3.2 uses Assumption B.1 and condition (1), and Theorem 3.3 uses Assumption B.1 and condition (2). The proofs in this appendix use the bounded versions of $\pi_t,\mu_t$ chosen at the beginning of Appendix B and equations (2.7), (3.2), (3.6), and (3.7) of the main text, recalled at the beginning of Appendix B.

In this appendix, $C$ denotes a positive constant that does not depend on $N$ and whose value may change from one occurrence to the next. The constant $C$ may depend on $K$, $\rho$, $a$, $c$, and the constants in the assumptions, and it is taken to be at least the constants in Assumptions B.1 and C.1, so the inequalities in these assumptions also hold with this $C$.

\subsection{Proof of Theorem 3.2}\label{sec:C.1}

This subsection proves Theorem 3.2 of Section 3.3. The goals of Theorem 3.2 are the two convergences in (3.8) and the coverage probability of the interval. We first split the difference between the estimator and the true value into three terms, namely the error in the observational stratum shares, the error in the stratum means, and their product. The error in the stratum shares is a sample average over the observational sample and gives the part $\bar\mu_1(X)-\bar\mu_0(X)-\tau$ of (3.4). Lemma C.2 splits the error in the stratum means into five terms. The two sample averages evaluated at the true functions give the sample average of the function in (3.3) and the remaining part of (3.4). The two terms due to plugging in the estimates tend to zero faster than $N^{-1/2}$ by Assumption 3.6(1)–(5), and so does the discrepancy from $\bar\mu_t(x)$ that arises when the estimates are held fixed and the sample averages are replaced by expectations, by Lemma C.1 of Appendix C.1.2 and Assumption 3.5(2)–(4). The product of the two errors is $O_p(N^{-1})$. We then apply the central limit theorem to the sum of the two resulting sample averages, and Lemma C.3, which uses Assumption 3.6(6)–(11), to the variance estimator. Finally, Slutsky's lemma gives the limit of the studentized estimator and the coverage probability of the interval. The proof for cross-fitting in Appendix C.2 differs in two places, the bounds on the two terms due to plugging in the estimates and the bound on the variance estimator.

\subsubsection{Events with probability approaching one}\label{sec:C.1.1}

In the proof below, we work on the event on which the following three conditions hold simultaneously for all $t,x$, and we call it the event of Appendix C.1.1. (E1) $\widehat P(T=t,X=x\mid E)\ge c/2$ and $\widehat P(X=x\mid O)\ge c/2$. (E2) The event of Assumption C.1(1) occurs, that is, an event with probability approaching one on which $|\widehat\nu_t(s,x)|\le C$ for all $s$. (E3) $\|\widehat\omega_t^\nu\|_{2,t,x}$ and $\|\widehat\omega_t^\pi\|_{2,t,x}$ are finite. Here $c$ is the constant in Assumption 3.2(2), and $C$ is the constant in the bound of Assumption C.1(1). The probability of (E1) approaches one. For example, $\widehat P(T=t,X=x\mid E)$ is a sample average with mean $P(T=t,X=x\mid E)\ge c$ and variance at most $1/(4n_E)$, and Assumption 3.1 bounds $n_E$ from below by a positive constant times $N$, so by Chebyshev's inequality the probability of $\widehat P(T=t,X=x\mid E)<c/2$ tends to zero. The same holds for $\widehat P(X=x\mid O)$. The probability of (E2) approaches one by Assumption C.1(1). For (E3), Minkowski's inequality gives $\|\widehat\omega_t^\nu\|_{2,t,x}\le\|\omega_t^\nu\|_{2,t,x}+\|\widehat\omega_t^\nu-\omega_t^\nu\|_{2,t,x}$, where the first term on the right-hand side is finite by Assumption 3.4 and the second tends to zero in probability by Assumption 3.5(1), so the probability that the norm is finite approaches one. The same holds for $\widehat\omega_t^\pi$. Since the probabilities of finitely many events all approach one, the probability that they hold simultaneously also approaches one. On this event, all the groups over which (3.5), (3.6), and the variance estimator average are nonempty. Since the probability of the complement tends to zero, convergence in probability and in distribution established on this event continue to hold without the restriction to the event.

\subsubsection{Product remainders}\label{sec:C.1.2}

This part states and proves Lemma C.1, which expresses through products of estimation errors the discrepancy from $\bar\mu_t(x)$ that arises when the estimates are held fixed and the two sample averages in (3.5) are replaced by expectations. The lemma is used in the proof of Lemma C.2 in Appendix C.1.3.

\renewcommand{\thefixedlemma}{C.1}
\begin{fixedlemma}[Product representation of the estimation error]\label{lemma:C.1}

Suppose that Assumptions 2.1–2.8, the integrability conditions in Appendix A, Assumptions 3.1–3.4, and Assumption B.1 hold, and fix $t,x$. With the sample values held fixed, suppose that $|\widehat\nu_t(s,x)|\le C$ for all $s$ and that $\|\widehat\omega_t^\nu\|_{2,t,x}$ and $\|\widehat\omega_t^\pi\|_{2,t,x}$ are finite. In (3.5),
\[
\begin{aligned}
\widehat{\bar\mu}_t(x)={}&\widehat{\mathbb E}[\widehat\mu_t(S,x)-\widehat\nu_t(S,x)\widehat\omega_t^\nu(M,x)+\widehat\pi_t(S,x)\widehat\omega_t^\pi(M,x)\mid T=t,x,E]\\
&+\widehat{\mathbb E}[\mathbf 1\{T=t\}\{Y\widehat\omega_t^\nu(M,x)-\widehat\omega_t^\pi(M,x)\}\mid x,O]
\end{aligned}
\]
replacing the two sample averages $\widehat{\mathbb E}$ by the corresponding expectations $\mathbb E$ and subtracting $\bar\mu_t(x)$ gives
\begin{equation}\tag{C.1}\label{eq:C.1}
\begin{aligned}
&\mathbb E\!\left[\frac{\{\pi_t(S,x)-\widehat\pi_t(S,x)\}\{\widehat\mu_t(S,x)-\mu_t(S,x)\}}{\pi_t(S,x)}\,\middle|\,T=t,x,E\right]\\
&\qquad+\mathbb E[\{\nu_t(S,x)-\widehat\nu_t(S,x)\}\{\widehat\omega_t^\nu(M,x)-\omega_t^\nu(M,x)\}\mid T=t,x,E]\\
&\qquad-\mathbb E[\{\pi_t(S,x)-\widehat\pi_t(S,x)\}\{\widehat\omega_t^\pi(M,x)-\omega_t^\pi(M,x)\}\mid T=t,x,E]
\end{aligned}
\end{equation}
\end{fixedlemma}

The proof of Lemma C.1 consists of calculations that treat the estimates as fixed functions, so the lemma applies to each realization of the full-sample estimates. The first term on the right-hand side is a product of the error in the selection probability and the error in the ratio, and the second and third terms are means of products of the errors in the nuisance functions and the errors in the weight functions.

\textbf{Proof of Lemma C.1.} Fix $t,x$ and hold the sample values fixed. Since $\widehat\omega_t^\nu(\cdot,x)$ and $\widehat\omega_t^\pi(\cdot,x)$ are square-integrable and hence integrable, Lemma B.1(2) turns the observational expectation $\mathbb E[\mathbf 1\{T=t\}\{Y\widehat\omega_t^\nu(M,x)-\widehat\omega_t^\pi(M,x)\}\mid x,O]$ into $\mathbb E[\nu_t(S,x)\widehat\omega_t^\nu(M,x)-\pi_t(S,x)\widehat\omega_t^\pi(M,x)\mid T=t,x,E]$. Collecting the terms multiplied by the same $\widehat\omega$ shows that the quantity obtained by replacing the two sample averages in (3.5) by expectations, minus $\bar\mu_t(x)$, equals
\begin{equation}\tag{C.2}\label{eq:C.2}
\begin{aligned}
&\mathbb E[\widehat\mu_t(S,x)\mid T=t,x,E]-\bar\mu_t(x)+\mathbb E[\{\nu_t(S,x)-\widehat\nu_t(S,x)\}\widehat\omega_t^\nu(M,x)\mid T=t,x,E]\\
&\qquad-\mathbb E[\{\pi_t(S,x)-\widehat\pi_t(S,x)\}\widehat\omega_t^\pi(M,x)\mid T=t,x,E]
\end{aligned}
\end{equation}
Each expectation is finite by $|\widehat\nu_t|\le C$, the bound $|\widehat\mu_t|\le C/a$ implied by $\widehat\pi_t\ge a$, $|\nu_t|\le C$, $\widehat\pi_t,\pi_t\le1$, and the square integrability of $\widehat\omega_t^\nu,\widehat\omega_t^\pi,\omega_t^\nu,\omega_t^\pi$.

We write $\widehat\omega_t^\nu=\omega_t^\nu+(\widehat\omega_t^\nu-\omega_t^\nu)$ and $\widehat\omega_t^\pi=\omega_t^\pi+(\widehat\omega_t^\pi-\omega_t^\pi)$ in the last two terms of (C.2). The parts $\widehat\omega_t^\nu-\omega_t^\nu$ and $\widehat\omega_t^\pi-\omega_t^\pi$ give the last two terms of (C.1). For the parts $\omega_t^\nu$ and $\omega_t^\pi$, omitting the arguments $(S,x)$ and $(M,x)$, iterated expectations and (3.2) give $\mathbb E[(\nu_t-\widehat\nu_t)\omega_t^\nu\mid T=t,x,E]=\mathbb E[(\nu_t-\widehat\nu_t)/\pi_t\mid T=t,x,E]$ and $\mathbb E[(\pi_t-\widehat\pi_t)\omega_t^\pi\mid T=t,x,E]=\mathbb E[(\pi_t-\widehat\pi_t)\mu_t/\pi_t\mid T=t,x,E]$. Together with $\bar\mu_t(x)=\mathbb E[\mu_t\mid T=t,x,E]$, the remaining terms are the conditional mean of a function of $S$, which $\nu_t=\pi_t\mu_t$ and $\widehat\nu_t=\widehat\pi_t\widehat\mu_t$ simplify as
\[
\begin{aligned}
&\widehat\mu_t-\mu_t+\frac{\pi_t\mu_t-\widehat\pi_t\widehat\mu_t}{\pi_t}-\frac{(\pi_t-\widehat\pi_t)\mu_t}{\pi_t}\\
&=\frac{(\pi_t-\widehat\pi_t)\widehat\mu_t}{\pi_t}-\frac{(\pi_t-\widehat\pi_t)\mu_t}{\pi_t}
=\frac{(\pi_t-\widehat\pi_t)(\widehat\mu_t-\mu_t)}{\pi_t}
\end{aligned}
\]
This is the function in the first term of (C.1). $\square$

\subsubsection{Linear expansion within strata}\label{sec:C.1.3}

\renewcommand{\thefixedlemma}{C.2}
\begin{fixedlemma}[Linear expansion within strata]\label{lemma:C.2}

Suppose that Assumptions 2.1–2.8, the integrability conditions in Appendix A, Assumptions 3.1–3.5, Assumption 3.6(1)–(5), Assumption B.1, and Assumption C.1(1) hold. Then, for each $t\in\{0,1\}$ and $x\in\mathcal X$,
\begin{equation}\tag{C.3}\label{eq:C.3}
\begin{aligned}
&\widehat{\bar\mu}_t(x)-\bar\mu_t(x)\\
&=\frac1{n_E}\sum_{i=1}^{n_E}\frac{\mathbf 1\{T_i=t,X_i=x\}}{P(T=t,X=x\mid E)}\{\mu_t(S_i,x)-\bar\mu_t(x)-\nu_t(S_i,x)\omega_t^\nu(M_i,x)+\pi_t(S_i,x)\omega_t^\pi(M_i,x)\}\\
&\qquad+\frac1{n_O}\sum_{j=1}^{n_O}\frac{\mathbf 1\{X_j=x\}}{P(X=x\mid O)}\mathbf 1\{T_j=t\}\{Y_j\omega_t^\nu(M_j,x)-\omega_t^\pi(M_j,x)\}+o_p(N^{-1/2})
\end{aligned}
\end{equation}
and both sums on the right-hand side are $O_p(N^{-1/2})$.
\end{fixedlemma}

\textbf{Proof of Lemma C.2.} Fix $t,x$. The goal is (C.3). We split the difference between $\widehat{\bar\mu}_t(x)$ and $\bar\mu_t(x)$ into five terms, namely the two sample averages evaluated at the true functions, the changes in the two sample averages due to plugging in the estimates, and the discrepancy from $\bar\mu_t(x)$ that arises when the estimates are held fixed and the sample averages are replaced by expectations. The sample averages of the true functions become the two sums in (C.3). The changes due to the estimates tend to zero faster than $N^{-1/2}$ by Assumption 3.6(1)–(5), and so does the discrepancy, by Lemma C.1 of Appendix C.1.2, which rewrites it as means of products of estimation errors, and by Assumption 3.5(2)–(4).

In what follows, we work on the event of Appendix C.1.1. All expectations below are finite on this event, so they can be split term by term. Indeed, (E2) and $\widehat\pi_t\ge a$ give $|\widehat\mu_t|\le C/a$, and the expectations in the experimental stratum are finite by (E2), (E3), Assumption 3.4, Assumption B.1, and the Cauchy–Schwarz inequality. The expectations in the observational stratum are finite by Lemma B.1(2) applied to $\widehat\omega_t^\nu(\cdot,x)$, $\widehat\omega_t^\pi(\cdot,x)$, $\omega_t^\nu(\cdot,x)$, and $\omega_t^\pi(\cdot,x)$, which are square-integrable and hence integrable. Subtracting and adding the terms with the true functions in each term of (3.5) gives
\begin{equation}\tag{C.4}\label{eq:C.4}
\begin{aligned}
\widehat{\bar\mu}_t(x)-\bar\mu_t(x)
={}&\underbrace{\widehat{\mathbb E}[\mu_t(S,x)-\bar\mu_t(x)-\nu_t(S,x)\omega_t^\nu(M,x)+\pi_t(S,x)\omega_t^\pi(M,x)\mid T=t,x,E]}_{\text{(I)}}\\
&+\underbrace{\widehat{\mathbb E}[\mathbf 1\{T=t\}\{Y\omega_t^\nu(M,x)-\omega_t^\pi(M,x)\}\mid x,O]}_{\text{(II)}}\\
&+\underbrace{\begin{aligned}&(\widehat{\mathbb E}-\mathbb E)[\widehat\mu_t(S,x)-\mu_t(S,x)\mid T=t,x,E]\\&-(\widehat{\mathbb E}-\mathbb E)[\widehat\nu_t(S,x)\widehat\omega_t^\nu(M,x)-\nu_t(S,x)\omega_t^\nu(M,x)\mid T=t,x,E]\\&+(\widehat{\mathbb E}-\mathbb E)[\widehat\pi_t(S,x)\widehat\omega_t^\pi(M,x)-\pi_t(S,x)\omega_t^\pi(M,x)\mid T=t,x,E]\end{aligned}}_{\text{(III)}}\\
&+\underbrace{\begin{aligned}&(\widehat{\mathbb E}-\mathbb E)[\mathbf 1\{T=t\}Y\{\widehat\omega_t^\nu(M,x)-\omega_t^\nu(M,x)\}\mid x,O]\\&-(\widehat{\mathbb E}-\mathbb E)[\mathbf 1\{T=t\}\{\widehat\omega_t^\pi(M,x)-\omega_t^\pi(M,x)\}\mid x,O]\end{aligned}}_{\text{(IV)}}\\
&+\underbrace{\begin{aligned}&\mathbb E[\widehat\mu_t(S,x)-\widehat\nu_t(S,x)\widehat\omega_t^\nu(M,x)+\widehat\pi_t(S,x)\widehat\omega_t^\pi(M,x)\mid T=t,x,E]\\&+\mathbb E[\mathbf 1\{T=t\}\{Y\widehat\omega_t^\nu(M,x)-\widehat\omega_t^\pi(M,x)\}\mid x,O]-\bar\mu_t(x)\end{aligned}}_{\text{(V)}}
\end{aligned}
\end{equation}
To verify (C.4), note that, by Lemma B.1(3) and $\mathbb E[\mu_t\mid T=t,x,E]=\bar\mu_t(x)$, the function $\mu_t-\nu_t\omega_t^\nu+\pi_t\omega_t^\pi$ has conditional mean $\bar\mu_t(x)$ given $T=t,X=x$ and the function $\mathbf 1\{T=t\}\{Y\omega_t^\nu-\omega_t^\pi\}$ has conditional mean zero given $X=x$, where we omit the arguments $(S,x)$ and $(M,x)$. Hence the sum of terms (I) and (III) is $(\widehat{\mathbb E}-\mathbb E)[\widehat\mu_t-\widehat\nu_t\widehat\omega_t^\nu+\widehat\pi_t\widehat\omega_t^\pi\mid T=t,x,E]$, and the sum of terms (II) and (IV) is $(\widehat{\mathbb E}-\mathbb E)[\mathbf 1\{T=t\}\{Y\widehat\omega_t^\nu-\widehat\omega_t^\pi\}\mid x,O]$. Adding term (V) cancels the two expectations and leaves the right-hand side of (3.5) minus $\bar\mu_t(x)$, which is the left-hand side of (C.4). Terms (I) and (II) are sample averages evaluated at the true functions, terms (III) and (IV) are the changes in the sample averages due to plugging in the estimates, and term (V) is the left-hand side of Lemma C.1 of Appendix C.1.2.

For terms (I) and (II), we use the following equality. Suppose that a function $f$ of an experimental observation satisfies $\mathbb E[f(Z_E)\mid T=t,x,E]=0$ and $\mathbb E[f(Z_E)^2\mid T=t,x,E]<\infty$. By the definition of the sample average, $\widehat{\mathbb E}[f(Z_E)\mid T=t,x,E]=n_E^{-1}\sum_{i=1}^{n_E}\mathbf 1\{T_i=t,X_i=x\}f(Z_{E,i})/\widehat P(T=t,X=x\mid E)$, so
\begin{equation}\tag{C.5}\label{eq:C.5}
\begin{aligned}
&\widehat{\mathbb E}[f(Z_E)\mid T=t,x,E]-\frac1{n_E}\sum_{i=1}^{n_E}\frac{\mathbf 1\{T_i=t,X_i=x\}f(Z_{E,i})}{P(T=t,X=x\mid E)}\\
&=\frac1{n_E}\sum_{i=1}^{n_E}\mathbf 1\{T_i=t,X_i=x\}f(Z_{E,i})\cdot\frac{P(T=t,X=x\mid E)-\widehat P(T=t,X=x\mid E)}{\widehat P(T=t,X=x\mid E)P(T=t,X=x\mid E)}
\end{aligned}
\end{equation}
Since the terms $\mathbf 1\{T_i=t,X_i=x\}f(Z_{E,i})$ are independent and identically distributed with mean $P(T=t,X=x\mid E)\mathbb E[f(Z_E)\mid T=t,x,E]=0$ and variance $P(T=t,X=x\mid E)\mathbb E[f(Z_E)^2\mid T=t,x,E]$, Assumption 3.1 and Chebyshev's inequality show that the first factor on the right-hand side is $O_p(N^{-1/2})$. Similarly, $\widehat P(T=t,X=x\mid E)-P(T=t,X=x\mid E)$ is $O_p(N^{-1/2})$, and the denominator is at least $c^2/2$ on the event of Appendix C.1.1, so the right-hand side of (C.5) is $O_p(N^{-1})$. For the same reason, the sum on the left-hand side of (C.5) is $O_p(N^{-1/2})$.

Term (I) is the sample average of $f(Z_E)=\mu_t(S,x)-\bar\mu_t(x)-\nu_t(S,x)\omega_t^\nu(M,x)+\pi_t(S,x)\omega_t^\pi(M,x)$. Its mean is zero by Lemma B.1(3), and it is square-integrable by $|\mu_t|,|\bar\mu_t|,|\nu_t|\le C$, $\pi_t\le1$, and Assumption 3.4. Hence, by (C.5), term (I) equals the first sum in (C.3) plus $O_p(N^{-1})$.

Term (II) is the sample average of $g(Z_O)=\mathbf 1\{T=t\}\{Y\omega_t^\nu(M,x)-\omega_t^\pi(M,x)\}$, whose mean is zero by Lemma B.1(3). By $|Y|\le C$ and Lemma B.1(1), $\mathbb E[g(Z_O)^2\mid x,O]\le2C^2\mathbb E[\pi_t(S,x)\omega_t^\nu(M,x)^2\mid T=t,x,E]+2\mathbb E[\pi_t(S,x)\omega_t^\pi(M,x)^2\mid T=t,x,E]<\infty$. The version of (C.5) with $n_E$, $\mathbf 1\{T_i=t,X_i=x\}$, and $P(T=t,X=x\mid E)$ replaced by $n_O$, $\mathbf 1\{X_j=x\}$, and $P(X=x\mid O)$ shows that term (II) equals the second sum in (C.3) plus $O_p(N^{-1})$, and this sum is $O_p(N^{-1/2})$.

The three lines of term (III) and the two lines of term (IV) are, up to sign, the quantities that Assumption 3.6(1)–(5) requires to be $o_p(N^{-1/2})$, so terms (III) and (IV) are $o_p(N^{-1/2})$.

We bound term (V). On the event of Appendix C.1.1, $|\widehat\nu_t(s,x)|\le C$ for all $s$, and $\|\widehat\omega_t^\nu\|_{2,t,x}$ and $\|\widehat\omega_t^\pi\|_{2,t,x}$ are finite, so the premises of Lemma C.1 of Appendix C.1.2 hold for every set of sample values. Term (V) is the quantity obtained by replacing the two sample averages in (3.5) by expectations and subtracting $\bar\mu_t(x)$, that is, the left-hand side of Lemma C.1, so it equals the sum of the three terms in (C.1). For the first term, we bound the error in the ratio. Writing the difference between $\widehat\mu_t=\widehat\nu_t/\widehat\pi_t$ and $\mu_t=\nu_t/\pi_t$ over a common denominator and using $\widehat\pi_t\ge a$, $\pi_t\ge a$, and $|\nu_t|\le C$ gives, pointwise and with the arguments $(S,x)$ omitted,
\begin{equation}\tag{C.6}\label{eq:C.6}
|\widehat\mu_t-\mu_t|=\left|\frac{\widehat\nu_t-\nu_t}{\widehat\pi_t}+\frac{\nu_t(\pi_t-\widehat\pi_t)}{\widehat\pi_t\pi_t}\right|\le\frac{|\widehat\nu_t-\nu_t|}a+\frac{C|\widehat\pi_t-\pi_t|}{a^2}
\end{equation}
Minkowski's inequality gives $\|\widehat\mu_t-\mu_t\|_{2,t,x}\le a^{-1}\|\widehat\nu_t-\nu_t\|_{2,t,x}+Ca^{-2}\|\widehat\pi_t-\pi_t\|_{2,t,x}$, so $\pi_t\ge a$ and the Cauchy–Schwarz inequality give
\begin{equation}\tag{C.7}\label{eq:C.7}
\begin{aligned}
&\left|\mathbb E\!\left[\frac{\{\pi_t(S,x)-\widehat\pi_t(S,x)\}\{\widehat\mu_t(S,x)-\mu_t(S,x)\}}{\pi_t(S,x)}\,\middle|\,T=t,x,E\right]\right|\\
&\le\frac{\|\widehat\pi_t-\pi_t\|_{2,t,x}\|\widehat\mu_t-\mu_t\|_{2,t,x}}a\le C\|\widehat\pi_t-\pi_t\|_{2,t,x}\{\|\widehat\pi_t-\pi_t\|_{2,t,x}+\|\widehat\nu_t-\nu_t\|_{2,t,x}\}
\end{aligned}
\end{equation}
which is $o_p(N^{-1/2})$ by Assumption 3.5(2). For the second term, iterated expectations given $M$ and the Cauchy–Schwarz inequality give
\begin{equation}\tag{C.8}\label{eq:C.8}
\begin{aligned}
&|\mathbb E[\{\nu_t(S,x)-\widehat\nu_t(S,x)\}\{\widehat\omega_t^\nu(M,x)-\omega_t^\nu(M,x)\}\mid T=t,x,E]|\\
&=|\mathbb E[\mathbb E[\nu_t(S,x)-\widehat\nu_t(S,x)\mid M,T=t,x,E]\{\widehat\omega_t^\nu(M,x)-\omega_t^\nu(M,x)\}\mid T=t,x,E]|\\
&\le\|\mathbb E[\widehat\nu_t(S,x)-\nu_t(S,x)\mid M,T=t,x,E]\|_{2,t,x}\|\widehat\omega_t^\nu-\omega_t^\nu\|_{2,t,x}
\end{aligned}
\end{equation}
Conditioning on $S$ instead bounds it in the same way by $\|\widehat\nu_t-\nu_t\|_{2,t,x}\|\mathbb E[\widehat\omega_t^\nu(M,x)-\omega_t^\nu(M,x)\mid S,T=t,x,E]\|_{2,t,x}$, so its absolute value is at most the smaller of the two products, which is $o_p(N^{-1/2})$ by Assumption 3.5(3). The third term is $o_p(N^{-1/2})$ by the same bound with $\nu_t,\widehat\nu_t,\omega_t^\nu,\widehat\omega_t^\nu$ replaced by $\pi_t,\widehat\pi_t,\omega_t^\pi,\widehat\omega_t^\pi$ and by Assumption 3.5(4). Hence term (V) is $o_p(N^{-1/2})$.

Terms (I) and (II) in (C.4) equal the two sums in (C.3) plus $O_p(N^{-1})$, and terms (III), (IV), and (V) are $o_p(N^{-1/2})$, so (C.3) holds. $\square$

\subsubsection{Asymptotic linear representation}\label{sec:C.1.4}

Following the outline in Appendix C.1, this part proves the first convergence in (3.8) by decomposing the error, bounding each term, and establishing convergence in distribution, in this order. We work on the event of Appendix C.1.1. Substituting $\widehat P(X=x\mid O)=P(X=x\mid O)+\{\widehat P(X=x\mid O)-P(X=x\mid O)\}$ and $\widehat{\bar\mu}_1(x)-\widehat{\bar\mu}_0(x)=\bar\mu_1(x)-\bar\mu_0(x)+[\{\widehat{\bar\mu}_1(x)-\bar\mu_1(x)\}-\{\widehat{\bar\mu}_0(x)-\bar\mu_0(x)\}]$ into $\widehat\tau$ in (3.6) and $\tau=\sum_{x\in\mathcal X}P(X=x\mid O)\{\bar\mu_1(x)-\bar\mu_0(x)\}$ and expanding gives
\begin{equation}\tag{C.9}\label{eq:C.9}
\begin{aligned}
\widehat\tau-\tau
={}&\underbrace{\sum_{x\in\mathcal X}\{\widehat P(X=x\mid O)-P(X=x\mid O)\}\{\bar\mu_1(x)-\bar\mu_0(x)\}}_{\text{(i)}}\\
&+\underbrace{\sum_{x\in\mathcal X}P(X=x\mid O)[\{\widehat{\bar\mu}_1(x)-\bar\mu_1(x)\}-\{\widehat{\bar\mu}_0(x)-\bar\mu_0(x)\}]}_{\text{(ii)}}\\
&+\underbrace{\sum_{x\in\mathcal X}\{\widehat P(X=x\mid O)-P(X=x\mid O)\}[\{\widehat{\bar\mu}_1(x)-\bar\mu_1(x)\}-\{\widehat{\bar\mu}_0(x)-\bar\mu_0(x)\}]}_{\text{(iii)}}
\end{aligned}
\end{equation}
Term (i) is the estimation error in the stratum shares, term (ii) the estimation error in the stratum means, and term (iii) their product.

Since $\widehat P(X=x\mid O)=n_O^{-1}\sum_{j=1}^{n_O}\mathbf 1\{X_j=x\}$, we have $\sum_{x\in\mathcal X}\widehat P(X=x\mid O)g(x)=n_O^{-1}\sum_{j=1}^{n_O}g(X_j)$ for every function $g$ on $\mathcal X$. Applying this to $g=\bar\mu_1-\bar\mu_0$ and subtracting $\sum_{x\in\mathcal X}P(X=x\mid O)\{\bar\mu_1(x)-\bar\mu_0(x)\}=\tau$ gives
\begin{equation}\tag{C.10}\label{eq:C.10}
\text{(i)}=\frac1{n_O}\sum_{j=1}^{n_O}\{\bar\mu_1(X_j)-\bar\mu_0(X_j)-\tau\}
\end{equation}

We apply Lemma C.2 at $t=1$ and $t=0$, multiply each by $P(X=x\mid O)$, sum over $x$, and subtract the result for $t=0$ from that for $t=1$. The first sums in (C.3) then give $n_E^{-1}\sum_{i=1}^{n_E}\psi_E(Z_{E,i})$ by (3.3). In the second sums, $\mathbf 1\{X_j=x\}$ leaves only the term with $x=X_j$, and together with (C.10) they give $n_O^{-1}\sum_{j=1}^{n_O}\psi_O(Z_{O,j})$ by (3.4). The remainders in Lemma C.2 stay $o_p(N^{-1/2})$ after multiplication by $P(X=x\mid O)\le1$ and summation over finitely many terms, so
\begin{equation}\tag{C.11}\label{eq:C.11}
\text{(i)}+\text{(ii)}=\frac1{n_E}\sum_{i=1}^{n_E}\psi_E(Z_{E,i})+\frac1{n_O}\sum_{j=1}^{n_O}\psi_O(Z_{O,j})+o_p(N^{-1/2})
\end{equation}

The difference $\widehat P(X=x\mid O)-P(X=x\mid O)$ is the sample average of the independent and identically distributed terms $\mathbf 1\{X_j=x\}-P(X=x\mid O)$, which have mean zero, and its variance is at most $1/(4n_O)$. Assumption 3.1 bounds $n_O$ from below by a positive constant times $N$, so by Chebyshev's inequality this difference is $O_p(N^{-1/2})$. By Lemma C.2, $\widehat{\bar\mu}_t(x)-\bar\mu_t(x)=O_p(N^{-1/2})$, and since $\mathcal X$ is finite, term (iii) is $O_p(N^{-1})$.

Equations (C.9) and (C.11) and the bound on term (iii) give
\begin{equation}\tag{C.12}\label{eq:C.12}
\begin{aligned}
\sqrt N(\widehat\tau-\tau)
={}&\sqrt{\frac N{n_E}}\,\frac1{\sqrt{n_E}}\sum_{i=1}^{n_E}\psi_E(Z_{E,i})\\
&+\sqrt{\frac N{n_O}}\,\frac1{\sqrt{n_O}}\sum_{j=1}^{n_O}\psi_O(Z_{O,j})+o_p(1)
\end{aligned}
\end{equation}
By Lemma B.2, the $\psi_E(Z_{E,i})$ are independent and identically distributed with mean zero and finite variance, and so are the $\psi_O(Z_{O,j})$. The two samples are independent, so the central limit theorem in each sample, $N/n_E\to1/\rho$ and $N/n_O\to1/(1-\rho)$ of Assumption 3.1, and Slutsky's lemma show that the right-hand side of (C.12) converges in distribution to the normal distribution with mean zero and variance $\rho^{-1}\mathbb E[\psi_E(Z_E)^2\mid E]+(1-\rho)^{-1}\mathbb E[\psi_O(Z_O)^2\mid O]$, which is $\Sigma^2$ by (3.7). This proves the first convergence in (3.8). In particular, $\widehat\tau-\tau=O_p(N^{-1/2})$, a bound obtained without Lemma C.3.

\subsubsection{Consistency of the variance estimator}\label{sec:C.1.5}

\renewcommand{\thefixedlemma}{C.3}
\begin{fixedlemma}[Consistency of the variance estimator]\label{lemma:C.3}

Suppose that Assumptions 2.1–2.8, the integrability conditions in Appendix A, Assumptions 3.1–3.6, Assumption B.1, and Assumption C.1(1) hold. Then
\begin{equation}\tag{C.13}\label{eq:C.13}
\begin{gathered}
\frac1{n_E}\sum_{i=1}^{n_E}\{\widehat\psi_E(Z_{E,i})-\psi_E(Z_{E,i})\}^2\to_p0,\\
\frac1{n_O}\sum_{j=1}^{n_O}\{\widehat\psi_O(Z_{O,j})-\psi_O(Z_{O,j})\}^2\to_p0
\end{gathered}
\end{equation}
and $\widehat\Sigma^2\to_p\Sigma^2$.
\end{fixedlemma}

\textbf{Proof of Lemma C.3.} The goals are (C.13) and $\widehat\Sigma^2\to_p\Sigma^2$. We first show that (C.13) implies the consistency of $\widehat\Sigma^2$ and then show (C.13) separately for the experimental and the observational sample. The bounds for the two samples use Assumption 3.6(6)–(11) to bound sample averages of squared estimation errors.

The triangle inequality for the square roots of mean squares gives
\begin{equation}\tag{C.14}\label{eq:C.14}
\left|\Bigl\{\frac1{n_E}\sum_{i=1}^{n_E}\widehat\psi_E(Z_{E,i})^2\Bigr\}^{1/2}-\Bigl\{\frac1{n_E}\sum_{i=1}^{n_E}\psi_E(Z_{E,i})^2\Bigr\}^{1/2}\right|\le\Bigl\{\frac1{n_E}\sum_{i=1}^{n_E}\{\widehat\psi_E(Z_{E,i})-\psi_E(Z_{E,i})\}^2\Bigr\}^{1/2}
\end{equation}
By Lemma B.2 and the law of large numbers, $n_E^{-1}\sum_{i=1}^{n_E}\psi_E(Z_{E,i})^2\to_p\mathbb E[\psi_E(Z_E)^2\mid E]$, so (C.14), the first equation of (C.13), and the continuous mapping theorem give $n_E^{-1}\sum_{i=1}^{n_E}\widehat\psi_E(Z_{E,i})^2\to_p\mathbb E[\psi_E(Z_E)^2\mid E]$. The same argument with the second equation of (C.13) gives $n_O^{-1}\sum_{j=1}^{n_O}\widehat\psi_O(Z_{O,j})^2\to_p\mathbb E[\psi_O(Z_O)^2\mid O]$, and $N/n_E\to1/\rho$ and $N/n_O\to1/(1-\rho)$ of Assumption 3.1 and (3.7) then give $\widehat\Sigma^2\to_p\Sigma^2$.

We show the first equation of (C.13) for the experimental sample, working on the event of Appendix C.1.1. For an observation with $(T_i,X_i)=(t,x)$, only one term remains in (3.3) and in its plug-in version. Subtracting and adding $\widehat P(X=x\mid O)/\widehat P(T=t,X=x\mid E)$ times the expression in braces of (3.3) shows that, if $t=1$, $\widehat\psi_E(Z_{E,i})-\psi_E(Z_{E,i})$ equals
\begin{equation}\tag{C.15}\label{eq:C.15}
\begin{aligned}
&\underbrace{\frac{\widehat P(X=x\mid O)}{\widehat P(T=t,X=x\mid E)}[\{\widehat\mu_t-\mu_t\}-\{\widehat\nu_t\widehat\omega_t^\nu-\nu_t\omega_t^\nu\}+\{\widehat\pi_t\widehat\omega_t^\pi-\pi_t\omega_t^\pi\}]}_{\text{(I)}}\\
&\qquad-\underbrace{\frac{\widehat P(X=x\mid O)}{\widehat P(T=t,X=x\mid E)}\{\widehat{\bar\mu}_t-\bar\mu_t\}}_{\text{(II)}}\\
&\qquad+\underbrace{\Bigl\{\frac{\widehat P(X=x\mid O)}{\widehat P(T=t,X=x\mid E)}-\frac{P(X=x\mid O)}{P(T=t,X=x\mid E)}\Bigr\}\{\mu_t-\bar\mu_t-\nu_t\omega_t^\nu+\pi_t\omega_t^\pi\}}_{\text{(III)}}
\end{aligned}
\end{equation}
and if $t=0$, it equals the same expression with the opposite sign. Here the arguments of the functions are $(S_i,x)$ or $(M_i,x)$, which we omit together with the argument $x$ of $\widehat{\bar\mu}_t,\bar\mu_t$. The square of $\text{(I)}-\text{(II)}+\text{(III)}$ is at most $3\{\text{(I)}^2+\text{(II)}^2+\text{(III)}^2\}$. On the event of Appendix C.1.1, $\widehat P(T=t,X=x\mid E)\ge c/2$ and $\widehat P(X=x\mid O)\le1$, so the square of the ratio is at most $4c^{-2}$.

For term (I), we evaluate the function in square brackets at $(S_i,M_i)$, square it, multiply by $\mathbf 1\{T_i=t,X_i=x\}$, sum over $i$, and divide by $n_E$. By $\widehat P(T=t,X=x\mid E)\le1$, the result is at most $\widehat{\mathbb E}[\,\cdot\mid T=t,x,E]$ of the square of that function. The expression in square brackets is the sum of three differences, so its square is at most three times the sum of their squares, and it suffices to show that $\widehat{\mathbb E}[\,\cdot\mid T=t,x,E]$ of each squared difference tends to zero in probability. First, by (C.6) and because the square of a sum of two real numbers is at most twice the sum of their squares, $\{\widehat\mu_t-\mu_t\}^2\le2a^{-2}\{\widehat\nu_t-\nu_t\}^2+2C^2a^{-4}\{\widehat\pi_t-\pi_t\}^2$ at each observation, and by Assumption 3.6(6)(7), $\widehat{\mathbb E}[\,\cdot\mid T=t,x,E]$ of the right-hand side tends to zero in probability. Second, $\widehat\nu_t\widehat\omega_t^\nu-\nu_t\omega_t^\nu=\widehat\nu_t(\widehat\omega_t^\nu-\omega_t^\nu)+(\widehat\nu_t-\nu_t)\omega_t^\nu$ and $|\widehat\nu_t|\le C$ give $\{\widehat\nu_t\widehat\omega_t^\nu-\nu_t\omega_t^\nu\}^2\le2C^2(\widehat\omega_t^\nu-\omega_t^\nu)^2+2(\widehat\nu_t-\nu_t)^2(\omega_t^\nu)^2$ at each observation. For a positive number $r$, $|\widehat\nu_t-\nu_t|\le2C$ gives, at each observation,
\begin{equation}\tag{C.16}\label{eq:C.16}
(\widehat\nu_t-\nu_t)^2(\omega_t^\nu)^2\le r^2(\widehat\nu_t-\nu_t)^2+4C^2(\omega_t^\nu)^2\mathbf 1\{|\omega_t^\nu|>r\}
\end{equation}
which follows by bounding $(\omega_t^\nu)^2$ on the left-hand side by $r^2$ when $|\omega_t^\nu|\le r$ and $(\widehat\nu_t-\nu_t)^2$ by $4C^2$ when $|\omega_t^\nu|>r$. For any $\varepsilon>0$, Assumption 3.4 and the dominated convergence theorem allow us to choose $r$ such that $8C^2\mathbb E[\omega_t^\nu(M,x)^2\mathbf 1\{|\omega_t^\nu(M,x)|>r\}\mid T=t,x,E]<\varepsilon$. For this $r$, by the law of large numbers, $8C^2\widehat{\mathbb E}[\omega_t^\nu(M,x)^2\mathbf 1\{|\omega_t^\nu(M,x)|>r\}\mid T=t,x,E]<\varepsilon$ with probability approaching one, and by Assumption 3.6(7)(8), $2C^2\widehat{\mathbb E}[\{\widehat\omega_t^\nu(M,x)-\omega_t^\nu(M,x)\}^2\mid T=t,x,E]+2r^2\widehat{\mathbb E}[\{\widehat\nu_t(S,x)-\nu_t(S,x)\}^2\mid T=t,x,E]<\varepsilon$ with probability approaching one. Hence $\widehat{\mathbb E}[\{\widehat\nu_t(S,x)\widehat\omega_t^\nu(M,x)-\nu_t(S,x)\omega_t^\nu(M,x)\}^2\mid T=t,x,E]<2\varepsilon$ with probability approaching one. Third, we write $\widehat\pi_t\widehat\omega_t^\pi-\pi_t\omega_t^\pi=\widehat\pi_t(\widehat\omega_t^\pi-\omega_t^\pi)+(\widehat\pi_t-\pi_t)\omega_t^\pi$ and apply the same argument with $\widehat\pi_t\le1$, $|\widehat\pi_t-\pi_t|\le1$, and Assumption 3.6(6)(9).

Multiplying the square of term (II) by $\mathbf 1\{T_i=t,X_i=x\}$, summing over $i$, and dividing by $n_E$ gives a quantity at most $4c^{-2}\{\widehat{\bar\mu}_t(x)-\bar\mu_t(x)\}^2$, which tends to zero in probability by Lemma C.2. For term (III), the law of large numbers and Assumption 3.2(2) show that the difference between the ratios tends to zero in probability, and $n_E^{-1}\sum_{i=1}^{n_E}\mathbf 1\{T_i=t,X_i=x\}\{\mu_t-\bar\mu_t-\nu_t\omega_t^\nu+\pi_t\omega_t^\pi\}^2$ converges in probability to a finite limit, so the average of the square of term (III) tends to zero in probability. Summing over the finitely many $(t,x)$ gives the first equation of (C.13).

We show the second equation of (C.13) for the observational sample. For observation $j$, (3.4) and its plug-in version give
\begin{equation}\tag{C.17}\label{eq:C.17}
\begin{aligned}
\widehat\psi_O(Z_{O,j})-\psi_O(Z_{O,j})
={}&\underbrace{\{\widehat{\bar\mu}_1(X_j)-\bar\mu_1(X_j)\}-\{\widehat{\bar\mu}_0(X_j)-\bar\mu_0(X_j)\}-(\widehat\tau-\tau)}_{\text{(IV)}}\\
&+\underbrace{\mathbf 1\{T_j=1\}[Y_j\{\widehat\omega_1^\nu-\omega_1^\nu\}(M_j,X_j)-\{\widehat\omega_1^\pi-\omega_1^\pi\}(M_j,X_j)]}_{\text{(V)}}\\
&-\underbrace{\mathbf 1\{T_j=0\}[Y_j\{\widehat\omega_0^\nu-\omega_0^\nu\}(M_j,X_j)-\{\widehat\omega_0^\pi-\omega_0^\pi\}(M_j,X_j)]}_{\text{(VI)}}
\end{aligned}
\end{equation}
Here $\{\widehat\omega_t^\nu-\omega_t^\nu\}(M_j,X_j)$ denotes $\widehat\omega_t^\nu(M_j,X_j)-\omega_t^\nu(M_j,X_j)$. Since $\mathcal X$ is finite, Lemma C.2 and the bound $\widehat\tau-\tau=O_p(N^{-1/2})$, established in Appendix C.1.4 without Lemma C.3, show that term (IV) tends to zero in probability uniformly in $j$. We multiply the squares of terms (V) and (VI) by $\mathbf 1\{X_j=x\}$, sum over $j$, and divide by $n_O$. By $|Y_j|\le C$, which holds for all $j$ with probability one by Assumption B.1, and because the square of a sum of two real numbers is at most twice the sum of their squares, the result is at most $2C^2\widehat P(X=x\mid O)\widehat{\mathbb E}[\mathbf 1\{T=t\}\{\widehat\omega_t^\nu(M,x)-\omega_t^\nu(M,x)\}^2\mid x,O]+2\widehat P(X=x\mid O)\widehat{\mathbb E}[\mathbf 1\{T=t\}\{\widehat\omega_t^\pi(M,x)-\omega_t^\pi(M,x)\}^2\mid x,O]$, with $t=1$ for term (V) and $t=0$ for term (VI), and it tends to zero in probability by Assumption 3.6(10)(11). The left-hand side of (C.17) equals $\text{(IV)}+\text{(V)}-\text{(VI)}$, and the square of a sum of three real numbers is at most three times the sum of their squares, so its square is at most $3\{\text{(IV)}^2+\text{(V)}^2+\text{(VI)}^2\}$. Multiplying this by $\mathbf 1\{X_j=x\}$, summing over $j$, dividing by $n_O$, and summing over the finitely many $x$ gives the second equation of (C.13). $\square$

\subsubsection{Studentization and efficiency}\label{sec:C.1.6}

The second convergence in (3.8) is Lemma C.3. Suppose that $\Sigma^2>0$. Then $\widehat\Sigma\to_p\Sigma>0$ by the continuous mapping theorem, so $P(\widehat\Sigma>0)\to1$, and on $\{\widehat\Sigma>0\}$ the ratio $\sqrt N(\widehat\tau-\tau)/\widehat\Sigma$ equals $\{\sqrt N(\widehat\tau-\tau)/\Sigma\}(\Sigma/\widehat\Sigma)$, which converges in distribution to $\mathcal N(0,1)$ by Slutsky's lemma. On the same event, the interval contains $\tau$ if and only if $|\sqrt N(\widehat\tau-\tau)/\widehat\Sigma|\le z_{1-\alpha/2}$, so, since the distribution function of the standard normal distribution is continuous, the coverage probability converges to $1-\alpha$. $\square$

The rest of this part proves the efficiency claims of Theorems 3.2 and 3.3. Under the true law $P$, the two estimators have the same linear expansion (C.12), obtained in Appendix C.1.4 for the estimator of Theorem 3.2 and in Appendix C.2.4 for the estimator of Theorem 3.3. The argument below uses only (C.12), Lemmas B.2 and B.3, and (B.3), which Lemma B.4 gives along every smooth submodel, so it applies to both estimators.

Take any smooth submodel with score $(a_E,a_O)$. We find the joint limit of the two sums on the right-hand side of (C.12) and of the quantity in (B.12). By Lemma B.2 and the definition of the score, the $(\psi_E(Z_{E,i}),a_E(Z_{E,i}))$ are independent and identically distributed two-dimensional vectors with mean zero and finite second moments, and so are the $(\psi_O(Z_{O,j}),a_O(Z_{O,j}))$. By the multivariate central limit theorem, $n_E^{-1/2}\sum_i(\psi_E(Z_{E,i}),a_E(Z_{E,i}))$ converges in distribution to a normal distribution whose covariance matrix has entries $\mathbb E[\psi_E^2\mid E]$, $\mathbb E[\psi_Ea_E\mid E]$, and $\mathbb E[a_E^2\mid E]$, and the observational sum converges in distribution in the same way. The two samples are independent, and Assumption 3.1 gives $N/n_E\to1/\rho$ and $N/n_O\to1/(1-\rho)$, so by Slutsky's lemma the sum of the two terms on the right-hand side of (C.12) and the quantity in (B.12) converge jointly in distribution to a two-dimensional normal distribution. The limiting variance is $\rho^{-1}\mathbb E[\psi_E^2\mid E]+(1-\rho)^{-1}\mathbb E[\psi_O^2\mid O]=\Sigma^2$ for the sum of the terms on the right-hand side of (C.12) and $\rho\mathbb E[a_E^2\mid E]+(1-\rho)\mathbb E[a_O^2\mid O]=\|(a_E,a_O)\|_\rho^2$ for the quantity in (B.12). The coefficients of the experimental sums converge to $\sqrt{N/n_E}\to1/\sqrt\rho$ and $\sqrt{n_E/N}\to\sqrt\rho$, and those of the observational sums to $1/\sqrt{1-\rho}$ and $\sqrt{1-\rho}$, so the limiting covariance is
\begin{equation}\tag{C.18}\label{eq:C.18}
\frac1{\sqrt\rho}\sqrt\rho\,\mathbb E[\psi_Ea_E\mid E]+\frac1{\sqrt{1-\rho}}\sqrt{1-\rho}\,\mathbb E[\psi_Oa_O\mid O]=\mathbb E[\psi_Ea_E\mid E]+\mathbb E[\psi_Oa_O\mid O]
\end{equation}
By Slutsky's lemma, the $o_p(1)$ terms in (C.12) and in (B.4) do not change the limit. Hence $\sqrt N(\widehat\tau-\tau)$ and the log likelihood ratio on the left-hand side of (B.4) converge jointly in distribution to a two-dimensional normal distribution with mean $(0,-\|(a_E,a_O)\|_\rho^2/2)$, the variances above, and the covariance (C.18).

Le Cam's third lemma \citep{ref32} then shows that, under $P_{1/\sqrt N}$, $\sqrt N(\widehat\tau-\tau)$ converges in distribution to a normal distribution with mean equal to the covariance (C.18) and variance $\Sigma^2$. On the other hand, (B.3) of Lemma B.4 applied along $\varepsilon=1/\sqrt N$ gives
\begin{equation}\tag{C.19}\label{eq:C.19}
\sqrt N\{\tau(P_{1/\sqrt N})-\tau(P)\}\to\mathbb E[\psi_Ea_E\mid E]+\mathbb E[\psi_Oa_O\mid O]
\end{equation}
We decompose the estimation error under $P_{1/\sqrt N}$ as
\[
\sqrt N\{\widehat\tau-\tau(P_{1/\sqrt N})\}=\sqrt N\{\widehat\tau-\tau(P)\}-\sqrt N\{\tau(P_{1/\sqrt N})-\tau(P)\}
\]
The first term on the right-hand side converges in distribution to the normal distribution above, and the second converges to the constant in (C.19), so by Slutsky's lemma the left-hand side converges in distribution under $P_{1/\sqrt N}$ to a normal distribution with mean zero and variance $\Sigma^2$. This limit does not depend on the submodel, so $\widehat\tau$ is regular in the model $\mathcal P$ and, in particular, regular with respect to the subspace in Assumption B.2. This normal limit with mean zero and variance $\Sigma^2$ is the convolution of claim (4) of Appendix B.2.2 in which the other probability distribution is the point mass at zero, and its variance equals the efficiency bound $\Sigma^2$ of Theorem 3.1. Hence $\widehat\tau$ attains the efficiency bound. $\square$

\subsection{Lemmas and proof of Theorem 3.3}\label{sec:C.2}

This subsection proves Theorem 3.3 of Section 3.4. We first recall the stratum means for each fold.

The estimator of Theorem 3.3 uses, for each fold $k$, the stratum mean given by (3.9) of the main text,
\[
\begin{aligned}
\widehat{\bar\mu}_{t,k}(x)={}&\widehat{\mathbb E}_k[\widehat\mu_t^{(-k)}(S,x)-\widehat\nu_t^{(-k)}(S,x)\widehat\omega_t^{\nu,(-k)}(M,x)+\widehat\pi_t^{(-k)}(S,x)\widehat\omega_t^{\pi,(-k)}(M,x)\mid T=t,x,E]\\
&+\widehat{\mathbb E}_k[\mathbf 1\{T=t\}\{Y\widehat\omega_t^{\nu,(-k)}(M,x)-\widehat\omega_t^{\pi,(-k)}(M,x)\}\mid x,O]
\end{aligned}
\]

The proof of the claims other than efficiency follows the proof of Theorem 3.2 in Appendix C.1. The proofs in Appendix C.1.4 and in the first half of Appendix C.1.6 use only Lemma B.2, (C.3) of Lemma C.2, and (C.13) of Lemma C.3 on the event of Appendix C.1.1. If (C.3) and (C.13) are shown for the cross-fitted $\widehat{\bar\mu}_t(x)$ and $\widehat\Sigma^2$, the arguments of Appendix C.1.4 and of the first half of Appendix C.1.6 therefore give Theorem 3.3 without change. We show (C.3) by carrying out the proof of Lemma C.2 in Appendix C.1.3 fold by fold, and (C.13) by replacing the full-sample quantities by fold quantities in the proof of Lemma C.3 in Appendix C.1.5. Appendix C.2.1 gives the events with probability approaching one used in this proof. To bound the terms due to plugging in the estimates and the variance estimator, we use Lemma C.4 in place of Assumption 3.6, and Lemma C.5 then turns averages over folds into sums over the full samples.

\subsubsection{Events with probability approaching one}\label{sec:C.2.1}

We define the events with probability approaching one used in the proof of Theorem 3.3. To (E1) of Appendix C.1.1, that is, $\widehat P(T=t,X=x\mid E)\ge c/2$ and $\widehat P(X=x\mid O)\ge c/2$ for all $t,x$, we add the following three conditions for all $t,x,k$. (E4) $\widehat P_k(T=t,X=x\mid E)\ge c/2$ and $\widehat P_k(X=x\mid O)\ge c/2$. (E5) The event of Assumption C.1(2) occurs, that is, an event with probability approaching one on which $|\widehat\nu_t^{(-k)}(s,x)|\le C$ for all $s$, where $C$ is the constant in the bound of Assumption C.1(2). (E6) $\|\widehat\omega_t^{\nu,(-k)}\|_{2,t,x}$ and $\|\widehat\omega_t^{\pi,(-k)}\|_{2,t,x}$ are finite. We call the event on which (E1), (E4), (E5), and (E6) hold simultaneously the event of Appendix C.2.1. The probability of (E1) approaches one by Appendix C.1.1. For (E4), the fold sizes differ by at most one, so $|\mathcal I_{E,k}|\ge n_E/K-1$, and Assumption 3.1 bounds $|\mathcal I_{E,k}|$ from below by a positive constant times $N$. Since $\widehat P_k(T=t,X=x\mid E)$ is a sample average with mean $P(T=t,X=x\mid E)\ge c$ and variance at most $1/(4|\mathcal I_{E,k}|)$, by Chebyshev's inequality the probability of $\widehat P_k(T=t,X=x\mid E)<c/2$ tends to zero. The same holds for $\widehat P_k(X=x\mid O)$. The probability of (E5) approaches one by Assumption C.1(2). For (E6), Minkowski's inequality gives $\|\widehat\omega_t^{\nu,(-k)}\|_{2,t,x}\le\|\omega_t^\nu\|_{2,t,x}+\|\widehat\omega_t^{\nu,(-k)}-\omega_t^\nu\|_{2,t,x}$, where the first term on the right-hand side is finite by Assumption 3.4 and the second tends to zero in probability by Assumption 3.7, so the probability that the norm is finite approaches one. The same holds for $\widehat\omega_t^{\pi,(-k)}$. Since the probabilities of finitely many events all approach one, the probability that they hold simultaneously also approaches one. On this event, all the groups over which (3.9), (3.6), and the variance estimator average are nonempty. Since the probability of the complement tends to zero, convergence in probability and in distribution established on this event continue to hold without the restriction to the event. Below, we also use this event in place of the event of Appendix C.1.1 in the proofs of Appendix C.1.4 and of the first half of Appendix C.1.6, which use only property (E1) of the event of Appendix C.1.1.

\subsubsection{Conditional bounds for averages over a fold}\label{sec:C.2.2}

\renewcommand{\thefixedlemma}{C.4}
\begin{fixedlemma}[Averages over a fold]\label{lemma:C.4}

Suppose that Assumptions 2.1–2.8, the integrability conditions in Appendix A, Assumptions 3.1–3.4, Assumption 3.7, Assumption B.1, and Assumption C.1(2) hold. Then the following hold for each $t,x,k$, where $(\widehat{\mathbb E}_k-\mathbb E)[\,\cdot\mid T=t,x,E]=\widehat{\mathbb E}_k[\,\cdot\mid T=t,x,E]-\mathbb E[\,\cdot\mid T=t,x,E]$ and $(\widehat{\mathbb E}_k-\mathbb E)[\,\cdot\mid x,O]$ is defined in the same way for the observational sample.

\begin{enumerate}
\item $(\widehat{\mathbb E}_k-\mathbb E)[\widehat\mu_t^{(-k)}(S,x)-\mu_t(S,x)\mid T=t,x,E]$ is $o_p(N^{-1/2})$.
\item $(\widehat{\mathbb E}_k-\mathbb E)[\widehat\nu_t^{(-k)}(S,x)\widehat\omega_t^{\nu,(-k)}(M,x)-\nu_t(S,x)\omega_t^\nu(M,x)\mid T=t,x,E]$ is $o_p(N^{-1/2})$.
\item $(\widehat{\mathbb E}_k-\mathbb E)[\widehat\pi_t^{(-k)}(S,x)\widehat\omega_t^{\pi,(-k)}(M,x)-\pi_t(S,x)\omega_t^\pi(M,x)\mid T=t,x,E]$ is $o_p(N^{-1/2})$.
\item $(\widehat{\mathbb E}_k-\mathbb E)[\mathbf 1\{T=t\}Y\{\widehat\omega_t^{\nu,(-k)}(M,x)-\omega_t^\nu(M,x)\}\mid x,O]$ is $o_p(N^{-1/2})$.
\item $(\widehat{\mathbb E}_k-\mathbb E)[\mathbf 1\{T=t\}\{\widehat\omega_t^{\pi,(-k)}(M,x)-\omega_t^\pi(M,x)\}\mid x,O]$ is $o_p(N^{-1/2})$.
\item $\widehat{\mathbb E}_k[\{\widehat\pi_t^{(-k)}(S,x)-\pi_t(S,x)\}^2\mid T=t,x,E]$ tends to zero in probability.
\item $\widehat{\mathbb E}_k[\{\widehat\nu_t^{(-k)}(S,x)-\nu_t(S,x)\}^2\mid T=t,x,E]$ tends to zero in probability.
\item $\widehat{\mathbb E}_k[\{\widehat\omega_t^{\nu,(-k)}(M,x)-\omega_t^\nu(M,x)\}^2\mid T=t,x,E]$ tends to zero in probability.
\item $\widehat{\mathbb E}_k[\{\widehat\omega_t^{\pi,(-k)}(M,x)-\omega_t^\pi(M,x)\}^2\mid T=t,x,E]$ tends to zero in probability.
\item $\widehat{\mathbb E}_k[\mathbf 1\{T=t\}\{\widehat\omega_t^{\nu,(-k)}(M,x)-\omega_t^\nu(M,x)\}^2\mid x,O]$ tends to zero in probability.
\item $\widehat{\mathbb E}_k[\mathbf 1\{T=t\}\{\widehat\omega_t^{\pi,(-k)}(M,x)-\omega_t^\pi(M,x)\}^2\mid x,O]$ tends to zero in probability.
\end{enumerate}
\end{fixedlemma}

\textbf{Proof of Lemma C.4.} Fix $t,x,k$, and write $\mathcal G_k$ for the σ-field generated by the observations outside fold $k$. The estimates that leave out fold $k$ are measurable with respect to $\mathcal G_k$, and by Assumption 3.1 the observations in fold $k$ are independent of $\mathcal G_k$ and have the original distribution, so the conditional expectation given $\mathcal G_k$ of a function of an observation in fold $k$ and of these estimates is the expectation with the estimates held fixed defined in Section 3.3. The event of Appendix C.2.1 involves fold $k$ and need not be measurable with respect to $\mathcal G_k$, so we take conditional expectations given $\mathcal G_k$ only of quantities without the restriction to that event, and we use the share bounds in (E4) and the bound in (E5) only outside conditional expectations.

For claims (1)–(5), let $f_k$ be a function of an experimental observation and of the estimates that leave out fold $k$ whose mean square $\mathbb E[f_k(Z_E)^2\mid T=t,x,E]$, computed with the estimates held fixed, tends to zero in probability. We show that $(\widehat{\mathbb E}_k-\mathbb E)[f_k(Z_E)\mid T=t,x,E]=o_p(N^{-1/2})$. The definition of the sample average gives
\begin{equation}\tag{C.20}\label{eq:C.20}
\begin{aligned}
&(\widehat{\mathbb E}_k-\mathbb E)[f_k(Z_E)\mid T=t,x,E]\\
&=\frac1{\widehat P_k(T=t,X=x\mid E)}\cdot\frac1{|\mathcal I_{E,k}|}\sum_{i\in\mathcal I_{E,k}}\mathbf 1\{T_i=t,X_i=x\}\{f_k(Z_{E,i})-\mathbb E[f_k(Z_E)\mid T=t,x,E]\}
\end{aligned}
\end{equation}
The mean square is determined by the estimates alone, so the event that it is finite is measurable with respect to $\mathcal G_k$, and its probability approaches one. On this event, conditional on $\mathcal G_k$, the summands in (C.20) are independent with mean zero and second moment $P(T=t,X=x\mid E)\operatorname{Var}(f_k(Z_E)\mid T=t,x,E)\le\mathbb E[f_k(Z_E)^2\mid T=t,x,E]$, so the conditional expectation given $\mathcal G_k$ of $N$ times the square of their average over $\mathcal I_{E,k}$ is at most $(N/|\mathcal I_{E,k}|)\mathbb E[f_k(Z_E)^2\mid T=t,x,E]$. This bound tends to zero in probability because $N/|\mathcal I_{E,k}|\to K/\rho$ by Assumption 3.1. A nonnegative random variable $U$ whose conditional expectation given $\mathcal G_k$ tends to zero in probability itself tends to zero in probability, because, for $\varepsilon>0$ and $\delta\in(0,1)$, the conditional Markov inequality gives $P(U>\varepsilon)\le\mathbb E[\min\{1,\mathbb E[U\mid\mathcal G_k]/\varepsilon\}]\le\delta+P(\mathbb E[U\mid\mathcal G_k]>\delta\varepsilon)$. Applied to $N$ times the squared average multiplied by the indicator of the event, this shows that $\sqrt N$ times the average tends to zero in probability. Using $\widehat P_k(T=t,X=x\mid E)\ge c/2$ in (E4) outside the conditional expectation, (C.20) gives $(\widehat{\mathbb E}_k-\mathbb E)[f_k(Z_E)\mid T=t,x,E]=o_p(N^{-1/2})$. For a function $g_k$ of an observation from the observational sample such that $\mathbb E[g_k(Z_O)^2\mid x,O]$ tends to zero in probability, the same argument gives $(\widehat{\mathbb E}_k-\mathbb E)[g_k(Z_O)\mid x,O]=o_p(N^{-1/2})$.

For claims (1)–(3), the mean squares of the three functions tend to zero in probability by the bounds of Appendices C.1.3 and C.1.5 with the estimates that leave out fold $k$. In claim (1), (C.6) and Minkowski's inequality give
\[
\|\widehat\mu_t^{(-k)}-\mu_t\|_{2,t,x}\le a^{-1}\|\widehat\nu_t^{(-k)}-\nu_t\|_{2,t,x}+Ca^{-2}\|\widehat\pi_t^{(-k)}-\pi_t\|_{2,t,x}
\]
and the right-hand side tends to zero in probability by Assumption 3.7. In claim (2), we write
\[
\widehat\nu_t^{(-k)}\widehat\omega_t^{\nu,(-k)}-\nu_t\omega_t^\nu=\widehat\nu_t^{(-k)}(\widehat\omega_t^{\nu,(-k)}-\omega_t^\nu)+(\widehat\nu_t^{(-k)}-\nu_t)\omega_t^\nu
\]
On the event in (E5), the norm of the first part is at most $C\|\widehat\omega_t^{\nu,(-k)}-\omega_t^\nu\|_{2,t,x}$, and taking expectations in (C.16) with $\widehat\nu_t^{(-k)}$ in place of $\widehat\nu_t$ gives, for every $r>0$,
\[
\|(\widehat\nu_t^{(-k)}-\nu_t)\omega_t^\nu\|_{2,t,x}^2\le r^2\|\widehat\nu_t^{(-k)}-\nu_t\|_{2,t,x}^2+4C^2\mathbb E[\omega_t^\nu(M,x)^2\mathbf 1\{|\omega_t^\nu(M,x)|>r\}\mid T=t,x,E]
\]
For any $\varepsilon>0$, Assumption 3.4 and the dominated convergence theorem allow us to choose $r$ so that the second term is less than $\varepsilon/2$, and for this $r$, by Assumption 3.7, the first term is less than $\varepsilon/2$ with probability approaching one. Since the probability of (E5) approaches one, the mean square of the function in claim (2) tends to zero in probability. Claim (3) follows in the same way with $\widehat\pi_t^{(-k)}\le1$ and $|\widehat\pi_t^{(-k)}-\pi_t|\le1$. The conclusion above applied to these three functions gives claims (1)–(3).

For claims (4) and (5), $|Y|\le C$, Lemma B.1(1), and $\pi_t\le1$ bound the mean squares of the two observational functions by $C^2\|\widehat\omega_t^{\nu,(-k)}-\omega_t^\nu\|_{2,t,x}^2$ and $\|\widehat\omega_t^{\pi,(-k)}-\omega_t^\pi\|_{2,t,x}^2$, which tend to zero in probability by Assumption 3.7, so the observational version of the conclusion above gives claims (4) and (5).

We show claims (6)–(9). For claim (6), let $f_k(Z_E)=\{\widehat\pi_t^{(-k)}(S,x)-\pi_t(S,x)\}^2$. The conditional expectation given $\mathcal G_k$ of the nonnegative random variable $|\mathcal I_{E,k}|^{-1}\sum_{i\in\mathcal I_{E,k}}\mathbf 1\{T_i=t,X_i=x\}f_k(Z_{E,i})$ is $P(T=t,X=x\mid E)\|\widehat\pi_t^{(-k)}-\pi_t\|_{2,t,x}^2$, which tends to zero in probability by Assumption 3.7, so by the conditional Markov argument above this random variable tends to zero in probability. Since $\widehat{\mathbb E}_k[f_k(Z_E)\mid T=t,x,E]$ is this random variable divided by $\widehat P_k(T=t,X=x\mid E)$, it tends to zero in probability by (E4). Claims (7)–(9) follow from the same argument with $\widehat\pi_t^{(-k)}-\pi_t$ replaced by $\widehat\nu_t^{(-k)}-\nu_t$, $\widehat\omega_t^{\nu,(-k)}-\omega_t^\nu$, and $\widehat\omega_t^{\pi,(-k)}-\omega_t^\pi$. For claims (10) and (11), we use that the conditional expectation given $\mathcal G_k$ of $|\mathcal I_{O,k}|^{-1}\sum_{j\in\mathcal I_{O,k}}\mathbf 1\{X_j=x\}\mathbf 1\{T_j=t\}\{\widehat\omega_t^{\nu,(-k)}(M_j,x)-\omega_t^\nu(M_j,x)\}^2$ is at most $P(X=x\mid O)\mathbb E[\pi_t(S,x)\{\widehat\omega_t^{\nu,(-k)}(M,x)-\omega_t^\nu(M,x)\}^2\mid T=t,x,E]\le\|\widehat\omega_t^{\nu,(-k)}-\omega_t^\nu\|_{2,t,x}^2$ by Lemma B.1(1) and $\pi_t\le1$, and apply the same argument. The same holds for $\widehat\omega_t^{\pi,(-k)}$. $\square$

\subsubsection{Averages over the folds}\label{sec:C.2.3}

\renewcommand{\thefixedlemma}{C.5}
\begin{fixedlemma}[Sums of averages over the folds]\label{lemma:C.5}

Suppose that Assumptions 3.1 and 3.2 hold, and fix $t,x$. If a function $f$ of an experimental observation satisfies $\mathbb E[f(Z_E)\mid T=t,x,E]=0$ and $\mathbb E[f(Z_E)^2\mid T=t,x,E]<\infty$, then
\begin{equation}\tag{C.21}\label{eq:C.21}
\frac1K\sum_{k=1}^K\widehat{\mathbb E}_k[f(Z_E)\mid T=t,x,E]=\frac1{n_E}\sum_{i=1}^{n_E}\frac{\mathbf 1\{T_i=t,X_i=x\}f(Z_{E,i})}{P(T=t,X=x\mid E)}+O_p(N^{-1})
\end{equation}
and the sum on the right-hand side is $O_p(N^{-1/2})$. If a function $g$ of an observation from the observational sample satisfies $\mathbb E[g(Z_O)\mid x,O]=0$ and $\mathbb E[g(Z_O)^2\mid x,O]<\infty$, the version of (C.21) with the sample averages over $\mathcal I_{E,k}$, $n_E$, $\mathbf 1\{T_i=t,X_i=x\}$, and $P(T=t,X=x\mid E)$ replaced by those over $\mathcal I_{O,k}$, $n_O$, $\mathbf 1\{X_j=x\}$, and $P(X=x\mid O)$ also holds.
\end{fixedlemma}

\textbf{Proof of Lemma C.5.} We work on the event on which (E4) holds, whose probability approaches one by Appendix C.2.1. The argument for (C.5), applied to the observations in fold $k$, whose number is at least a positive constant times $N$ by Assumption 3.1, shows that $\widehat{\mathbb E}_k[f(Z_E)\mid T=t,x,E]$ equals $|\mathcal I_{E,k}|^{-1}\sum_{i\in\mathcal I_{E,k}}\mathbf 1\{T_i=t,X_i=x\}f(Z_{E,i})/P(T=t,X=x\mid E)$ plus $O_p(N^{-1})$. The fold sizes differ by at most one, so $|n_E-K|\mathcal I_{E,k}||<K$ and $|1/(K|\mathcal I_{E,k}|)-1/n_E|\le1/(|\mathcal I_{E,k}|n_E)\le CN^{-2}$ for large $N$. Since $\sum_{i\in\mathcal I_{E,k}}\mathbf 1\{T_i=t,X_i=x\}f(Z_{E,i})=O_p(N^{1/2})$ by Chebyshev's inequality, averaging over $k$ replaces these normalized sums over the folds by the sum on the right-hand side of (C.21) up to $O_p(N^{-3/2})$, which gives (C.21). The sum on the right-hand side of (C.21) is $O_p(N^{-1/2})$ by Chebyshev's inequality, and the observational version follows in the same way. $\square$

\subsubsection{Proof of Theorem 3.3}\label{sec:C.2.4}

\textbf{Proof of Theorem 3.3.} We first show (C.3). Fix $t,x$ and work on the event of Appendix C.2.1. Replacing, in (C.4) of Appendix C.1.3, the full-sample $\widehat{\mathbb E}$ and the full-sample estimates by $\widehat{\mathbb E}_k$ over fold $k$ and the estimates that leave out fold $k$ splits $\widehat{\bar\mu}_{t,k}(x)-\bar\mu_t(x)$ into five terms, which we also call (I)–(V). The calculation that verifies (C.4) uses only the linearity of sample averages and expectations, Lemma B.1(3), and $\mathbb E[\mu_t(S,x)\mid T=t,x,E]=\bar\mu_t(x)$, so it applies to fold $k$ without change, and all expectations are finite on the event of Appendix C.2.1.

Averaging these five terms over $k=1,\ldots,K$ gives $\widehat{\bar\mu}_t(x)-\bar\mu_t(x)$ on the left-hand side. Term (I) is the sample average over fold $k$ of the function $f(Z_E)=\mu_t(S,x)-\bar\mu_t(x)-\nu_t(S,x)\omega_t^\nu(M,x)+\pi_t(S,x)\omega_t^\pi(M,x)$ of Appendix C.1.3, which has mean zero and a finite second moment, so by Lemma C.5 the average of term (I) over $k$ equals the first sum in (C.3) plus $O_p(N^{-1})$. Term (II) is the sample average over fold $k$ of the function $g(Z_O)=\mathbf 1\{T=t\}\{Y\omega_t^\nu(M,x)-\omega_t^\pi(M,x)\}$ of Appendix C.1.3, so by the observational version of Lemma C.5 its average over $k$ equals the second sum in (C.3) plus $O_p(N^{-1})$. Both sums are $O_p(N^{-1/2})$. The three lines of term (III) are, up to sign, the three quantities in Lemma C.4(1)–(3), and the two lines of term (IV) are, up to sign, the two quantities in Lemma C.4(4)(5), so terms (III) and (IV) are $o_p(N^{-1/2})$.

Term (V) is the quantity obtained by replacing the two sample averages in (3.9) by expectations and subtracting $\bar\mu_t(x)$. On the event of Appendix C.2.1, (E5) and (E6) make the premises of Lemma C.1 hold for the estimates that leave out fold $k$, for every set of sample values. Since the proof of Lemma C.1 consists of calculations that treat the estimates as fixed functions, term (V) equals the sum of the three expectations obtained from (C.1) by replacing $\widehat\pi_t,\widehat\nu_t,\widehat\mu_t,\widehat\omega_t^\nu,\widehat\omega_t^\pi$ by $\widehat\pi_t^{(-k)},\widehat\nu_t^{(-k)},\widehat\mu_t^{(-k)},\widehat\omega_t^{\nu,(-k)},\widehat\omega_t^{\pi,(-k)}$. By the bounds (C.7) and (C.8) of Appendix C.1.3, the absolute values of the three expectations are at most a constant times the three quantities of Assumption 3.5(2)–(4) written with the estimates that leave out fold $k$, which are $o_p(N^{-1/2})$ by Assumption 3.7. Since $K$ is fixed, the averages of terms (III), (IV), and (V) over $k$ are also $o_p(N^{-1/2})$, and (C.3) holds, with both sums on the right-hand side $O_p(N^{-1/2})$ as shown above. The argument of Appendix C.1.4 applied to this (C.3) gives the first convergence in (3.8) and $\widehat\tau-\tau=O_p(N^{-1/2})$ for the cross-fitted estimator, and the latter is used below to bound term (IV) of (C.17).

Finally, we show (C.13). The part of the proof in Appendix C.1.5 that derives $\widehat\Sigma^2\to_p\Sigma^2$ from (C.13) holds without change. For the experimental sample, for observations $i\in\mathcal I_{E,k}$, the version of (C.15) of Appendix C.1.5 with $\widehat\mu_t,\widehat\nu_t,\widehat\pi_t,\widehat\omega_t^\nu,\widehat\omega_t^\pi$ replaced by $\widehat\mu_t^{(-k)},\widehat\nu_t^{(-k)},\widehat\pi_t^{(-k)},\widehat\omega_t^{\nu,(-k)},\widehat\omega_t^{\pi,(-k)}$ holds. We write $f_k$ for the function in square brackets in term (I) of (C.15). By $|\mathcal I_{E,k}|\le n_E$ and $\widehat P_k(T=t,X=x\mid E)\le1$,
\[
\begin{aligned}
&\frac1{n_E}\sum_{i\in\mathcal I_{E,k}}\mathbf 1\{T_i=t,X_i=x\}f_k(S_i,M_i)^2\\
&=\frac{|\mathcal I_{E,k}|}{n_E}\widehat P_k(T=t,X=x\mid E)\widehat{\mathbb E}_k[f_k(S,M)^2\mid T=t,x,E]\le\widehat{\mathbb E}_k[f_k(S,M)^2\mid T=t,x,E]
\end{aligned}
\]
As in Appendix C.1.5, we split the square of $f_k$ into the squares of three differences, use Lemma C.4(6)–(9) in place of Assumption 3.6(6)–(9), and apply the law of large numbers within fold $k$ to the sample average of $\omega_t^\nu(M,x)^2\mathbf 1\{|\omega_t^\nu(M,x)|>r\}$ in (C.16), which shows that $\widehat{\mathbb E}_k[f_k(S,M)^2\mid T=t,x,E]$ tends to zero in probability. The bounds on terms (II) and (III) of (C.15) are the same as in Appendix C.1.5, and summing over the finitely many $(t,x,k)$ gives the first equation of (C.13). For the observational sample, for observations $j\in\mathcal I_{O,k}$, the version of (C.17) with $\widehat\omega_t^\nu,\widehat\omega_t^\pi$ replaced by $\widehat\omega_t^{\nu,(-k)},\widehat\omega_t^{\pi,(-k)}$ holds, and the bound on term (IV) of (C.17) is the same as in Appendix C.1.5. We multiply the squares of terms (V) and (VI) of (C.17) by $\mathbf 1\{X_j=x\}$, sum over $j\in\mathcal I_{O,k}$, and divide by $n_O$. By $|\mathcal I_{O,k}|\le n_O$ and $\widehat P_k(X=x\mid O)\le1$, the result is at most
\[
2C^2\widehat{\mathbb E}_k[\mathbf 1\{T=t\}\{\widehat\omega_t^{\nu,(-k)}(M,x)-\omega_t^\nu(M,x)\}^2\mid x,O]+2\widehat{\mathbb E}_k[\mathbf 1\{T=t\}\{\widehat\omega_t^{\pi,(-k)}(M,x)-\omega_t^\pi(M,x)\}^2\mid x,O]
\]
with $t=1$ for term (V) and $t=0$ for term (VI), and it tends to zero in probability by Lemma C.4(10)(11). Summing over the finitely many $(x,k)$ gives the second equation of (C.13). Since (C.3) and (C.13) hold, the arguments of Appendix C.1.4 and of the first half of Appendix C.1.6 give the claims of Theorem 3.3 on the asymptotic distribution and the variance estimator. The efficiency claim of Theorem 3.3 is shown in Appendix C.1.6. $\square$

\section{Supplementary materials}\label{sec:D}

This appendix describes the two supplementary materials that accompany this paper, which will be made available at \url{https://kkawato.github.io/}.
Supplementary Material 1 treats the design of Remark 2.1, in which the two samples share a pretreatment proxy of the characteristic instead of a post-treatment measurement. It states the assumptions and the identification result for this design, gives the efficient influence function and an estimator without sample splitting, reports a simulation study, proves the identification result and the form of the efficient influence function, and verifies completeness in the additive-error case of its example and in its simulation design. It also gives conditions for identification when the two samples record both a pretreatment proxy and a shared post-treatment measurement.

Supplementary Material 2 describes how the estimators are computed. For the simulation studies of Section 5 and of Supplementary Material 1, it gives the population functions of the designs, the computation of the proposed and the comparison estimators in each replication, and the evaluation. For the empirical illustration of Section 6, it gives the selection rules, the construction of the samples, the computation of the estimators, and the assumptions that the construction satisfies. It also reports additional results, namely the penalties selected in the simulation study and in the empirical illustration, the ranges of the selection probabilities under the three selection rules, descriptive statistics of the Job Corps variables, the rank condition behind completeness in the empirical illustration, the results with two, four, and five earnings categories, and the results under two other selection rules. It also describes the computation of the bounds of Section 4, the influence functions of their ends, and the confidence intervals, and it compares the standard errors with the dispersion of the estimated ends in the simulation study. It ends with R code that implements the computation of the point estimators in the simulation studies and the empirical illustration.


\begin{thebibliography}{28}
\bibitem[Ai and Chen(2003)]{ref1}
Ai, C., and X. Chen (2003). Efficient estimation of models with conditional moment restrictions containing unknown functions. \emph{Econometrica} 71, 1795–1843.

\bibitem[Andrews(2017)]{ref2}
Andrews, D. W. K. (2017). Examples of $L^2$-complete and boundedly-complete distributions. \emph{Journal of Econometrics} 199, 213–220.

\bibitem[Ashenfelter(1978)]{ref3}
Ashenfelter, O. (1978). Estimating the effect of training programs on earnings. \emph{The Review of Economics and Statistics} 60(1), 47–57.

\bibitem[Athey, Chetty, and Imbens(2025)]{ref4}
Athey, S., R. Chetty, and G. W. Imbens (2025). The experimental selection correction estimator: Using experiments to remove biases in observational estimates. NBER Working Paper 33817. \url{https://www.nber.org/papers/w33817}

\bibitem[Athey et al.(2026)]{ref5}
Athey, S., R. Chetty, G. W. Imbens, and H. Kang (2026). The surrogate index: Combining short-term proxies to estimate long-term treatment effects more rapidly and precisely. \emph{The Review of Economic Studies} 93, 2284–2312.

\bibitem[Blundell, Chen, and Kristensen(2007)]{ref6}
Blundell, R., X. Chen, and D. Kristensen (2007). Semi-nonparametric IV estimation of shape-invariant Engel curves. \emph{Econometrica} 75, 1613–1669.

\bibitem[Carrasco, Florens, and Renault(2007)]{ref7}
Carrasco, M., J.-P. Florens, and E. Renault (2007). Linear inverse problems in structural econometrics: Estimation based on spectral decomposition and regularization. In J. J. Heckman and E. E. Leamer (eds.), \emph{Handbook of Econometrics}, Vol. 6B, 5633–5751. Elsevier.

\bibitem[Chen(2007)]{ref8}
Chen, X. (2007). Large sample sieve estimation of semi-nonparametric models. In J. J. Heckman and E. E. Leamer (eds.), \emph{Handbook of Econometrics}, Vol. 6B, 5549–5632. Elsevier.

\bibitem[Chen and Christensen(2018)]{ref9}
Chen, X., and T. M. Christensen (2018). Optimal sup-norm rates and uniform inference on nonlinear functionals of nonparametric IV regression. \emph{Quantitative Economics} 9, 39–84.

\bibitem[Chen and Pouzo(2012)]{ref10}
Chen, X., and D. Pouzo (2012). Estimation of nonparametric conditional moment models with possibly nonsmooth generalized residuals. \emph{Econometrica} 80, 277–321.

\bibitem[Chernozhukov et al.(2018)]{ref11}
Chernozhukov, V., D. Chetverikov, M. Demirer, E. Duflo, C. Hansen, W. Newey, and J. Robins (2018). Double/debiased machine learning for treatment and structural parameters. \emph{The Econometrics Journal} 21(1), C1–C68. \url{https://doi.org/10.1111/ectj.12097}

\bibitem[Cui et al.(2024)]{ref12}
Cui, Y., H. Pu, X. Shi, W. Miao, and E. Tchetgen Tchetgen (2024). Semiparametric proximal causal inference. \emph{Journal of the American Statistical Association} 119, 1348–1359.

\bibitem[Darolles et al.(2011)]{ref13}
Darolles, S., Y. Fan, J.-P. Florens, and E. Renault (2011). Nonparametric instrumental regression. \emph{Econometrica} 79, 1541–1565.

\bibitem[d'Haultfœuille(2010)]{ref14}
d'Haultfœuille, X. (2010). A new instrumental method for dealing with endogenous selection. \emph{Journal of Econometrics} 154, 1–15.

\bibitem[d'Haultfœuille(2011)]{ref15}
d'Haultfœuille, X. (2011). On the completeness condition in nonparametric instrumental problems. \emph{Econometric Theory} 27, 460–471.

\bibitem[Fan, Sherman, and Shum(2014)]{ref18}
Fan, Y., R. Sherman, and M. Shum (2014). Identifying treatment effects under data combination. \emph{Econometrica} 82, 811–822.

\bibitem[Galichon and Henry(2026)]{ref19}
Galichon, A., and M. Henry (2026). An econometrician's guide to optimal transport. arXiv:2604.04227v2. \url{https://arxiv.org/abs/2604.04227v2}

\bibitem[Heckman, LaLonde, and Smith(1999)]{ref20}
Heckman, J. J., R. J. LaLonde, and J. A. Smith (1999). The economics and econometrics of active labor market programs. In O. Ashenfelter and D. Card (eds.), \emph{Handbook of Labor Economics}, Vol. 3A, 1865–2097. Elsevier.

\bibitem[Horowitz(2011)]{ref21}
Horowitz, J. L. (2011). Applied nonparametric instrumental variables estimation. \emph{Econometrica} 79(2), 347–394.

\bibitem[Imbens and Manski(2004)]{ref22}
Imbens, G. W., and C. F. Manski (2004). Confidence intervals for partially identified parameters. \emph{Econometrica} 72, 1845–1857.

\bibitem[LaLonde(1986)]{ref23}
LaLonde, R. J. (1986). Evaluating the econometric evaluations of training programs with experimental data. \emph{American Economic Review} 76(4), 604–620.

\bibitem[Miao, Geng, and Tchetgen Tchetgen(2018)]{ref24}
Miao, W., Z. Geng, and E. J. Tchetgen Tchetgen (2018). Identifying causal effects with proxy variables of an unmeasured confounder. \emph{Biometrika} 105, 987–993.

\bibitem[Newey(1990)]{ref26}
Newey, W. K. (1990). Semiparametric efficiency bounds. \emph{Journal of Applied Econometrics} 5, 99–135.

\bibitem[Newey and Powell(2003)]{ref28}
Newey, W. K., and J. L. Powell (2003). Instrumental variable estimation of nonparametric models. \emph{Econometrica} 71, 1565–1578.

\bibitem[Schennach(2016)]{ref29}
Schennach, S. M. (2016). Recent advances in the measurement error literature. \emph{Annual Review of Economics} 8, 341–377.

\bibitem[Schochet, Burghardt, and Glazerman(2001)]{ref30}
Schochet, P. Z., J. Burghardt, and S. Glazerman (2001). \emph{National Job Corps Study: The Impacts of Job Corps on Participants' Employment and Related Outcomes}. Final report, June 2001. Mathematica Policy Research.

\bibitem[Tchetgen Tchetgen et al.(2024)]{ref31}
Tchetgen Tchetgen, E. J., A. Ying, Y. Cui, X. Shi, and W. Miao (2024). An introduction to proximal causal inference. \emph{Statistical Science} 39, 375–390.

\bibitem[van der Vaart(1998)]{ref32}
van der Vaart, A. W. (1998). \emph{Asymptotic Statistics}. Cambridge University Press.
\end{thebibliography}
\end{document}